%% file: main.tex
\documentclass[final,5p,times,twocolumn]{elsarticle}

\newif\ifanonymous
\anonymousfalse  % Set to \anonymoustrue for anonymous submission

\newif\ifdedicatedtitlepage
\dedicatedtitlepagefalse  % Set to \dedicatedtitlepagetrue for dedicated title page

\newif\ifaiuse
\aiusefalse  % Set to \aiusefalse to hide the AI declaration

\usepackage{xcolor}
\usepackage[utf8]{inputenc}
\usepackage[T1]{fontenc}
\usepackage[english]{babel}
\usepackage{csquotes}
\usepackage{graphicx}
\usepackage{svg}
\usepackage{epstopdf}
\usepackage[final]{microtype}
\usepackage{etoolbox}
\usepackage{catchfile}
\usepackage{booktabs}
\usepackage{array}
\usepackage{tabularx}
\usepackage{multirow}
\newcolumntype{L}[1]{>{\raggedright\arraybackslash}p{#1}}
\newcolumntype{Y}{>{\raggedright\arraybackslash}X}
\usepackage{rotating}
\usepackage{subfiles}  
\usepackage[backref=page]{hyperref}
\usepackage{placeins}
\usepackage{caption}
\usepackage{cuted}
\usepackage{algorithm}
\usepackage{amsmath}
\usepackage{algpseudocode}
\usepackage{amssymb}
\usepackage{bm}
\usepackage{subcaption}
\usepackage{tikz}
\usetikzlibrary{shapes.geometric, arrows.meta, positioning, fit, calc}
\usepackage{sansmath}
\usepackage{ragged2e}
\usepackage{enumitem}
\usepackage{pifont}
\usepackage{dirtree}
\usepackage[nolist]{acronym}
\usepackage{stfloats} % For bottom float support in two-column
\input{acronyms}

\setlist[itemize]{before=\RaggedRight, parsep=0pt, itemsep=2pt}
\setlist[enumerate]{before=\RaggedRight, parsep=0pt, itemsep=2pt}

\definecolor{SNcolor}{RGB}{0,0,0}         % black       — SN ON
\definecolor{NoSNcolor}{RGB}{230,159,0}   % amber/gold  — SN OFF

\newcommand{\cmark}{\ding{51}}
\newcommand{\xmark}{\ding{55}}
\newcommand{\pmark}{\raisebox{0.1pt}{\scriptsize(\ding{51})}}

\newcommand{\Cdi}{C_{\mathrm{di}}}
\newcommand{\Cwx}{C_{\mathrm{wx}}}
\newcommand{\Cbb}{C_{\mathrm{bb}}}
\newcommand{\Cboxpl}{C_{\text{box-pl}}}
\newcommand{\Clift}{C_{\mathrm{lift}}}

\newcommand{\Omass}{O_{\mathrm{mass}}}

\newcommand{\lperf}{\lambda_{\mathrm{perf}}}
\newcommand{\lDPP}{\lambda_{\mathrm{DPP}}}
\newcommand{\ldi}{\lambda_{\mathrm{di}}}
\newcommand{\lwx}{\lambda_{\mathrm{wx}}}
\newcommand{\lbb}{\lambda_{\mathrm{bb}}}
\newcommand{\llift}{\lambda_{\mathrm{lift}}}
\newcommand{\lboxpl}{\lambda_{\mathrm{box-pl}}}

\newsavebox{\snRightBox}

\begin{document}

\begin{frontmatter}

%% Title
\title{GLUE: Coordinating Pre-Trained Generative Models for System-Level Design}

%% Authors
\ifanonymous
    \author{A. Nonymous\corref{cor1}}
    \author{R. E. Dacted}
    \author{N. O. Body}
    \cortext[cor1]{Corresponding author (location unknown)}
    \affiliation{organization={Department of Engineering, University of [REDACTED]},
                addressline={[REDACTED]},
                postcode={[REDACTED]},
                city={[REDACTED]},
                country={Earth (probably)}}
\else
    \author[eth]{Tim Aebersold\corref{cor1}}
    \ead{taebersold@ethz.ch}

    \author[eth]{Soheyl Massoudi}
    \ead{smassoudi@ethz.ch}

    \author[eth]{Mark D. Fuge}
    \ead{mafuge@ethz.ch}

    \cortext[cor1]{Corresponding author}

    \affiliation[eth]{organization={Department of Mechanical and Process Engineering, ETH Zurich},
                addressline={Tannenstrasse 3},
                postcode={8092},
                city={Zurich},
                country={Switzerland}}
\fi

\ifdedicatedtitlepage
    %% No abstract/keywords on title page
\else
    %% Abstract
    \subfile{sections/00_abstract}
\fi

%% Keywords 

\end{frontmatter}

\ifdedicatedtitlepage
    \clearpage

    %% Abstract on main content page
    \subfile{sections/00_abstract}
    \vspace{0.5cm}

    \twocolumn  % Force restart of two-column layout on new page
\fi

\setlength{\stripsep}{-20pt}
\begin{strip}
\centering
\includegraphics[width=0.85\textwidth]{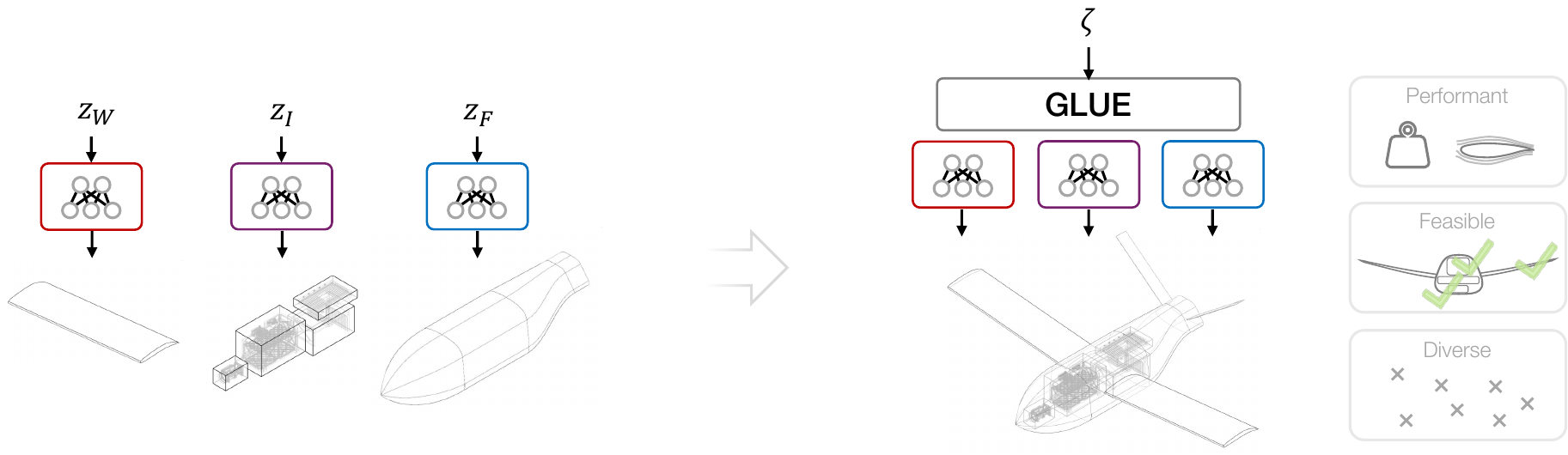}
\vspace{0.4cm}
\captionof{figure}[GLUE methodology overview]{\textbf{Given}: Frozen, pre-trained specialized models with latents $z_i$. 
\textbf{Contribution}: Coordination with \ac{GLUE} models from system-level latent $\zeta$.}
\label{fig:intro}
\vspace{30pt}
\end{strip}

%% Main Content

\subfile{sections/01_introduction}

\subfile{sections/02_relatedwork}

\subfile{sections/03_methodology}

\subfile{sections/04_results}

\subfile{sections/05_conclusion}

% Code and Data Availability
\section*{Code and Data Availability}
 
\ifanonymous
    \noindent{Code and data will be made publicly available upon acceptance.}
\else
    \noindent{Our code will be made publicly accessible shortly under:\\}
    \noindent{\url{github.com/IDEALLab/glue}}
\fi

\ifaiuse
\section*{AI Use Declaration}
During the preparation of this work, the authors made targeted use of large language model (LLM)--based tools to assist coding and manuscript preparation. 
All AI-generated work was carefully reviewed and edited by the authors, who take full responsibility for the content of the published article.
\fi

%% Bibliography
%% Standard Elsevier style
\FloatBarrier
\bibliographystyle{elsarticle-num}
\bibliography{references}
\FloatBarrier
\newpage

%% Appendices
\appendix

\subfile{appendix/dfjm_hyperparameters}

\subfile{appendix/subsystem_models}

\subfile{appendix/architecture_details}

\subfile{appendix/aalm_scheme}

\subfile{appendix/uav_design_problem}

\subfile{appendix/dpp_loss}

\subfile{appendix/aircraft_cases}

\subfile{appendix/benchmarking_all_cases}

\subfile{appendix/qualitative_comparison}

\subfile{appendix/opt_algos}

\subfile{appendix/ddjm}

\subfile{appendix/geometry_layer}

\end{document}

%% file: acronyms.tex
\begin{acronym}[MDD-GAN] % Use longest acronym for proper spacing
    
\acro{2D}{Two-Dimensional}
\acro{3D}{Three-Dimensional}
\acro{ALM}{Augmented Lagrangian Method}
\acro{BO}{Bayesian Optimization}
\acro{CAD}{Computer-Aided Design}
\acro{CF}{Common Features}
\acro{CFD}{Computational Fluid Dynamics}
\acro{CMA-ES}{Covariance Matrix Adaptation Evolution Strategy}
\acro{CPU}{Central Processing Unit}
\acro{cVAE}{Conditional Variational Autoencoder}
\acro{DDPM}{Denoising Diffusion Probabilistic Model}
\acro{DD}{Data-Driven}
\acro{DF}{Data-Free}
\acro{DPP}{Determinantal Point Process}
\acro{EMA}{Exponential Moving Average}
\acro{FEM}{Finite Element Method}
\acro{FLOP}{Floating Point Operation}
\acro{GAN}{Generative Adversarial Network}
\acro{GD}{Gradient Descent}
\acro{GLUE}{Generative Latent Unification of Expertise-Informed Engineering Models}
\acro{GNN}{Graph Neural Network}
\acro{GP}{Gaussian Process}
\acrodefplural{GP}[GPs]{Gaussian Processes}
\acro{GPU}{Graphics Processing Unit}
\acro{HPO}{Hyperparameter Optimization}
\acro{iGD}{inner Gradient Descent}
\acro{KL}{Kullback-Leibler}
\acro{LHS}{Latin Hypercube Sampling}
\acro{LLM}{Large Language Model}
\acro{LR}{Learning Rate}
\acro{MDD-GAN}{Multi-Discriminator Generative Adversarial Network}
\acro{MDO}{Multidisciplinary Design Optimization}
\acro{MI}{Mutual Information}
\acro{MLP}{Multi-Layer Perceptron}
\acro{MSE}{Mean Squared Error}
\acro{OT-GAN}{Optimal Transport Generative Adversarial Network}
\acro{PA-DPP}{Performance-Augmented Determinantal Point Process}
\acro{SDF}{Signed Distance Function}
\acro{SN}{Spectral Normalization}
\acro{TuRBO}{Trust Region Bayesian Optimization}
\acro{UAV}{Unmanned Aerial Vehicle}
\acro{VAE}{Variational Autoencoder}
\acro{WRB}{Wide Residual Block}

\end{acronym}

%% file: sections/00_abstract.tex
\begin{abstract}
    Engineering complex systems (aircraft, buildings, vehicles) requires coordinating geometric and performance couplings across subsystems. 
    As generative models proliferate for specialized domains, a key research gap is how to coordinate frozen, pre-trained submodels to generate full-system designs that are feasible, diverse, and high-performing.
    We introduce \ac{GLUE}, which orchestrates pre-trained, frozen generators while enforcing system-level feasibility, optimality, and diversity. Compatible models must be end-to-end differentiable with a smooth, well-behaved latent-to-output mapping.
    We propose and benchmark \textit{(i)} data-driven \ac{GLUE} models trained on pre-generated system-level designs and \textit{(ii)} a data-free \ac{GLUE} model trained on a differentiable geometry layer. 
    On a UAV design problem with five coupling constraints, we find that data-driven approaches yield diverse, high-performing designs but require large datasets to satisfy constraints reliably.
    The \textit{data-free} approach is competitive with Bayesian optimization and gradient-based optimization in performance and feasibility while training a full generative model in only $\sim$10 min on an RTX 4090 \acs{GPU}, requiring more than two orders of magnitude fewer geometry evaluations and \acsp{FLOP} than the data-driven method. We identify equality constraint satisfaction as a key difficulty and remaining limitation, and ablate approaches that improve this for the data-free approach.
    As a first step toward scaling generative design to complex, real-world engineering systems, this work explores how unmodified, domain-informed submodels can be integrated into a modular generative workflow.
\end{abstract}

%% file: sections/01_introduction.tex
\section{Introduction}

Generative models have been applied to engineering design problems ranging from architecture \cite{paraGenerativeLayoutModeling2020} to aerospace \cite{chenBezierGANAutomaticGeneration2021} and drug discovery \cite{sanchez-lengelingInverseMolecularDesign2018}. Once trained, their fast inference supports design-space exploration and decision-making in ways that are difficult to match with traditional optimization algorithms alone \cite{balmerBridgeVAE}. 
Many such models compress designs into interpretable, low-dimensional latent spaces ($z$), which makes high-dimensional trade spaces more tractable.

In practice, these generative models are almost always purpose-trained for a specific domain and often for a single component type (\textit{e.g.}, wings, fuselages, or structural parts). Domain-informed priors in the parametrization, architecture, and loss typically improve performance and validity on that narrow task. Yet real-world engineering problems (\textit{e.g.}, an aircraft) are inherently multi-component and multi-domain. This requires coordinating coupled interfaces, constraints, and objectives across subsystems: the system optimum rarely equals the sum of subsystems optimized in isolation.

Current approaches to multi-component systems largely rely on generative models trained end-to-end for the entire system, which enforces a single architecture and loss formulation. 
This makes it difficult to distribute work across teams or incorporate pre-trained third-party models. As generative modeling is applied to increasingly complex real-world engineering systems, we inevitably need methods to combine several pre-trained, specialized generative models (see Figure~\ref{fig:intro}). 
To preserve validity, certifications, or bounds provided by the original model creators, the pre-trained subsystem models ideally remain frozen. The critical gap is thus a coordination mechanism that operates on \textit{pre-trained} domain-informed subsystem models \textit{without modification}. 
It should expose a system-level latent $\zeta$ that offers a low-dimensional, human-interpretable space for design exploration, and it should synthesize designs that satisfy three key criteria:

\begin{itemize}[label={-}]
    \item \textit{Feasibility}: Satisfy all coupling constraints and system-level requirements within engineering tolerances.
    \item \textit{System-level optimality}: Optimize performance objectives across the entire system.
    \item \textit{Diversity}: Generate varied designs across the feasible design space. 
\end{itemize}

\paragraph{Key Contributions}

To address this coordination challenge, we make the following contributions.

\begin{enumerate}
    \item We introduce \ac{GLUE} models, which orchestrate pre-trained, frozen generative models across subsystems, exposing a low-dimensional system-level latent space $\zeta$.
    We propose and compare two training paradigms:
    \begin{itemize}[label={-}]
        \item \textit{\ac{DD}} \ac{GLUE} models train on feasible, high-performing designs generated by conventional optimization algorithms.
        \item \textit{\ac{DF}} \ac{GLUE} models learn orchestration directly by backpropagating gradients from a differentiable, parallelizable geometry layer.
    \end{itemize}
    \item We benchmark both paradigms on a multi-component \ac{UAV} design problem, demonstrating that \ac{DF}-\ac{GLUE}:
    \begin{enumerate}[label=\textit{\alph*)}]
        \item Learns constraints with over two orders of magnitude greater computational efficiency than \ac{DD}-\ac{GLUE}.
        \item Is near-competitive with gradient-based and Bayesian optimization in optimality and feasibility while being more robust to local minima.
        \item Enables explicit, user-controlled trade-offs between diversity and optimality.
    \end{enumerate}
    \item For \ac{DF}-\ac{GLUE}, we ablate strategies for improved satisfaction of (inherently hard-to-learn) equality constraints.
\end{enumerate}

%% file: sections/02_relatedwork.tex
\section{Related Work}
\label{section:relatedwork}

We \textit{(i)} briefly introduce generative modeling for engineering design, including \acp{UAV}, \textit{(ii)} draw parallels between the architectures used in \ac{MDO} and the current state of generative modeling for multi-domain systems, and \textit{(iii)} examine training strategies that target feasible, performant, and diverse designs in generative models for engineering.

\subsection{Generative Models for Engineering Design}

Generative models have shown promising results across a wide range of engineering design applications~\cite{regenwetterDeepGenerativeModels2022}.
Here, we distinguish between two roles of models.
\textit{Neural surrogate models} predict performance from geometry, replacing costly simulations with near-instant evaluation~\cite{zuoFastSparseFlow2022,karaliDesignDeepLearning2020,taoApplicationDeepLearning2019, regenwetter2023framed}, and have been extensively applied to \ac{MDO} \cite{benamaraMultifidelityPODSurrogateassisted2017,kontogiannisGeneralizedMethodologyMultidisciplinary2020,yaoSurrogateBasedMultistagemultilevel2012,massoudiIntegratedApproachDesigning2024}. However, surrogates lack two key advantages of generative models: \textit{(i)} fast inverse design to explore trade spaces and \textit{(ii)} a structured, interpretable, low-dimensional latent space. In this paper, we focus solely on \textit{generative models}, which directly create novel geometries based on random noise, performance requirements or user-specified conditions.

In engineering, design problems are usually high-dimensional (\textit{i.e.}, a design is described by a large number of variables).
\acsp{GAN}~\cite{goodfellowGenerativeAdversarialNetworks2014} and \acsp{VAE} \cite{kingmaAutoEncodingVariationalBayes2022}, applied extensively in engineering \cite{chenBezierGANAutomaticGeneration2021, cobbDiverseSystemLevelDesign2022,regenwetterDeepGenerativeModels2022}, address this by compressing design features into interpretable, low-dimensional latent spaces. 
Further improvements include latent-space disentanglement~\cite{CompressingLatentSpace2023,chenInfoGANInterpretableRepresentation2016} and optimal-transport \acsp{GAN}~\cite{genevayLearningGenerativeModels2017,arjovskyWassersteinGAN2017} to reduce adversarial training instability.
More recently, diffusion models~\cite{hoDenoisingDiffusionProbabilistic2020} have shown promising results in engineering~\cite{mazeDiffusionModelsBeat2023, dinizOptimizingDiffusionDiffuse2024}.
For an overview of further model architectures and applications to engineering, we refer to \cite{regenwetterDeepGenerativeModels2022}.
We build on these prior contributions, both for our subsystem generators (Section~\ref{sec:subsystem-models}, \ref{ap:subsystem_models}) and for the data-driven \acs{GLUE} variants we evaluate (Section~\ref{sec:method-datadriven}).

\subsection{Multidisciplinary Design Optimization}
\label{sec:mdo}

Extending generative modeling to multi-component systems places our work within the scope of \ac{MDO}, originally popularized by Cramer \textit{et al.}~\cite{cramerProblemFormulationMultidisciplinary1994} and Sobieszczanski-Sobieski~\cite{sobieszczanski-sobieskiMultidisciplinaryDesignOptimization1995}. \ac{MDO} coordinates coupled disciplines so that components are co-optimized rather than tuned in isolation, and has been widely applied \cite{martinsMultidisciplinaryDesignOptimization2013}, including to \ac{UAV} design \cite{waltherIntegrationAspectsCollaborative2020,kimOPENVSPBASEDAERODYNAMIC}.

\subsubsection{Monolithic and Distributed MDO}
Martins and Lambe~\cite{martinsMultidisciplinaryDesignOptimization2013} contrast \textit{monolithic} \ac{MDO} (all design variables optimized within a single framework) with \textit{distributed} \ac{MDO} (decomposition into discipline-level subproblems with explicit or implicit coordination). We can use this distinction to draw direct parallels for generative models.

\subsection{Monolithic and Distributed Generative Modeling}

\label{sec:dist-advantages}
 
\begin{figure}[t!]
    \centering
    \begin{subfigure}{0.45\columnwidth}
        \centering
        \includegraphics[width=\linewidth]{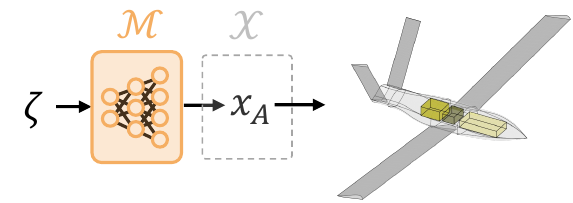}
        \vspace{0.5mm}
        \caption{Monolithic}
        \label{fig:monolithic_approach}
    \end{subfigure}
    \hfill
    \begin{subfigure}{0.48\columnwidth}
        \centering
        \includegraphics[width=\linewidth]{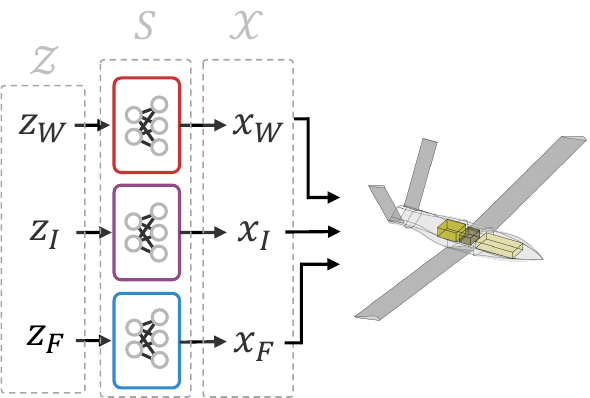}
        \caption{Distributed}
        \label{fig:distributed_approach}
    \end{subfigure}
    \caption[Monolithic \textit{vs.} distributed generative modeling]{Monolithic and distributed generative modeling. Approach (b) enables specialized models but requires latent coordination.}
    \label{fig:monolithic_vs_distributed}
\end{figure}

We interpret this analogy as follows (see Figure~\ref{fig:monolithic_vs_distributed}). A \textit{monolithic generative model} is a \textit{single} model trained on the full design space across all disciplines and components.
For example, Cobb \textit{et al.}~\cite{cobbDiverseSystemLevelDesign2022} train a single \acs{VAE} for an entire race car. So far, for multi-component systems, the overwhelming focus has been on this approach~\cite{parrottMachineLearningSurrogates2023, regenwetterBikeBenchBicycleDesign2025}.

In contrast, the \textit{distributed} generative approach employs \textit{multiple} generative models for individual subsystems or disciplines, then coordinates between them.
Many existing models, \textit{e.g.}, for airfoils~\cite{chenBezierGANAutomaticGeneration2021, dinizOptimizingDiffusionDiffuse2024} or structural components~\cite{mazeDiffusionModelsBeat2023, berzinsGeometryInformedNeuralNetworks2024}, can be interpreted as subsystem models ($S$ in Figure~\ref{fig:monolithic_vs_distributed}). The distributed approach offers two major advantages:

\begin{enumerate}
    \item \textit{Domain-Specific Priors and Architectures.}
Specialized models embed domain knowledge in architectures, losses, or parametrization~\cite{chenBezierGANAutomaticGeneration2021,dinizOptimizingDiffusionDiffuse2024,berzinsGeometryInformedNeuralNetworks2024}, which can improve generative model performance~\cite{feltenEngiBenchFrameworkDataDriven2025}.
Distributed modeling retains these domain-informed tweaks for each submodel.
In contrast, monolithic generative approaches are tied to a single architecture and loss formulation, which is especially restrictive when subsystems require different modeling approaches.

\item \textit{Organizational Advantages.} Industrial practice typically distributes design tasks to specialized engineering groups that retain control over their procedures and in-house expertise~\cite{Kroo1997LargeScaleMDO, martinsMultidisciplinaryDesignOptimization2013}. Distributed generative modeling naturally aligns with this practice: specialized teams independently train domain-specific subsystem models without having to agree on a single architecture or training recipe, and can reuse third-party models whose data and training costs were borne elsewhere.

\end{enumerate}

For these two reasons, we believe the distributed generative modeling approach is critically underappreciated. 

\subsubsection{Existing Distributed Modeling}
Earlier approaches \cite{chenSynthesizingDesignsInterpart2019, dongOneMoreContextual2025, talabot2025partsdf} to distributed generative modeling require all subsystem models to be trained jointly, meaning one cannot leverage the advantages of distributed modeling described in Section \ref{sec:dist-advantages}. Instead, here we assume \textit{pre-trained} generative submodels $S_i$ and focus on coordination \textit{without modification}. The resulting system-level designs must be system-level feasible, optimal, and diverse. To this end, we now examine three training regimes. 

\subsubsection{Data-Driven Training}
\label{sec:data_driven_training}

Findings from recent work \cite{regenwetterBikeBenchBicycleDesign2025, chenSynthesizingDesignsInterpart2019} indicate that purely data-driven models can struggle with constraint satisfaction. A partial remedy is to inject labeled negative data, \textit{i.e.}, designs that violate constraints. Regenwetter \textit{et al.} show that this approach can improve the feasibility of synthesized designs \cite{regenwetterConstrainingGenerativeModels2024}.

\subsubsection{Scoring-Augmented Data-Driven Training}

Scoring-augmented approaches extend data-driven training with score-based loss terms targeting performance, diversity, or feasibility. Scores come from closed-form checks, surrogate models, or full simulators. For example, Chen and Ahmed extend on \ac{DPP}-based diversity scoring \cite{GDPPLearningDiverseGenerations2019} to generate novel, high-performing designs \cite{chenPaDGANLearningGenerate2020, chenMOPaDGANReparameterizingEngineering2021}. Regenwetter \textit{et al.} show that such scoring-augmented generative models beat purely data-driven ones in validity and optimality and start to close the gap to optimizers like NSGA-II~\cite{debFastElitistMultiobjective2002} and EPO \cite{mahapatraMultiTaskLearningUser}.

\subsubsection{Data-Free Training}

More recently, fully data-free methods train generative models purely from scoring (without using any data), \textit{e.g.}, for topology optimization \cite{chandrasekhar2021tounn, joglekar2024dmf}.
Most relevant to our work, Berzins \textit{et al.} applied this to single-part shape generation~\cite{berzinsGeometryInformedNeuralNetworks2024}. To balance constraint satisfaction, optimality, and diversity terms in a single loss, they use adaptive \ac{ALM} \cite{basirAdaptiveAugmentedLagrangian2023}, which replaces hand-tuned per-constraint penalties with three global hyperparameters $(\alpha,\gamma,\varepsilon)$. The \ac{ALM} schemes we used are summarized in \ref{ap:alm_schemes}.

\subsection{Differentiable Scoring}
For scoring-augmented and data-free training, differentiable scores are especially attractive because they provide gradients for backpropagation in training. Diversity scores such as \ac{DPP} \cite{chenPaDGANLearningGenerate2020} are differentiable when implemented carefully. 
For feasibility and performance, two main routes exist: \textit{(i)} differentiable surrogate models, which require substantial training data but are broadly applicable, and \textit{(ii)} native differentiable simulators. The latter have been extensively explored, \textit{e.g.}, in aerodynamics (SU2 \cite{economonSU2OpenSourceSuite2016}, DAFoam \cite{heAerodynamicDesignOptimization2018,heDAFoamOpenSourceAdjoint2020}), solid mechanics (JAX-\ac{FEM} \cite{xueJAXFEMDifferentiableGPUaccelerated2023}), and rigid-body dynamics (JaxSim \cite{ferretti_accelerated_optimization_2025}, MuJoCo MJX \cite{todorov2012mujoco}). 

Additionally, differentiable B-Rep--based \ac{CAD} tools have recently emerged \cite{banovicAlgorithmicDifferentiationOpen2018,banovic2024pythonocc_ad,cascaval2022differentiable,prasad2022nurbsdiff}, alongside growing interest in neural implicit \acp{SDF} as flexible, robust, and differentiable \ac{CAD} representations \cite{park2019deepsdf,hao2020dualsdf,vasu2022hybridsdf}. 
The methods proposed here are in principle agnostic to \ac{CAD} representation type.

%% file: sections/03_methodology.tex
\section{Methodology}
\label{section:methodology}

First, we provide a brief overview of our multi-component \ac{UAV} test case and its constituent pre-trained subsystem models. Then, we detail the baseline optimization algorithms (I.) and data-driven (II.) and data-free (III.) \ac{GLUE} training paradigms.

\subsection{UAV Design Problem}

We use a fixed-wing \ac{UAV} design problem as our case study. With coupled multi-component subsystems and high-dimensional design spaces, \ac{UAV} design captures key coordination challenges of multi-component engineering design. The aerospace domain also offers extensive generative model literature \cite{chenBezierGANAutomaticGeneration2021, dinizOptimizingDiffusionDiffuse2024, heyraninobariRANGEGANDesignSynthesis2021, cobbDesignUnmannedAir2022, sungBlendedNetBlendedWing2025}, data, and benchmarks, and \acp{UAV} are small and affordable enough for eventual real-world validation.

We use a classical \ac{UAV} decomposition into fuselage (outer skin), wings (lifting surfaces), and internal components. 
Internals are approximated by axis-aligned bounding boxes as placeholders for subsystems, with detailed component models (battery, electronics, \textit{etc.}) left to future work.  

We formulate the \ac{UAV} design task as a constrained optimization problem in a unified latent space $\mathcal{Z} \subset \mathbb{R}^{22}$ that collects all subsystem latents and coupling variables (see Figure~\ref{fig:explainer_graph}). The full design space $\mathcal{X} \subset \mathbb{R}^{324}$ contains the full aircraft geometry parameters. Subsystem models $S_i : \mathcal{Z}_i \to \mathcal{X}_i$ and the differentiable geometry layer $\mathcal{G} : \mathcal{X} \to \mathbb{R}^{n_o + n_c}$ map latents to a single mass objective ($n_o = 1$) and constraints ($n_c = 5$). See \ref{ap:uav_design_problem} for a detailed account of $\mathcal{Z}$ and $\mathcal{X}$ and the formal problem statement. Performance issues related to aerodynamic stability and aircraft dynamics are omitted in this work. 

The constraints are: $\textit{(i)}$ wing--fuselage interface fit ($\Cbb$), $\textit{(ii)}$ consistent wing orientation (dihedral: $\Cdi$) and $\textit{(iii)}$ position ($\Cwx$) on the fuselage, $\textit{(iv)}$ all internals contained inside the fuselage ($\Cboxpl$), and $\textit{(v)}$ sufficient lift in level flight ($\Clift$). 
Note that all five constraints require explicit coordination between subsystems. 
Additionally, the objective (minimize $\Omass$) dictates that the fuselage must wrap tightly around the internals without violating $\Cboxpl$. 
This implies that internals need to be arranged in a way that allows for a small fuselage. This coupling yields a nonconvex objective landscape with numerous local minima.

\subsection{Subsystem Models}
\label{sec:subsystem-models}

\begin{table}[!htbp]
\centering
\caption[Overview of subsystem models]{High-level comparison of subsystem models.}
\label{tab:subsystem_overview}
\footnotesize
\begin{tabular}{llll}
\toprule
 & \textbf{Wing} / \textbf{Airfoil} \cite{chenBezierGANAutomaticGeneration2021} & \textbf{Fuselage} & \textbf{Internals} \\
\midrule
\textbf{Model} & MLP / CEBGAN & VAE & Autoreg. MLP \\
\textbf{\boldmath $\text{Dim}(c_i)$} & 1 / 3 & - &  $N_{box}$ \\
\textbf{\boldmath $\text{Dim}(z_i)$} & 2 / - & 4 & 4 \\
\textbf{\boldmath $\text{Dim}(x_i)$} & 3 / 385 & 15 & $6 \times N_{box}$ \\
\midrule
\textbf{Losses} & \begin{tabular}[t]{@{}l@{}}Lift Accuracy\\Diversity\\ / \\OT \cite{genevayLearningGenerativeModels2017,feydyInterpolatingOptimalTransport} \end{tabular} & \begin{tabular}[t]{@{}l@{}}Recon.\\KL~Div.\\LV \cite{CompressingLatentSpace2023}\end{tabular} & \begin{tabular}[t]{@{}l@{}}Overlap\\Compactness\\Accurate Vol.\\Adjacent Div.\end{tabular} \\
\midrule
\textbf{Data} &  Online / 995 (Opt.) & 551 (Feas.) &  Online \\
\midrule
\textbf{Cleanup} &  - / - & - & Grad. Desc. \\
\bottomrule
\end{tabular}
\end{table}

For the wing, we wrap the airfoil generator of Chen \textit{et al.} \cite{chenBezierGANAutomaticGeneration2021} in a higher-level wing model. 
For the fuselage and internals, we use custom models to directly match our use case. 
Table~\ref{tab:subsystem_overview} summarizes the three subsystem models. 
They differ substantially to address the unique challenges of each subsystem. 
Complete descriptions are provided in \ref{ap:subsystem_models}.

Here, these subsystem models are \textit{pretrained} once and treated as \textit{frozen} during coordination. 
Our method can interface with pretrained subsystem models \textit{without} knowledge about their architectures, training data, or domain-specific tweaks. 
However, we require differentiability (support for automatic differentiation) and a smooth, continuous mapping $z_i \to x_i$ (more on this in Section~\ref{sec:smoothness-ablation}).
Table~\ref{tab:compat} classifies common subsystem model and evaluator classes by their compatibility with \ac{DF}-\ac{GLUE}.

\input{tables/compatibility}

\input{figures/explainers/explainer_graph.tex}

\subsection{Geometry Layer}
All geometry-layer functionality is implemented in PyTorch \cite{paszkePyTorchImperativeStyle2019}, enabling native parallel execution and automatic differentiation. 
Details are given in \ref{ap:geometry_layer}. The current implementation is tailored to this aircraft parametrization.
We now discuss three methods for latent space coordination in detail. 

\subsection{Method I: Optimization Algorithms}
\label{sec:method_i_optimization}

Optimization algorithms correspond to Approach I in Figure~\ref{fig:explainer_graph}, optimizing directly in the unified latent space $\mathcal{Z}$.
These optimization baselines establish the performance ceiling that data-driven and data-free \ac{GLUE} models are benchmarked against. They also serve as sources of training data for the data-driven \ac{GLUE} models.
We explicitly avoid optimization in the full design space $\mathcal{X}$ (324-D), which would correspond to the monolithic setting we aim to avoid. We consider two optimizer classes.

\subsubsection{Bayesian Optimization}

\ac{BO} is robust to local minima and hence an attractive benchmark for both \ac{GLUE} training regimes.
However, vanilla \ac{BO} scales poorly to high dimensions, which may bias benchmarking toward \ac{GLUE}.
Specialized variants~\cite{lethamReExaminingLinearEmbeddings,erikssonHighDimensionalBayesianOptimization2021} address this via search-space compression, but we expect little benefit in our already compressed $\mathcal{Z}$. We therefore use \ac{TuRBO}~\cite{erikssonScalableGlobalOptimization2020}, which builds local trust regions with lightweight \acp{GP}. 

Additionally, convergence slows sharply as we add constraints, likely because the feasible hypervolume shrinks multiplicatively with each new constraint. For our five-constraint aircraft problem, standard \ac{TuRBO} is effectively unusable. To make \ac{BO} more competitive, we use \ac{iGD} for equality constraints with known convex structure ($\Cwx$, $\Cdi$, $\Clift$), yielding {TuRBO-iGD}. This accelerates convergence because \ac{TuRBO} now only enforces two remaining constraints and operates in a lower-dimensional subspace (15-D vs.\ 22-D). More details can be found in \ref{ap:opt_algos}.

% \subsubsection{Evolutionary Algorithms}

% Since there are many evolutionary algorithms, we restrict attention to \ac{CMA-ES}, which Hansen \textit{et al.} identify as a top performer on a 20D problem similar to ours~\cite{hansen2010comparing}. On our constrained, strongly coupled \ac{UAV} problem, however, \ac{CMA-ES} shows very slow convergence and fails to produce competitive designs. This is consistent with results by Omidvar and Li \cite{omidvar2010comparative}, who report substantial performance degradation of \ac{CMA-ES} (and other evolutionary schemes) with increasing dimensionality.

\subsubsection{Gradient Descent}

With this method, we perform gradient descent in the unified latent space $\mathcal{Z}$ using derivatives from the differentiable geometry layer. We minimize:

\begin{equation}
\mathcal{L}
= O(z)
+ \sum_{k=1}^{n_c} \left( \lambda_k\, C_k(z)
+ \frac{\mu_k}{2}\, C_k(z)^2 \right)
\end{equation}

with $\lambda_k$ and $\mu_k$ updated based on constraint violations $C_k$ (see \ref{ap:alm_schemes}). Here, $O(z)$ is simply $\Omass$. Among all optimizers tested, \ac{ALM}-\ac{GD} achieves the best combination of feasibility and mass, but remains sensitive to local minima, leading to a large set of diverse, locally optimal designs. Hyperparameters are hand-tuned to balance optimality, constraint satisfaction, and compute cost. Full settings and convergence criteria are given in \ref{ap:opt_algos}.
We now turn to the \ac{GLUE} coordination models themselves.

\subsection{Method II: Data-Driven \acs{DD}-\acs{GLUE}}
\label{sec:method-datadriven}

We propose a \ac{GLUE} model $\Phi$ that maps a system-level latent $\zeta$ to subsystem latents ${z_i}$ and coupling variables $x_C$. Here, $\zeta$ is the unified latent code for the full aircraft. \ac{DD}-\ac{GLUE} corresponds to Approach~II in Figure~\ref{fig:explainer_graph}. To span common model architectures, we employ four data-driven \ac{GLUE} models, the first two of which leverage negative data \cite{regenwetterConstrainingGenerativeModels2024}: 

\begin{description}

\item[\textbf{cVAE}]  A Conditional \acs{VAE} (\acs{cVAE}) with label conditioning.

\item[\textbf{MDD-GAN}] A Multi-Discriminator \acs{GAN} (\acs{MDD-GAN}) with a 3-class discriminator (fake, feasible, infeasible)~\cite{regenwetterConstrainingGenerativeModels2024}.

\item[\textbf{OT-GAN}] An \acs{OT-GAN} using Sinkhorn divergences~\cite{genevayLearningGenerativeModels2017}.

\item[\textbf{DDPM}] A \acs{DDPM} with a cosine noise schedule~\cite{hoDenoisingDiffusionProbabilistic2020}.

\end{description}

For implementation details, see \ref{ap:ddjm}. 
For fair comparison, all data-driven models have similar or greater capacity than the 4.7M-parameter data-free \ac{GLUE} model. For details on the \ac{DD}-\ac{GLUE} models, see Table~\ref{tab:sota_hyperparameters}.

\subsubsection{Training Data and Augmentation}

We train \ac{DD}-\ac{GLUE} on designs optimized with \ac{ALM}-\ac{GD}, as it is the most compute-efficient of the tested optimizers and yields high-quality, feasible samples. To improve constraint learning efficiency, the dataset is augmented with Gaussian jitter $\epsilon \sim \mathcal{N}(0,\sigma_{\text{aug}}^2)$. Each perturbed sample is classified as feasible or infeasible via the geometry layer. The negative-data models (cVAE, MDD-GAN) use both labels during training. Augmentation parameters ($\sigma_{\text{aug}}$, dataset multiplier $N$) are tuned per model via hyperparameter optimization (see \ref{ap:ddjm}).

All data-driven \ac{GLUE} models are hyperparameter-optimized so that quality differences relative to the data-free approach are not due to poor tuning (see~\ref{ap:ddjm}).
As a third approach, we now describe training \ac{GLUE} directly on the differentiable geometry layer.

\subsection{Method III: Data-Free \acs{GLUE}}
\label{sec:df-glue}
The core idea of \ac{DF}-\ac{GLUE} is to sample in $\zeta$, propagate these samples through the \ac{GLUE} model $\Phi$, the frozen subsystem models $S_i$, and geometry layer $\mathcal{G}$, which evaluates performance, constraint, and diversity scores. These scores are then backpropagated directly into $\Phi$ (III in Figure~\ref{fig:explainer_graph}).
Note that \textit{data-free} refers to \ac{GLUE} training only, as the frozen subsystem models may or may not have been trained on data.

\paragraph{Loss Formulation}{
    
Equation~\ref{eq:dfglued-loss} shows the \ac{DF}-\ac{GLUE} training loss used in Algorithm~\ref{alg:glued_training}

\begingroup
\begin{align}
&\mathcal{L}_{\text{total}} = \lperf \left( \lambda_m \Omass + \lambda_d O_d \right) \nonumber \\
&\hspace{2pt} + \lbb \Cbb + \lwx \Cwx + \ldi \Cdi + \llift \Clift + \lboxpl \Cboxpl \nonumber   \\
&\hspace{2pt} + \lambda_{\mathrm{DPP},z_W} \text{DPP}(z_W) + \lambda_{\mathrm{DPP},z_F} \text{DPP}(z_F) + \lambda_{\mathrm{DPP},z_I} \text{DPP}(z_I) \nonumber \\
&\hspace{2pt} + \lDPP \text{DPP}(\mathbf{x}) \nonumber\\
&\hspace{2pt} + \lambda_{\mathrm{MI}} \mathcal{L}_{\text{MI}} 
\label{eq:dfglued-loss}
\end{align}
\endgroup

where $\lbb, \lwx, \ldi, \llift$, and $\lboxpl$ are constraint multipliers updated via an \ac{ALM} scheme (see \ref{ap:alm_schemes}). 
We use a mass-only objective with $\lambda_m = 1$ and $\lambda_d = 0$ (no drag). 
The optional mutual information loss $\mathcal{L}_{\text{MI}}$ is used only in ablations (Section~\ref{sec:smoothness-ablation}) and is not shown in Algorithm~\ref{alg:glued_training}.
Diversity is enforced via a \ac{DPP} loss, either in design space $\mathcal{X}$ with a single $\lDPP$ (benchmarking) or in latent space $\mathcal{Z}$ with separate $\lambda_{\mathrm{DPP},z_W}, \lambda_{\mathrm{DPP},z_F}, \lambda_{\mathrm{DPP},z_I}$ (ablations) to control subsystem-wise diversity.
These weights are set manually and tune the trade-off between performance and diversity.
}

\begin{algorithm}[H]
    \caption{DF-GLUE Training}{DF-GLUE Training}
    \label{alg:glued_training}
    \begin{algorithmic}
    \State \textbf{Input:} System-level latent $\zeta$
    \State \textbf{Models:} Coordinator model $\Phi$ with parameters $\theta_\Phi$, subsystem models $\{S_i\}$, differentiable scoring layer $\mathcal{G}$
    \State \textbf{Parameters:} Lagrange multipliers $\lambda_k$, performance weights $\lambda_p$
    \For{epoch in epochs}
        \State $(\{z_i\}, x_C) \gets \Phi(\zeta)$ \Comment{Coordinator forward pass}
        \For{each subsystem $i$}
            \State $x_i \gets S_i(z_i)$ \Comment{Subsystem forward pass}
        \EndFor
        \State $\mathbf{x} \gets \{x_i, x_C\}$ \Comment{Instantiate designs}
        \State $O_p, C_k \gets \mathcal{G}(\mathbf{x})$ \Comment{Evaluate designs}
        \State $\mathcal{L}_{\text{div}} \gets \text{DPP}(\mathbf{z})$ \Comment{Compute diversity loss}
        \State $\mathcal{L}_{\text{feas}} \gets \sum_k \lambda_k C_k$ \Comment{Compute feasibility loss}
        \State $\mathcal{L}_{\text{perf}} \gets \sum_p \lambda_p O_p$ \Comment{Compute performance loss}
        \State Backpropagate $\mathcal{L}_{\text{feas}} + \mathcal{L}_{\text{perf}} + \mathcal{L}_{\text{div}}$ and update $\theta_\Phi$
        \State $\lambda_k \gets \text{ALM}(\lambda_k, C_k)$ \Comment{Update multipliers}
        \EndFor
    \end{algorithmic}
\end{algorithm}

\paragraph{Architecture}{For \ac{DF}-\ac{GLUE}, we use an \ac{MLP}-based architecture with dedicated heads for each subsystem latents and placement. 
We use sigmoid and tanh activations to keep outputs within the bounds prescribed by the subsystem models, with custom \ac{MLP} initialization to prevent initial gradient vanishing. 
Details on the architecture and ablations are provided in \ref{app:df-architecture_details}.}

\paragraph{Experimental Details}{We use a system-level latent dimension of $\dim(\zeta) = 2$ for visualization experiments and $\dim(\zeta) = 4$ otherwise.
We train with a batch size of 1296 for 2500 epochs. Unless mentioned otherwise, operating conditions are fixed to case~1 (see \ref{ap:aircraft_cases}). 
In principle, \ac{GLUE} can be trained over broad condition ranges for true inverse design, which we leave for future work.}
With all three coordination methods in place, we now turn to results.

%% file: tables/compatibility.tex
\begin{table}[!t]
\centering
\caption{Compatibility with \ac{DF}-\ac{GLUE} (requires end-to-end differentiability)}
\label{tab:compat}
\footnotesize
\renewcommand{\arraystretch}{1.15}
\begin{tabularx}{\columnwidth}{@{} c | c @{\hskip 4pt} L{3.1cm} Y L{1.4cm} @{}}
\toprule
\multicolumn{1}{@{} c}{} & & & \textbf{Challenge} & \textbf{Examples} \\
\midrule
\multirow{6}{*}[-3pt]{\rotatebox[origin=c]{90}{\footnotesize\textit{Subsystem Models}}}
& \cmark & VAE, GAN, neural SDF
  & --- & \cite{park2019deepsdf,talabot2025partsdf,chenBezierGANAutomaticGeneration2021} \\[1pt]
& \pmark & Diffusion, flow matching
  & $T$-step backprop & \cite{hoDenoisingDiffusionProbabilistic2020,bagazinski2023shipgen} \\[1pt]
& \pmark & VQ-VAE
  & Discrete codebook & \cite{xu2022skexgen} \\[1pt]
& \xmark & Autoregressive
  & Discrete generation & \cite{etesam2025gearformer,wu2021deepcad} \\[1pt]
& \xmark & Variable-structure
  & Discrete topology & \cite{zhao2020robogrammar,xu2024brepgen} \\[1pt]
& \xmark & Black-box / API
  & --- & --- \\
\midrule
\multirow{4}{*}[-2pt]{\rotatebox[origin=c]{90}{\footnotesize\textit{Evaluators}}}
& \cmark & Geometry checking
  & --- & \cite{tracy2023dcol,prasad2022nurbsdiff} \\[1pt]
& \cmark & RANS CFD, lin.\ FEM, EM
  & --- & \cite{economonSU2OpenSourceSuite2016,heDAFoamOpenSourceAdjoint2020} \\[1pt]
& \pmark & DNS CFD, nonlin. FEM
  & Ongoing research & \cite{xueJAXFEMDifferentiableGPUaccelerated2023,ataei2024xlb} \\[1pt]
& \xmark & Crash, combustion
  & Non-smooth physics & --- \\
\bottomrule
\end{tabularx}
\end{table}

%% file: figures/explainers/explainer_graph.tex
\begin{figure*}[!ht]
    \centering
    \begin{tikzpicture}
        \node[anchor=south west,inner sep=0] (image) at (0,0) {\includegraphics[width=0.9\textwidth]{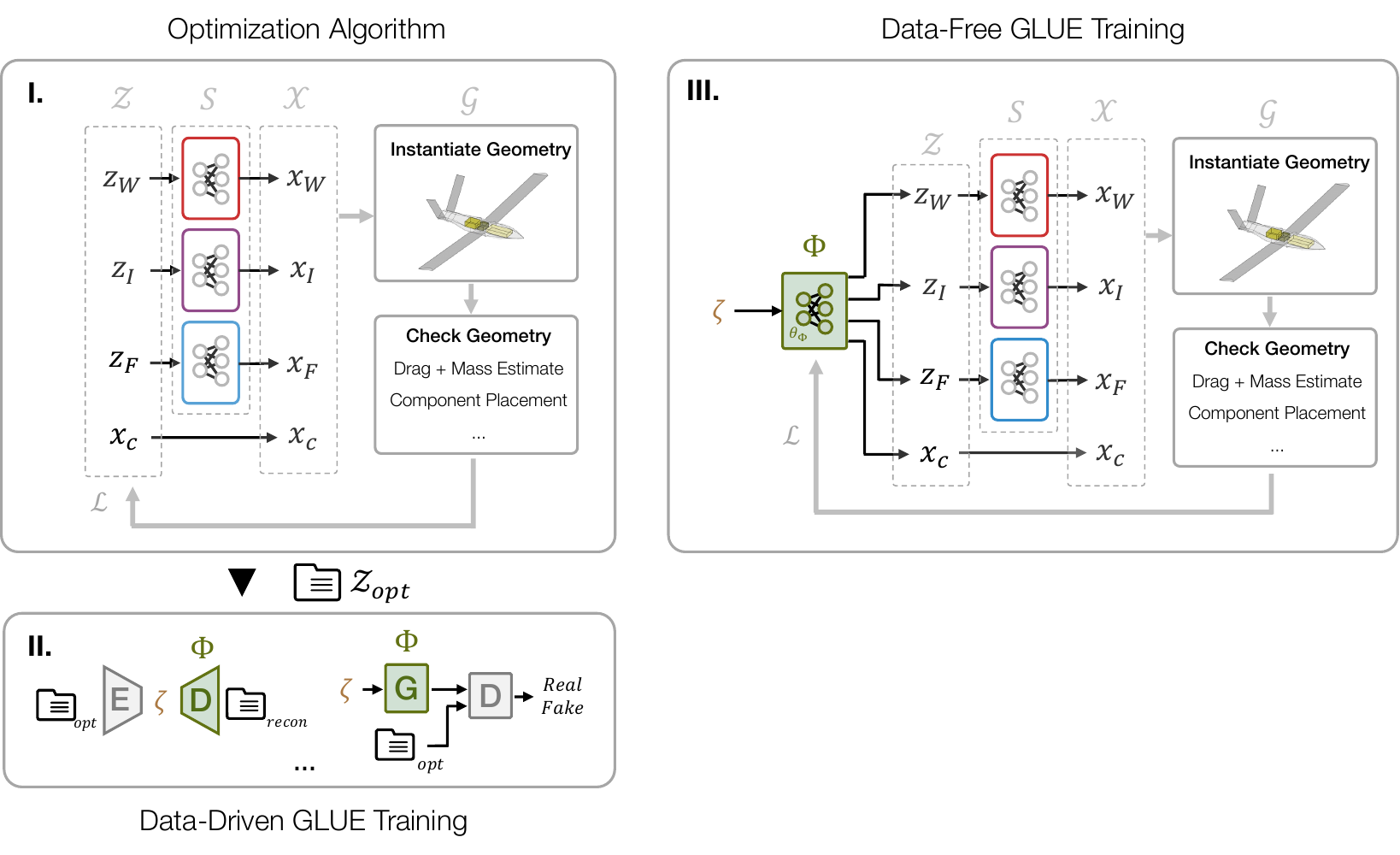}};
        \node[anchor=south east, xshift=-14mm, yshift=6mm, align=left] at (image.south east) {
            \fontsize{8}{10}\selectfont\sffamily
            \begin{tabular}{@{}r@{\ -- }l@{}}
                $\zeta$ & System-level latent \quad  \quad $\Phi$ -- \ac{GLUE} model \\
                $S$ & Pre-trained subsystem models $S_i$ \\
                $\mathcal{Z}$ & Latent codes $z_i$ for $S_i$ \\
                $\mathcal{X}$ & Parametric design variable space \\
                $\mathcal{G}$ & Differentiable geometry layer \\
                $\mathcal{L}$ & Loss (performance, feasibility, diversity)
            \end{tabular}
        };
    \end{tikzpicture}
    
    \caption[Overview of coordination approaches]{\textbf{I:} Optimization algorithm for creation of dataset of optimized designs ($\mathcal{Z}_{opt}$) (\textit{e.g.}, gradient descent or Bayesian optimization). 
    \textbf{II:} Data-driven \ac{GLUE} models (\acs{GAN}, \acs{VAE}, \acs{DDPM}, ...) trained on large datasets obtained using {I} (optimization algorithms). 
    \textbf{III:} Data-Free \ac{GLUE}. Here, gradient descent on a loss $\mathcal{L} = \mathcal{L}_{feas} +  \mathcal{L}_{perf} + \mathcal{L}_{div}$ is used to train \ac{GLUE} model directly to map $\zeta$ to $\mathcal{Z}$.}
    \label{fig:explainer_graph}
\end{figure*}

%% file: sections/04_results.tex
\section{Results}
\label{section:results}

\subsection{Evaluation Metrics}

\paragraph{Feasibility} A sample is feasible if inequality constraints ($\Cboxpl$ and $\Cbb$) satisfy $\le 0$, and equality constraints satisfy their constraint-specific tolerances. 
These are $\Cwx\le 0.02\cdot b$ (2\% of wing span), $\Cdi\le \{0.5^\circ,\,2^\circ\}$ (front/rear), and $\Clift\le 0.02 \cdot L_{\text{req}}$ (2\% of lift requirement). 
Feasibility is assessed with an additional numerical tolerance of $10^{-3}$. 

\paragraph{Single-Objective Optimality} In this study, we minimize $\Omass$ only. The objective is made dimensionless by normalizing with internals mass, \textit{i.e.}, the reported figures do not directly correspond to absolute aircraft mass.

\paragraph{Diversity} We measure design variety via a \ac{DPP} score in $[0,\infty)$, where $0$ indicates high diversity and $\infty$ identical samples. Note that \ac{DPP} scores are not directly comparable across experiments with different feature spaces or \ac{DPP} length scales. However, relative DPP rankings between methods are stable across the $\sigma_{\text{DPP}}$ values we tested. For implementation details, see~\ref{ap:dpp_loss}. 

\paragraph{Compute Efficiency} We report three complementary cost metrics (see Table~\ref{tab:runtime-stats}). \textit{Geometry evaluations} count forward passes through the subsystem models $S_i$ and the geometry layer $\mathcal{G}$. \textit{\acp{FLOP}} measure total compute including backpropagation. \textit{Wall-clock time} reports end-to-end runtime. For \ac{DD}-\ac{GLUE} variants (II.), we separate data generation (via \ac{ALM}-\ac{GD}) and model training.

\input{figures/visual_designs_compact/visual_designs_compact.tex}

\subsection{Visual Comparison}

\label{sec:vis-comparison}

We start with a qualitative look at selected samples in Figure~\ref{fig:visual_designs_compact}. \ac{ALM}-\ac{GD} (I.) is prone to local minima, \textit{e.g.}, stacked internals instead of a single row (the middle sample in Figure~\ref{fig:visual_designs_compact}) prevent further fuselage shrinkage and yield suboptimal $\Omass$. \ac{DD}-\ac{GLUE} variants (II., shown here only \acs{OT-GAN}) are trained on \ac{ALM}-\ac{GD} data and therefore tend to reproduce similar designs. 
\ac{DF}-\ac{GLUE} (III.) consistently collapses to the best observed optimum, despite also relying on gradient descent. We expect \ac{DD}-\ac{GLUE} would match this high-optimality, low-diversity regime given a similarly collapsed training set. 
A more detailed visual comparison can be found in~\ref{ap:qualitative_comparison}.

% Moved up, before "basin connectivity" section do this lands on the same page
\begin{figure*}[!hb]
    \centering
    \includegraphics[width=0.95\textwidth]{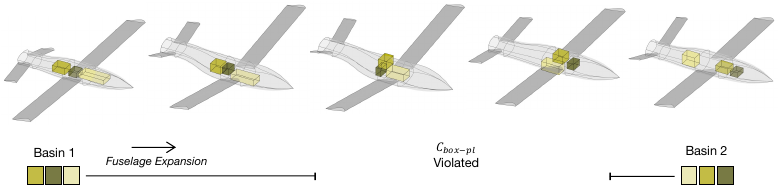}
    \caption[Traversing non-feasible, non-optimal design region to achieve diversity]{Traversing non-feasible, non-optimal design region to achieve diversity.}
    \label{fig:transition_explainer}
\end{figure*}

\subsection{Basin Connectivity}

In our experiments, \ac{GLUE} demonstrates the capacity to \textit{bridge distinct design basins} by briefly passing through infeasible or suboptimal regions. This holds when multiple basins exist in the underlying data or diversity is explicitly incentivized (\textit{e.g.}, via $\lambda_{\mathrm{DPP},z_I}$). In our UAV case, transitions either remain feasible but highly suboptimal, or stay near-optimal with a short, strongly infeasible leap. In practice, training yields an intermediate path, governed by hyperparameters that balance objective and constraint penalties. Figure~\ref{fig:transition_explainer} illustrates this behavior. During internals reordering, the model slightly inflates the fuselage to lower penalty while still tolerating a small $\Cboxpl$ violation.

\subsection{Benchmarking}

\label{sec:opt-feas-div}

\begin{figure*}[!t]
    \centering
    \begin{tikzpicture}
        \node[anchor=south west, inner sep=0] (image) at (0,0) {\includegraphics[width=\textwidth]{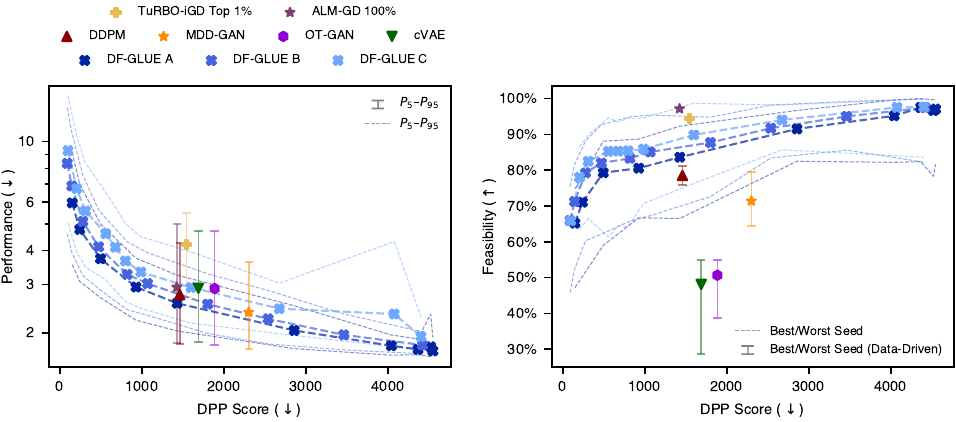}};
        \node[anchor=north east, align=center, xshift=0.0cm, yshift=0.48cm] at (image.north east) {
             \resizebox{0.45\textwidth}{!}{
                 \setlength{\tabcolsep}{3pt}
                 \renewcommand{\arraystretch}{1.1}
                 \begin{tabular}{lcc}
                 & $\lperf$ & $\lDPP$ \\
                 \midrule
                 \ac{DF}-\ac{GLUE} \textbf{A} & 0.25 & $[10.0, 5.0, 2.0, 1.0, 0.75, 0.5, 0.35, 0.25, 0.15, 0.1]$ \\
                 \ac{DF}-\ac{GLUE} \textbf{B} & 0.10 &  $[10.0, 5.0, 2.0, 1.0, 0.5, 0.4, 0.3, 0.25, 0.2, 0.1]$ \\
                 \ac{DF}-\ac{GLUE} \textbf{C} & 0.05 & $[5.0, 2.0, 1.0, 0.5, 0.4, 0.3, 0.25, 0.2, 0.15, 0.1, 0.05]$ \\
                 \end{tabular}
             }
        };
    \end{tikzpicture}
    \setlength{\abovecaptionskip}{-5pt}
    \newcommand{\mycaption}{\input{figures/opt_vs_diversity/case_1_b/caption.txt}}
    \caption[Optimality-feasibility-diversity benchmark for case 1]{\input{figures/opt_vs_compute/case_1_b/caption.txt}}
    \label{fig:opt_vs_diversity_case_1_b}
\end{figure*}

\paragraph{Experiment Details}{\ac{GLUE} is assessed using 10 seeds and 1000 uniform $\zeta$ samples per seed. We report $P5/P95$ percentiles across the resulting 10k designs. Feasibility is reported for the mean, best, and worst seed out of 10. Figure~\ref{fig:opt_vs_diversity_case_1_b} shows a single condition set (case~1, see~\ref{ap:aircraft_cases}). For a high-level cross-case comparison, see~\ref{ap:benchmarking_all_cases}. Since \ac{TuRBO}-\ac{iGD} retains its 200 best designs per run, we score it on its top 1\% for fairness. \ac{DD}-\ac{GLUE} variants use the full training set to maximize constraint satisfaction. \acs{MDD-GAN} is the only exception, as it reaches best feasibility at 80\% (see Fig.~\ref{fig:opt_vs_compute_case_1_b}). 
We suspect this may result from the interplay between augmentation hyperparameters ($\sigma_{aug}$ and $N$, see~\ref{ap:ddjm}) and dataset structure, which together determine the positive-to-negative data ratio. 
}

\paragraph{Findings}{As a baseline, \ac{ALM}-\ac{GD} can find strong designs but is sensitive to local minima, as indicated by the high objective variance across \(\sim\)70k runs and low \ac{DPP} score, which implies high diversity.
As seen in Figure~\ref{fig:opt_vs_diversity_case_1_b}, \ac{TuRBO}-\ac{iGD} achieves poor optimality on this high-dimensional, tightly constrained setting due to the fundamental shortcomings of \ac{BO}. 
On the other hand, \ac{DF}-\ac{GLUE} with $\lDPP=0$ consistently converges to the best observed optimum across seeds and achieves near-perfect feasibility. 
Sweeping $\lDPP$ and $\lperf$ (see table in Figure~\ref{fig:opt_vs_diversity_case_1_b}) traces a Pareto front indicating the trade-off between diversity on one hand and optimality and feasibility on the other. 
At equal diversity, \ac{DF}-\ac{GLUE} is near-optimizer-feasible and exceeds all data-driven feasibility. 
The \ac{DD}-\ac{GLUE} variants roughly replicate optimality and diversity of the underlying \ac{ALM}-\ac{GD} data, which is to be expected. 
Among the data-driven models, \acs{DDPM} is strongest in feasibility, achieving similar feasibility to \ac{DF}-\ac{GLUE}. 
However, the \acs{DDPM} achieves this at significantly higher computational cost than \ac{DF}-\ac{GLUE}, as detailed in the next section.}

% Placed before the section on computational efficiency to spread floats better
\input{tables/runtime_stats.tex}

\subsection{Computational Efficiency}

\begin{figure*}[!ht]
    \centering
    \includegraphics[width=\textwidth]{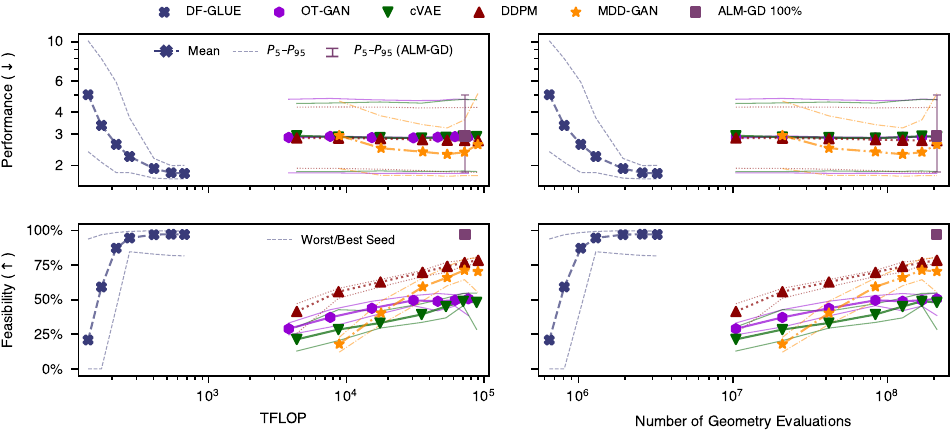}
    \setlength{\abovecaptionskip}{-5pt}
    \newcommand{\mycaption}{\input{figures/opt_vs_compute/case_1_b/caption.txt}}
    \caption[Optimality, feasibility, and constraint violation \textit{vs.} computational cost]{\input{figures/opt_vs_compute/case_1_b/caption.txt}}
    \label{fig:opt_vs_compute_case_1_b}
\end{figure*}

\label{sec:opt-feas-comp}

We now examine how each method scales with compute and geometry evaluations.

\paragraph{Experiment Details}{We use the same sampling protocol and summary statistics as in Section~\ref{sec:opt-feas-div} (10 seeds, 1000 $\zeta$ samples/seed, $P5/P95$ for performance and mean/best/worst-seed feasibility, conditions from case 1). Figure~\ref{fig:opt_vs_compute_case_1_b} shows optimality and feasibility with varying training budget. \ac{DF}-\ac{GLUE} is trained for 600--2500 epochs ($\lDPP=0$). The \ac{DD}-\ac{GLUE} variants are trained on \ac{ALM}-\ac{GD} subsets from 5\% to 100\% of the $\sim$70k-design dataset. The shown \ac{ALM}-\ac{GD} datapoint corresponds to the full dataset.}

\paragraph{Findings}{We find that in this \ac{UAV} study, optimality is rather easy to learn for the data-driven models. Even when trained on 5\% of data ($\sim$3.5k designs), the distribution of optimality closely matches the underlying dataset. In contrast, learning our coupled, non-convex constraints is highly data-intensive. For all data-driven models, feasibility scales extremely weakly with increasing compute budgets. The gradient-driven \ac{DF}-\ac{GLUE}, however, \textit{reaches high feasibility using orders of magnitude less compute} (see Figure~\ref{fig:opt_vs_compute_case_1_b}). Notably, \ac{DF}-\ac{GLUE} produces a full generative model in less wall-clock time than any of the \ac{DD}-\ac{GLUE} variants (see Table~\ref{tab:runtime-stats}, where \ac{DD}-\ac{GLUE} models are shown at 100\% dataset training). Despite not requiring forward passes through the geometry layer, \ac{DD}-\ac{GLUE} training remains expensive due to the large datasets involved. For producing a single design, \ac{ALM}-\ac{GD} remains the most cost-effective approach at $<$3k geometry evaluations per run.}

\subsection{Discussion}
\label{sec:discussion}

Having presented the results, we now interpret these findings across the three method classes.

\paragraph{Bayesian Optimization Limitations}{\ac{BO} fails to produce high-quality designs despite our modifications (\ac{TuRBO}-\ac{iGD}). We attribute this to \ac{BO}'s fundamental limitations in constrained optimization problems, which also underscores that the \ac{UAV} problem is a non-trivial benchmark.}

\paragraph{Data-Driven Models}{On our \ac{UAV} design problem, \ac{DD}-\ac{GLUE} variants learn optimal, diverse solutions from relatively few samples, but exhibit poor ability to recover hard-to-learn constraints (Section~\ref{sec:opt-feas-div}) even with extensive datasets (Section~\ref{sec:opt-feas-comp}). This is consistent with prior observations that learning feasibility purely from data is non-trivial (see Section~\ref{sec:data_driven_training}).}

\subsubsection[DF-GLUE]{Data-Free GLUE}

\paragraph{Optimization Robustness}{Given $\lDPP=0$, we observe that \ac{DF}-\ac{GLUE} consistently converges to the best feasible solution found by any method, with reduced sensitivity to local minima compared to \ac{ALM}-\ac{GD} (see Sections~\ref{sec:vis-comparison}, \ref{sec:opt-feas-div}, \ref{sec:opt-feas-comp}). We attribute this to reduced sensitivity to local minima in over-parametrized network parameter space (4.7M parameters) versus latent-space optimization in $\mathcal{Z}$ (22-D).}

\paragraph{Tractable Constraint Learning}{\ac{DF}-\ac{GLUE} learns feasibility far more efficiently than \ac{DD}-\ac{GLUE} variants, requiring orders of magnitude fewer geometry evaluations and \acp{FLOP} (see Sec.~\ref{sec:opt-feas-comp} and Table~\ref{tab:runtime-stats}).

We attribute this to \textit{information efficiency}. Data-driven training keeps only final converged \ac{ALM}-{GD} designs, discarding optimizer trajectories. Additionally, it lacks access to local gradients. \ac{DF}-\ac{GLUE} instead backpropagates through differentiable objectives, maximizing learning signal and fully utilizing each forward pass.

Additionally, we see headroom for further speedups. We find that training convergence speed is insensitive to increases in batch size. This implies that \ac{DF}-\ac{GLUE} training is not bottlenecked by information acquisition through forward passes, but rather training dynamics (\ac{ALM} weight growth and primal learning-rate schedule). We hypothesize that studying these dynamics could compress the current \ac{DF}-\ac{GLUE} training times even further at equal solution quality.}

\paragraph{Explicit Diversity Control}{A further advantage of \ac{DF}-\ac{GLUE} is explicit, user-specified control of the trade-off between diversity and robust convergence to the best observed optimum (with $\lDPP=0$). Data-driven models lack this explicit control. A similar result could be achieved with dataset filtering, but this would further degrade compute efficiency.}

\input{figures/smoothness-ablation/latent_sweep_comparison.tex}

\subsection[Data-Free GLUE: Ablation Studies]{Data-Free \ac{GLUE}: Ablation Studies}

We now shift focus to what we identify as a fundamental challenge for \ac{GLUE}: equality constraints define a zero-measure set in $\mathcal{Z}$, making their exact satisfaction by a generative model producing diverse samples practically impossible without dedicated mitigation strategies. This challenge manifests in two concurrent symptoms: (a)~\textit{low feasibility}: \ac{GLUE} cannot reliably produce designs that satisfy equality constraints, and (b)~\textit{mode collapse}: the growing equality penalty ($\lambda \gg 1$) overpowers optimality, diversity, and other constraints, trapping the optimizer once feasibility is reached (Fig.~\ref{fig:mode_collapse_explainer}, baseline).

Focusing on \ac{DF}-\ac{GLUE}, we ablate two categories of mitigation strategies (applied throughout the results above): (\textit{i})~\ac{GLUE}-side strategies (tolerancing, smoothing of equality constraints, and a dedicated inner optimizer) and (\textit{ii})~properties of the subsystem models that affect constraint learning tractability, particularly for equality constraints. For both experiments, see \ref{ap:dfjm_hyperparams} for hyperparameters and \ref{ap:dpp_loss} for \ac{DPP} scoring details.

\subsubsection{\ac{GLUE} Training Strategies}
\label{sec:mode-collapse-ablation}

At the \ac{GLUE} training level (\textit{i}), we use three approaches to mitigate mode collapse:

\paragraph{Inner Optimizer}{A dedicated inner optimizer (Adam) backpropagates selected equality constraints only through the output layers controlling their relevant variables (see \ref{app:df-architecture_details}, Fig.~\ref{fig:inner-adam-explainer}). While effective, this requires prior knowledge of variable-constraint dependencies.} 

\paragraph{Equality Constraint Tolerancing} We soften equalities into tolerance bands so the optimizer can make incremental progress without immediate large penalty spikes (see Fig.~\ref{fig:mode_collapse_explainer}).
We use $\Cwx$ within 2\% of span and $\Cdi \le \{0.5^\circ/2^\circ\}$ (front/rear dihedral).

\paragraph{Smoothing of Equality Constraints} We replace the ReLU baseline with an L1-Huber activation function for equality constraints ($\phi_{eq}$). This allows the optimizer to pursue optimality, diversity, and other constraints without incurring a large loss spike at the constraint boundary (see Fig.~\ref{fig:mode_collapse_explainer}).

\input{figures/mode-collapse-ablation/mode-collapse-ablation.tex}

\input{tables/ablation_mode_collapse.tex}

\paragraph{Results} The baseline in Table~\ref{tab:mode-collapse} exhibits clear signs of mode collapse: a DPP score of 50k and $\Omass$ of 23.8, indicating the growing equality penalty dominates the loss at the expense of diversity and optimality. The inner optimizer is the dominant mitigation, reducing DPP by two orders of magnitude and lifting feasibility by 10--15\%. Tolerancing and Huber smoothing provide further gains, with all three strategies combined yielding the best optimality ($\Omass = 4.15$) and diversity (DPP $= 7$), while the zero-tolerance variant trades slightly higher feasibility for less diversity due to its stricter training target.

Having addressed the \ac{GLUE}-side training strategies (category \textit{i}), we now turn to \textit{ii}: the effect of subsystem model properties on equality constraint learning.

\subsubsection{Effect of Subsystem Model Properties}
\label{sec:smoothness-ablation}

\begin{figure*}[!t]
    \centering
    %% ── Right column: constraint plots + subcaption + table ─────────────────
    %% Save into a box so we can measure its total height and match the left.
    %% (\newsavebox is declared in main.tex preamble to avoid figure* measurement bug)
    \begin{lrbox}{\snRightBox}%
        \begin{minipage}[t]{0.63\textwidth}%
            %% Subcaption for (b)–(d): measured offsets (lm=10.34%, pw=25.43%, gap=6.44%)
            \footnotesize\itshape\noindent%
            \vphantom{$\xrightarrow{S_F}$}%
            \hspace*{0.103\linewidth}%
            \makebox[0.254\linewidth][c]{(b)~$C_{wx}$ (wing pos.)}%
            \hspace*{0.064\linewidth}%
            \makebox[0.254\linewidth][c]{(c)~$C_{di}$ (dihedral)}%
            \hspace*{0.064\linewidth}%
            \makebox[0.254\linewidth][c]{(d)~$C_{lift}$ (lift)}%
            \\[5pt]
            \includegraphics[width=\linewidth]{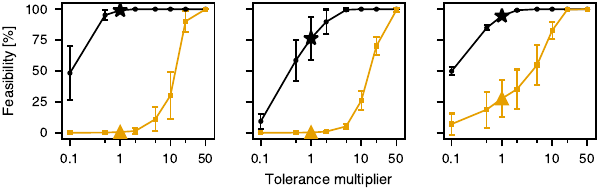}%
            \\[7pt]
            %% ── Table: ★ / ▲ match the markers at tol=1× in the plots ──
            \normalfont\footnotesize\noindent\hspace*{0.12\linewidth}
            \begin{tabular}{c c c c c c c c}
            \toprule
            & SN & $\Omass$ ($\downarrow$) & $\Cwx$ ($\uparrow$) & $\Cdi$ ($\uparrow$) & $\Clift$ ($\uparrow$) & $\Cbb$ ($\uparrow$) & $\Cboxpl$ ($\uparrow$) \\
            \midrule
            \textcolor{NoSNcolor}{$\blacktriangle$} & \xmark & 11.08 & 0.6\% & 0.2\% & 27.7\% & 74.8\% & 75.6\% \\
            \textcolor{SNcolor}{$\bigstar$}         & \cmark & \textbf{4.54} & \textbf{99.1}\% & \textbf{76.4}\% & \textbf{94.5}\% & \textbf{100\%} & \textbf{96.1}\% \\
            \bottomrule
            \end{tabular}
        \end{minipage}%
    \end{lrbox}%
    %% ── Left column: natural aspect ratio, fig_h=4.53cm matched to right column ─
    \begin{minipage}[t]{0.23\textwidth}%
        \vspace{3pt}%  ← tune this to vertically position (a)
        %% Subcaption for (a): axis centre at 60.3% of PDF width
        \footnotesize\itshape\noindent\hspace*{0.40\linewidth}(a)~$z_{F,0} \xrightarrow{S_F} x_{F,9}$%
        \\[5pt]
        \includegraphics[width=\linewidth]{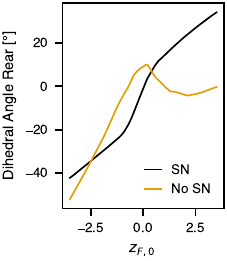}%
    \end{minipage}%
    \hspace{0.08\textwidth}%
    \usebox{\snRightBox}%
    \caption[Subsystem model smoothness and constraint feasibility vs.\ tolerance]{(a)~Mapping $z_{F,0} \xrightarrow{S_F} x_{F,9}$ is smoothed and meaningful with \ac{SN}. (b)--(d)~Per-constraint feasibility vs.\ tolerance multiplier, where \textcolor{SNcolor}{$\bigstar$}~(\ac{SN}) and \textcolor{NoSNcolor}{$\blacktriangle$}~(No~\ac{SN}) mark the nominal 1$\times$ tolerance.}
    \label{fig:smoothness-fuselage-model}
\end{figure*}

\paragraph{Experiment}{
We compare \ac{GLUE} training with two versions of the fuselage subsystem model $S_{\!F}$, differing only in whether \ac{SN} was enabled during training. For each $S_{\!F}$ variant, we train and evaluate \ac{GLUE} across multiples of the nominal equality constraint tolerances from Table~\ref{tab:mode-collapse}. \ac{SN} has three notable effects on $S_{\!F}$: first, it constrains the Lipschitz constant of each layer, yielding a smoother mapping $z_F \to x_F$ (Fig.~\ref{fig:smoothness-fuselage-model}(a)), second, it aids the least volume loss~\cite{CompressingLatentSpace2023} in producing a more disentangled latent space, and third, it slightly reduces the output diversity of $S_{\!F}$. We believe all three effects contribute to improved equality constraint learning. \ac{SN} was not applied to the wing or internals subsystem models. It was also not applied to the \ac{GLUE} model itself, where it would hinder rapid traversal through infeasible regions needed for basin bridging (Fig.~\ref{fig:transition_explainer}).
}

\paragraph{Results}{
Figures~\ref{fig:smoothness-fuselage-model}(b)--(d) show per-equality-constraint feasibility as a function of the tolerance multiplier. All three equality constraints ($\Cwx$, $\Cdi$, $\Clift$) improve markedly with \ac{SN}: at nominal tolerance (1$\times$, marked by $\bigstar$ and $\blacktriangle$), $\Cwx$ jumps from 0.6\% to 99.1\% and $\Cdi$ from 0.2\% to 76.4\%, indicating that \ac{SN} enables comparable equality constraint feasibility at substantially tighter tolerances.

The associated degradation in $\Omass$ and inequality constraints ($\Cbb$, $\Cboxpl$) parallels the mode collapse effect from Sec.~\ref{sec:mode-collapse-ablation}, where hard-to-satisfy equality constraints crowd out other objectives and constraints. Figure~\ref{fig:latent_sweep_comparison} shows this qualitatively: without \ac{SN} on $S_{\!F}$, sweeping the \ac{GLUE} latent space $\zeta$ reveals high fuselage model diversity but degraded system-level optimality. With \ac{SN}, the traversal yields better-optimized designs.
}

\subsubsection{Discussion}
\label{sec:ablation-discussion}

Both ablation categories tackle the same root cause: the zero-measure feasibility set of equality constraints makes them disproportionately hard to satisfy under \ac{ALM}-based training. At the subsystem model level, attributes such as output smoothness, disentanglement, and diversity directly affect equality constraint tractability, as the \ac{SN} ablation demonstrates. Subsystem models should be as smooth and disentangled as possible while maintaining sufficient output diversity. The combination of \ac{SN} and least volume loss~\cite{CompressingLatentSpace2023} used here is one approach to achieve this but not the only one (e.g.~\cite{higginsBetaVAE2017, hoffmanJacobianRegularization2019}). How critical these attributes are depends on the multi-component problem, including the number and tightness of equality constraints, their interactions, and the \ac{GLUE} architecture. Loosely constrained problems may be insensitive to subsystem model smoothness, while tightly constrained ones benefit substantially (Fig.~\ref{fig:smoothness-fuselage-model}(b)--(d)).

However, when working with pre-trained or off-the-shelf subsystem models, these attributes cannot always be influenced. In that case, the \ac{GLUE} training strategies become the primary lever. The inner optimizer provides the largest gains but requires prior knowledge of constraint-variable dependencies. Tolerancing and smoothing are broadly applicable fallbacks.

More broadly, the problem should be parametrized to avoid superfluous equality constraints. Additionally, tolerances should be set as generously as the problem allows. We leave inference-time projection or few-step refinement as further techniques worth exploring.

%% file: figures/visual_designs_compact/visual_designs_compact.tex
\begin{figure}[b!]
    \centering
    \includegraphics[width=\columnwidth]{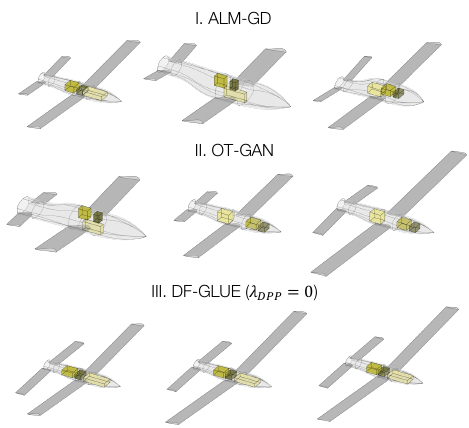}
    \caption[Exemplar aircraft designs for methods I, II and III.]{Exemplar aircraft designs for methods I, II and III.}
    \label{fig:visual_designs_compact}
\end{figure}

%% file: figures/opt_vs_diversity/case_1_b/caption.txt
Trade-offs between optimality, feasibility, and diversity for \ac{ALM}-\ac{GD}, \ac{TuRBO}-\ac{iGD}, data-driven \ac{GLUE} models (\acs{cVAE}, \acs{OT-GAN}, \acs{MDD-GAN}, \acs{DDPM}) and data-free \ac{GLUE} model (\ac{DF}-\ac{GLUE}). \ac{DF}-\ac{GLUE} is swept across multiple hyperparameter combinations to showcase the full range of accessible trade-offs between diversity, optimality, and feasibility. For generative models, each marker summarizes 10 seeds $\times$ 1000 samples/seed. Vertical bars show 5th to 95th percentile in performance across designs. Feasibility bars show best/worst feasibility out of 10 seeds. DPP on $\mathcal{X}$ (39-D) with $\sigma_{\text{DPP}} = 0.5$.

%% file: figures/opt_vs_compute/case_1_b/caption.txt
Comparison of optimality, feasibility, and mean constraint violation against computational cost. \ac{DF}-\ac{GLUE} ($\lambda_{\text{DPP}} = 0$) attains high feasibility and best observed optimality at orders of magnitude lower cost than \ac{DD}-\ac{GLUE} training. For \ac{DF}-\ac{GLUE}, each marker summarizes 10 seeds for different amounts of training epochs. For data-driven models, each marker summarizes 10 seeds at given data subset (5~\% -- 100~\%) for training.

%% file: tables/runtime_stats.tex
\begin{table*}[!hb]
    \centering
    \caption[Compute and wall-clock time comparison]{Compute cost and wall-clock times across three method categories. TFLOP values are profiled via PyTorch profiler \cite{paszkePyTorchImperativeStyle2019} (matrix-multiply operations). Inference cost is not included.}
    \label{tab:runtime-stats}
    \footnotesize
    \begin{tabular}{lcccccc}
    \toprule
    \multicolumn{7}{l}{\textbf{I. Optimization (Per Run)}} \\
    \cmidrule(r){1-7}
    Method & Hardware & & & Geom.\ Evals (Avg $\pm$ SD) & TFLOP (Avg $\pm$ SD) & Wall-clock (Avg $\pm$ SD) \\
    \midrule
    \ac{TuRBO}-\ac{iGD} & 2 \acs{CPU}, 2GB & & & & & 8h36m $\pm$ 1h12m \\
    \ac{ALM}-\ac{GD} & 1 \acs{CPU}, 4GB & & & 2.9k $\pm$ 0.7k & 0.0010 $\pm$ 0.0002 & 54m $\pm$ 18m \\
    \midrule
    \multicolumn{7}{l}{\textbf{II. Data-Driven \ac{GLUE} (Data Generation + Model Training)}} \\
    \cmidrule(r){1-7}
    Method & Hardware & Data Gen Geom.\ Evals & Data Gen TFLOP & & Training TFLOP & Wall-clock \\
    \midrule
    \acs{DDPM} & RTX 4090 & 209M $\pm$ 0 & 72.5 $\pm$ 0 & & 17.8 $\pm$ 0 & 48m21s $\pm$ 1m54s \\
    \acs{MDD-GAN} & RTX 4090 & 209M $\pm$ 0 & 72.5 $\pm$ 0 & & 17.4 $\pm$ 0.02 & 32m21s $\pm$ 11s \\
    \acs{OT-GAN} & RTX 4090 & 209M $\pm$ 0 & 72.5 $\pm$ 0 & & 4.1 $\pm$ 0 & 23m35s $\pm$ 20s \\
    \acs{cVAE} & RTX 4090 & 209M $\pm$ 0 & 72.5 $\pm$ 0 & & 15.4 $\pm$ 0 & 13m10s $\pm$ 7s \\
    \midrule
    \multicolumn{7}{l}{\textbf{III. Data-Free \ac{GLUE} (Training Only)}} \\
    \cmidrule(r){1-7}
    Method & Hardware & & & Training Geom.\ Evals & Training TFLOP & Wall-clock \\
    \midrule
    \acs{DF}-\acs{GLUE} & RTX 4090 & & & 3.2M $\pm$ 0 & 0.67 $\pm$ 0 & 10m29s $\pm$ 5s \\
    \bottomrule
    \end{tabular}
\end{table*}

%% file: figures/smoothness-ablation/latent_sweep_comparison.tex
\begin{figure*}[!b]
    \centering
    \begin{subfigure}[b]{0.495\textwidth}
        \centering
        \includegraphics[width=\linewidth,keepaspectratio]{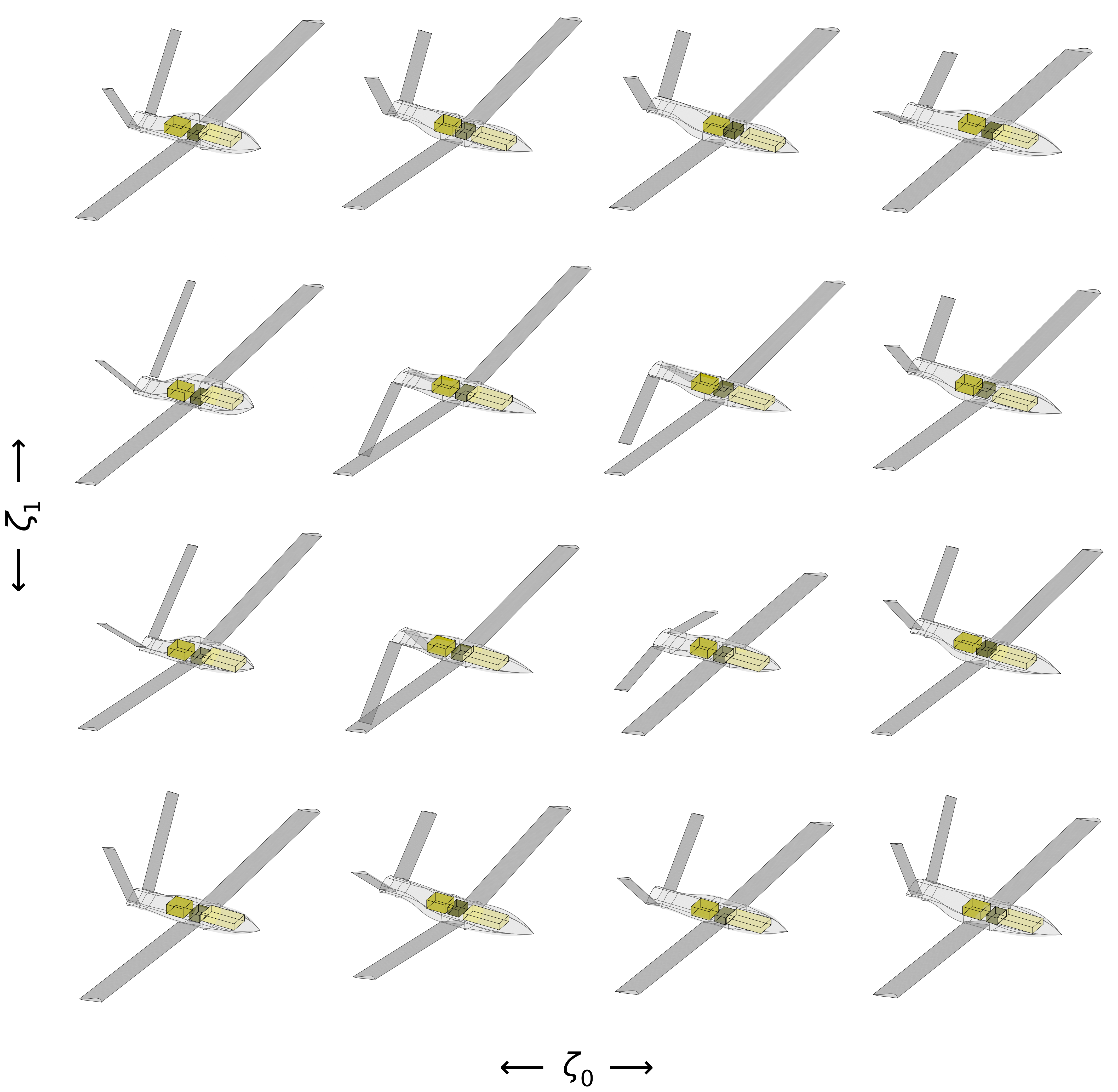}
        \caption{With \ac{SN} for $S_{\!F}$}
        \label{fig:latent_sweep_with_sn}
    \end{subfigure}
    \hfill
    \begin{subfigure}[b]{0.495\textwidth}
        \centering
        \includegraphics[width=\linewidth,keepaspectratio]{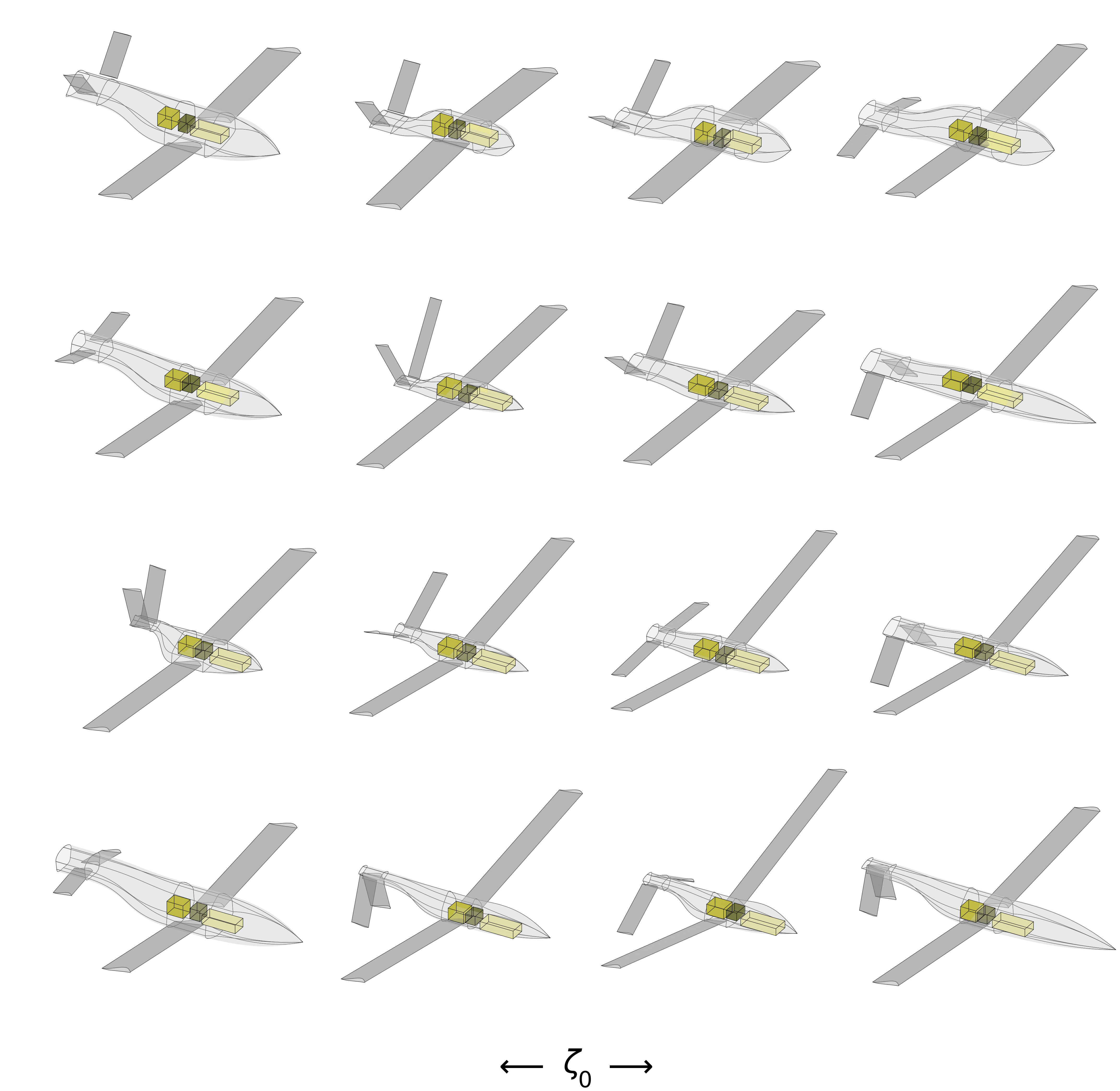}
        \caption{Without \ac{SN} for $S_{\!F}$}
        \label{fig:latent_sweep_without_sn}
    \end{subfigure}
    \caption[Latent space sweep comparison: with vs. without \ac{SN}]{Sweep of system-level latent space $\zeta$ comparing fuselage model \ac{SN} effects. (a) With \ac{SN} enabled, the smooth subsystem model allows improved constraint satisfaction and optimality. (b) Without \ac{SN}, we observe increased diversity in fuselage model outputs.}
    \label{fig:latent_sweep_comparison}
\end{figure*}

%% file: figures/mode-collapse-ablation/mode-collapse-ablation.tex
\begin{figure}[H]
    \centering
    \includegraphics[width=0.9\columnwidth]{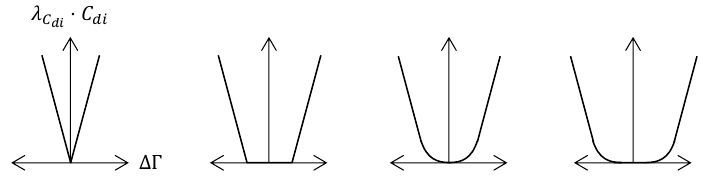}
    \vspace{0em}
    \begin{minipage}{0.9\columnwidth}
        \centering
        \begin{minipage}{0.2499\columnwidth}
            \footnotesize    
            \raggedright
            \hspace{0.5em}
            Baseline
        \end{minipage}%
        \begin{minipage}{0.2499\columnwidth}
            \footnotesize    
            \centering
            \hspace{0.5em}
            Tolerance
        \end{minipage}%
        \begin{minipage}{0.2499\columnwidth}
            \footnotesize    
            \centering
            \hspace{0.5em}
            Huber
        \end{minipage}%
        \begin{minipage}{0.2499\columnwidth}
            \footnotesize    
            \raggedleft
            Tol. + Huber
        \end{minipage}
    \end{minipage}
    \caption[Tolerance and smoothing of equality constraints]{Tolerance and smoothing of equality constraints.}
    \label{fig:mode_collapse_explainer}
\end{figure}

%% file: tables/ablation_mode_collapse.tex
\begin{table}[h!]
     \centering
     \small
     \caption[Mode collapse ablation study for \ac{DF}-\ac{GLUE}]{Mode collapse ablation study for case 1b. Each variant with 10 seeds and 1000 samples per seed. \ac{GLUE} model with $\text{dim}(\zeta) = 2$. High DPP scores and $\Omass$ are indicative of mode collapse. Baseline (first row) shows severe mode collapse. DPP on $z_F$ (4-D) with $\sigma_{\text{DPP}} = 0.05$. IO = Inner optimizer, Tol.\ = Tolerance, $\phi_{\text{eq}}$ = equality constraint loss smoothing function.}
     \footnotesize
     \begin{tabular}{c c c c c r}
     \toprule
     IO & Tol. & $\phi_{\text{eq}}$ 
          & Feas. ($\uparrow$) & DPP ($\downarrow$) & $\Omass$ ($\downarrow$) \\
     \midrule
     \xmark  & \underline{Zero}    & \underline{ReLU} & 11.6 \% & 50106 & {23.82} \\
     \midrule
      \cmark & Normal   & Huber & 69.9 \% & \textbf{7}     & \textbf{4.15} \\
      \cmark & \underline{Zero}     & Huber & \textbf{75.5 \%} & 54    & 4.20 \\
      \cmark & \underline{Zero}     & \underline{ReLU}  & {73.7 \%} & 112   & 4.30 \\
      \xmark  & Normal   & Huber & 59.6 \% & 9557  & 4.47 \\
      \xmark  & Normal   & \underline{ReLU}  & 55.8 \% & 19641 & 5.03 \\
     \bottomrule
     \end{tabular}%
     
     \label{tab:mode-collapse}
\end{table}

%% file: sections/05_conclusion.tex
\section{Conclusion}
\label{section:conclusion}

In this paper, we propose \ac{GLUE} models. These orchestrate \textit{frozen}, \textit{pre-trained} subsystem models (wings, fuselage, internals) to generate feasible, high-performing, and diverse system-level designs. 
We showcase this method on a high-dimensional, non-convex, tightly constrained \ac{UAV} problem that captures key coordination challenges of multi-component engineering design. 
Two training paradigms are compared: \textit{(i)} Data-Driven, which uses state-of-the-art generative architectures (\acs{cVAE}, \acs{MDD-GAN}, \acs{OT-GAN}, \acs{DDPM}) as \ac{GLUE} models and is trained on datasets of optimized designs, and \textit{(ii)} Data-Free, which uses a differentiable geometry layer to backpropagate constraint and objective gradients directly into the \ac{GLUE} model, eliminating the need for pre-generated datasets. 
Both paradigms are benchmarked against baseline latent-space optimization algorithms (\ac{TuRBO}-\ac{iGD} and \ac{ALM}-\ac{GD}).

When trained without an explicit diversity loss, \ac{DF}-\ac{GLUE} consistently converges to the best observed optimum, suggesting reduced sensitivity to local minima, especially compared to gradient-based optimization (\ac{ALM}-\ac{GD}). However, while \ac{DF}-\ac{GLUE} shows strong performance and promising feasibility, optimization algorithms still achieve slightly tighter constraint satisfaction. 
\ac{DF}-\ac{GLUE} is orders of magnitude more computationally efficient than data-driven \ac{GLUE} models, as it fully exploits gradient information from every forward pass.
Ablation studies show that equality constraints pose a fundamental challenge for \ac{GLUE}, which we attribute to their zero-measure feasibility set making them disproportionately hard to satisfy. Subsystem model attributes (smoothness, disentanglement, diversity) directly affect equality constraint tractability. \ac{GLUE} training strategies (tolerancing, smoothing, and inner optimizers) help mitigate this at the optimizer level, though robust handling of equality constraints remains an open challenge.

This work outlines a path toward modular, collaborative workflows in which specialized teams develop subsystem models for later integration without retraining. 
Our approach is especially attractive when subsystem development costs can be amortized across multiple projects or when reusing proprietary or open-source third-party models.

Some limitations remain. 
First, data-free training presumes differentiable simulators and constraint checkers, which may not be available in every domain. Differentiable surrogates can address this gap but reintroduce data-generation cost. 
Second, the suitability of our method with diffusion and flow-matching models for subsystems has yet to be explored. 
Third, the work presented here is limited to single-objective optimization. For multi-objective optimization, specialized methods \cite{pymoo,regenwetterBikeBenchBicycleDesign2025,massoudiIntegratedApproachDesigning2024,zhangLibMOONGradientbasedMultiObjective2024} are preferable to conflating competing objectives into a single weighted loss.
Fourth, extending the evaluation to problems with high-fidelity physics coupling would help clarify the method's applicability beyond the simplified benchmark presented in this work.
Fifth, the current \ac{MLP}-based \ac{GLUE} architecture is not well suited to variable-length subsystem conditions and outputs. Autoregressive or graph-based \ac{GLUE} architectures may be better suited for problems with a varying number of components.
Finally, results are limited to a single predefined parametrization. Extending to diverse design concepts and other engineering problems requires a generalizable geometry layer that can interface with differentiable models and simulators.

%% file: appendix/dfjm_hyperparameters.tex
\section[DF-GLUE Hyperparameters]{\ac{DF}-\ac{GLUE} Hyperparameters}
\label{ap:dfjm_hyperparams}

Table \ref{tab:dfjm_hyperparameters} lists the specific hyperparameters used for \ac{DF}-\ac{GLUE} across the different experiments presented in this work. 
The hyperparameters include the random seeds, sampling count, numerical tolerance, DPP kernel width $\sigma_{\text{DPP}}$, latent dimension size $\dim(\zeta)$, and the loss weights for \ac{DPP} diversity terms ($\lambda_{\text{DPP}}$), mutual information ($\lambda_{\text{MI}}$), and performance optimization ($\lambda_{\text{perf}}$).
Where weights are annealed during training, the schedule is denoted as $\text{start} \to \text{end}$.

\input{tables/dfjm_hyperparams.tex}

%% file: tables/dfjm_hyperparams.tex
\begin{table*}[ht]
    \centering
    \footnotesize
    \caption{\ac{DF}-\ac{GLUE} hyperparameters for the different experiments presented in the paper.}
    \label{tab:dfjm_hyperparameters}
    \begin{tabular}{lcccc}
    \toprule
     & \textbf{Opt-Feas-Div} & \textbf{Opt-Feas-Compute} & \textbf{SN Abl.} & \textbf{Mode Coll. Abl.} \\
     & (Figure \ref{fig:opt_vs_diversity_case_1_b}) & (Figure \ref{fig:opt_vs_compute_case_1_b}) & (Figure \ref{fig:smoothness-fuselage-model}) & (Table \ref{tab:mode-collapse}) \\
    \midrule
    $\dim(\zeta)$ & 4 & 4 & 2 & 2 \\
    \midrule
    Sampling Mode & Latin Hypercube & Latin Hypercube & Latin Hypercube & Latin Hypercube \\
    Batch Size & 1296 & 1296 & 1296 & 1296 \\
    Learning Rate & 0.001 & 0.001 & 0.001 & 0.001 \\
    LR Decay Multiplier & - & - & 0.9985 & 0.9985 \\
    \midrule
    ALM Mode & Pooled & Pooled & Hypercube & Hypercube \\
    ALM $\alpha$ & 0.9 & 0.9 & 0.9 & 0.9 \\
    ALM $C_{\text{cap}}$ & 500 & 500 & - & - \\
    ALM $\gamma$ & 0.005 & 0.005 & 0.005 & 0.005 \\
    \midrule
    $\lperf$ & Figure~\ref{fig:opt_vs_diversity_case_1_b} & 0.1 & 0.25 & 0.25 \\
    $\lambda_{\mathrm{DPP}, \mathcal{X}}$ & Figure~\ref{fig:opt_vs_diversity_case_1_b} & 0 & 0 & 0 \\
    $\lambda_{\mathrm{DPP},z_F}$ & 0 & 0 & 0.2 & 0.2 \\
    $\lambda_{\mathrm{DPP},z_W}$ & 0 & 0 & 0 & 0.1 \\
    $\lambda_{\mathrm{DPP},z_I}$ & 0 & 0 & 0 & $30 \to 0.4$ \\
    $\lambda_{\mathrm{MI}}$ & 0 & 0 & $8 \to 0.2$ & $8 \to 0.2$ \\
    \bottomrule
    \end{tabular}%
\end{table*}

%% file: appendix/subsystem_models.tex
\section{Subsystem Model Specifications}
\label{ap:subsystem_models}
Table \ref{tab:subsystem_comparison} provides a detailed comparison of the three subsystem models.

\subsection{Wing Model}

\begin{figure*}[!h]
    \centering
    \includegraphics[width=0.9\textwidth]{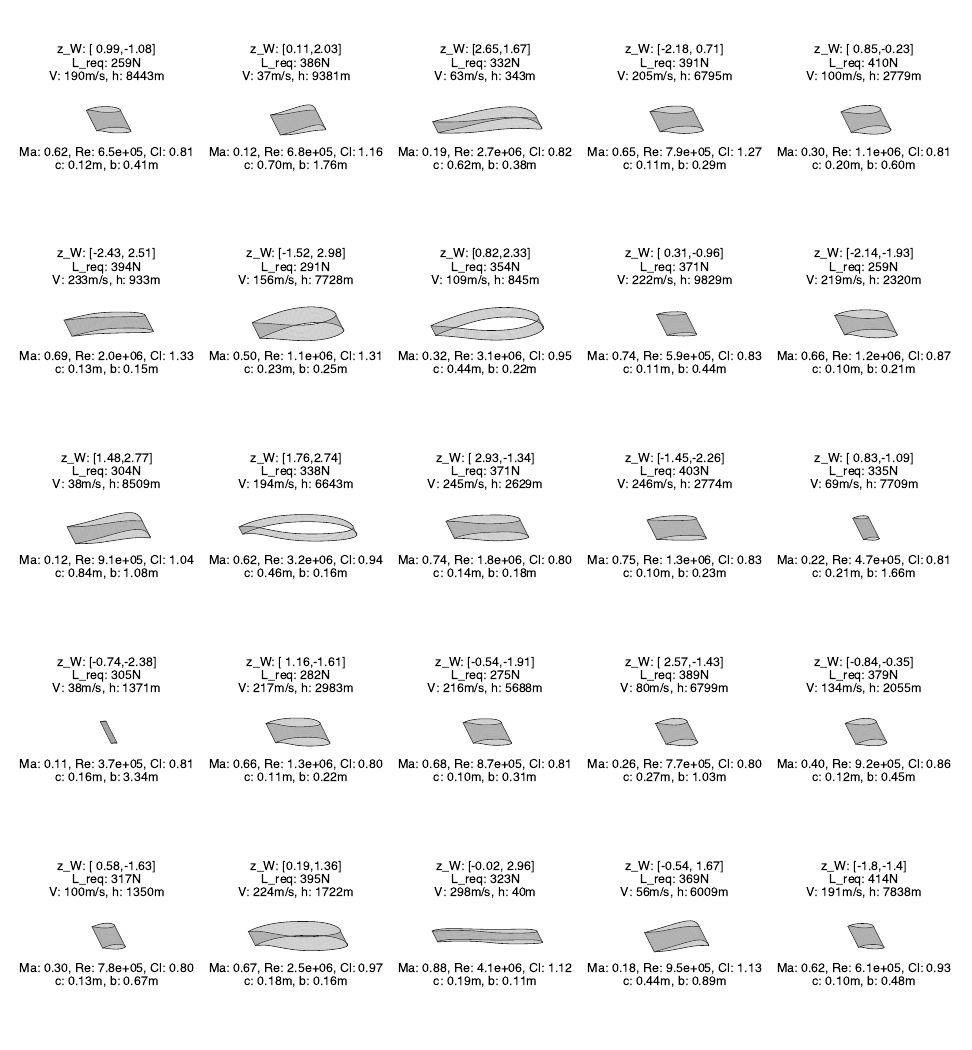}
    \caption[Wing model sample generation]{Wing model sample generation. The wing model maps flight requirements ($L_{\text{req}}, V, h$) and latent code $z_W$ to wing parameters ($c, b, C_L$). It also outputs conditions ($Ma, Re, C_L$) for the inner airfoil model, which generates the corresponding airfoil shape and angle of attack. The airfoil is then scaled by chord $c$ and extruded along span $b$.}
    \label{fig:wing_samples}
\end{figure*}

The wing model builds on existing work by Chen and Fuge \cite{chenBezierGANAutomaticGeneration2021}. The outer conditional generator takes a latent code $z \in \mathbb{R}^2$ together with flight requirements (velocity $V$, altitude $h$, wing lift requirement $L_{\text{req}}$) and produces the corresponding wing parameters. 
These parameters are chord $c$, span $b$, and coefficient of lift $C_L$. Training proceeds in two distinct phases: physics-based losses are applied during the initial epochs 0 to 150, after which diversity objectives are gradually introduced. A Q-network provides mutual information regularization to encourage meaningful latent representations. The resulting model satisfies lift requirements across a range of flight conditions while maintaining aerodynamic validity. The outer wing model provides conditions (Mach number, Reynolds number based on chord and velocity, and coefficient of lift $C_L$) to an inner airfoil model (BézierGAN). The inner airfoil model generates an airfoil shape (coordinate points and angle of attack). The airfoil is then scaled to the chord $c$ and extruded linearly along the span $b$.

\subsection{Airfoil Model}

For more information, please refer to \cite{chenBezierGANAutomaticGeneration2021}. We use the CEBGAN architecture trained on 995 optimized airfoil samples. The model maps 3 conditions (Mach number, Reynolds number, and coefficient of lift $C_l$) to 192 $\times$ 2 points (x, z coordinates of airfoil surface) and angle of attack. 
The wing model is trained to map ($z_W$, $L_{\text{req}}$) to appropriate airfoil parameters ($C_l$, $Re$, $Ma$) given environmental conditions (altitude, velocity). The airfoil is then scaled using chord $c$ and extruded linearly along the span $b$. These ($b, c$) are also outputs of the wing model.

\subsection{Fuselage Model}

\begin{figure*}[!h]
    \centering
    \includegraphics[width=0.6\textwidth]{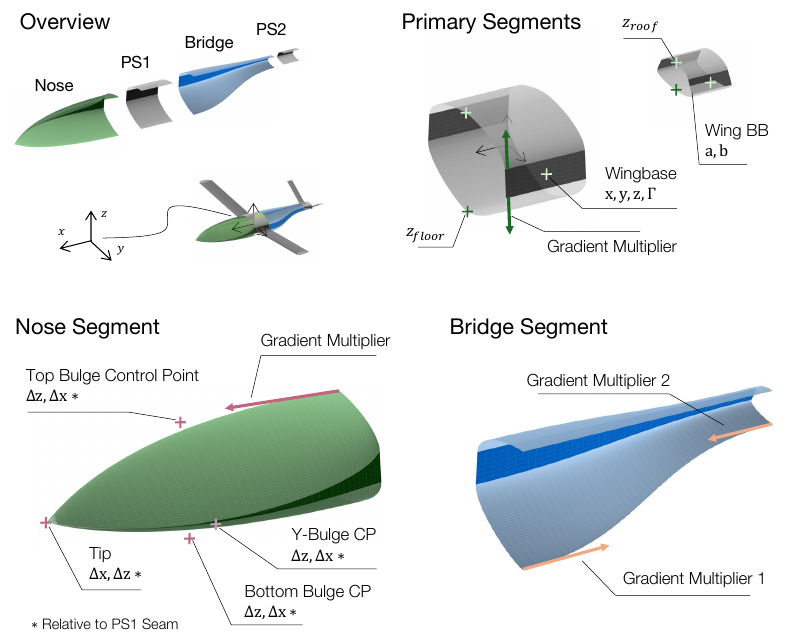}
    \vspace{0.2cm}
    \caption[Fuselage parametrization overview]{Fuselage parametrization overview showing primary segments, bridge, and nose components. The nose and bridge segments consist of three bicubic Bézier surface patches each. The primary segments consist of three linearly extruded cubic Bézier curves. The middle primary segment Bézier curves are perfectly straight and act as the wing interfaces.}
    \label{fig:fuselage_parametrization}
\end{figure*}

We employ a custom \acs{VAE} trained on parametric fuselage geometries, as illustrated in Figure \ref{fig:fuselage_parametrization}. The \acs{VAE} maps a four-dimensional latent code $z \in \mathbb{R}^4$ to 15 geometric parameters. Note that the full fuselage parametrization contains additional parameters that are fixed for simplicity (see Figure \ref{fig:fuselage_parametrization} for an overview). The training data consists of 551 valid designs selected from 200k candidates through rejection sampling (convexity, self-intersection, \textit{etc.}), yielding a 0.28\% acceptance rate. We employ \ac{SN} to ensure smooth $z_F \to x_F$ mappings and incorporate a minimum volume loss \cite{CompressingLatentSpace2023} to impose structure on the latent space. The resulting model performs unconditional generation with smooth interpolation properties.

Figure~\ref{fig:fuselage_sweep} illustrates the effect of the minimum volume loss on the latent space organization. The loss promotes disentanglement, resulting in principal axes that correspond to distinct geometric features: the first axis controls tail dihedral, the second tail location, the third main section width, and the fourth isotropic scaling. While other features (\textit{e.g.}, nose shape, front dihedral) also vary, the minimum volume loss effectively aligns the most dominant variations with the latent axes. Note that we fix the latent dimension to four without extensive dimensionality analysis.

\begin{figure*}[!h]
    \centering
    \includegraphics[width=0.9\textwidth]{figures/fuselage_sweep/fuselage_sweep.pdf}
    \caption[Fuselage latent space sweep]{Fuselage latent space sweep illustrating disentanglement. Each row shows variations along a single latent axis $z_i$ while holding others fixed at zero.}
    \label{fig:fuselage_sweep}
\end{figure*}

\subsection{Internals Component Placement Model}

We use a custom autoregressive model that serves as a simple placeholder for more detailed component models. The autoregressive model takes $z \in \mathbb{R}^4$ and volumes $\{V_i\}$ and produces box parameters $\{(\ell_i, w_i, h_i, x_i, y_i, z_i)\}_i$. The model is trained with variable sequence length, but for simplicity during optimization and \ac{GLUE} model training we fix the sequence length to three boxes. An embedding network processes placement history for spatial reasoning. We employ data-free training with geometric losses. These are: overlap, accurate volume (constraint to match each box's volume to the conditional requirement), compactness of the layout, and adjacent diversity (encouraging diverse configurations). An auxiliary network provides mutual information regularization. The inference procedure includes a cleanup step: after initial box prediction, a gradient descent procedure refines box positions to eliminate collisions and improve compactness. This cleanup step is crucial for ensuring physically valid configurations. Additionally, for the sake of simplicity, all boxes are placed with their centers at $y=0$ and are symmetric with regard to the $xz$ plane. 

In the long term, we are interested in more complex subsystem models for internal components and wish to build upon this simple box-based model.

\begin{table*}[h]
\centering
\caption[Detailed comparison of subsystem models]{Detailed comparison of subsystem models for wing, fuselage, and box arrangement (internals).}
\label{tab:subsystem_comparison}
\footnotesize
\begin{tabular}{@{}lp{4cm}p{4cm}p{3cm}@{}}
\toprule
\textbf{Aspect} & \textbf{Wing} & \textbf{Fuselage} & \textbf{Internals} \\
\midrule
\textbf{Model Type} & Conditional Generator & \acs{VAE} & Autoregressive \ac{MLP} \\
\textbf{Latent Dim} & 2 & 4 & 4 \\
\textbf{Output Dim} & 3 & 15 & $6 \times \text{seq\_len}$ \\
\textbf{Training Epochs} & 1,500 & 1,000 & 2,500 \\
\textbf{Batch Size} & 256 & 128 & 256 \\
\textbf{Learning Rate} & $10^{-3} \to 5 \times 10^{-4}$ & $2 \times 10^{-4}$ & $5 \times 10^{-4}$ \\
\midrule
Loss 1 & Lift Accuracy \newline ($\lambda=400$) & Reconstruction \ac{MSE} & Overlap Penalty \newline ($\lambda: 4 \to 20$) \\
Loss 2 & Reynolds Constraint \newline ($\lambda: 1 \to 1000$) & \ac{KL} Divergence \newline ($\lambda=0.04$) & Volume Deficit \newline ($\lambda: 5 \to 10$) \\
Loss 3 & \ac{PA-DPP} \newline ($\lambda: 0 \to 5 \times 10^{-5}$) & Least Volume \newline ($\lambda=0.12$) & Compactness \newline ($\lambda: 2 \to 4$) \\
Loss 4 & Mutual Information \newline ($\lambda=0.2$) & --- & Layout Diversity (\ac{DPP}) \newline ($\lambda=0.5$) \\
Loss 5 & --- & --- & Aspect Ratio Diversity \newline ($\lambda: 4 \to 1$) \\
Loss 6 & --- & --- & Mutual Information \newline ($\lambda=0.1$) \\
\midrule
\textbf{Total Active Losses} & 4 & 3 & 6 \\
\textbf{Ramp Schedule} & Yes (epoch 150) & No & Yes (epoch 500--1500) \\
\midrule
\textbf{Training Data} & Synthetic (online) & Filtered 200k $\to$ 551 & Synthetic (online) \\
\bottomrule
\end{tabular}%
\end{table*}

\subsection{Loss Formulations}
\subsubsection{Wing Model}

The total loss for the wing model is given by:
\begin{equation}
\mathcal{L}_{\text{wing}} = \lambda_{\mathrm{lift}} \mathcal{L}_{\text{lift}} + \lambda_{\mathrm{Re}} \mathcal{L}_{\text{Re}} + \lambda_{\mathrm{\acs{DPP}}} \mathcal{L}_{\text{\acs{PA-DPP}}} + \lambda_{\mathrm{\acs{MI}}} \mathcal{L}_{\text{\acs{MI}}}
\end{equation}

The lift accuracy loss is given by:
\begin{equation}
\mathcal{L}_{\text{lift}} = \text{\acs{MSE}}\left(\frac{L}{10^4}, \frac{L_{\text{req}}}{10^4}\right), \quad L = \frac{1}{2} \rho V^2 C_L c b
\end{equation}
where $C_L = C_\ell / (1 + C_\ell / (\pi e \cdot \text{AR}))$, $\text{AR} = b/c$, $e = 0.8$ (Oswald efficiency factor), and $\lambda_{\mathrm{lift}} = 400$.

To ensure that the Airfoil model (\cite{chenBezierGANAutomaticGeneration2021}) is only used with Reynolds numbers within the training range, we employ a penalty to enforce wing model outputs to be within this range.
\begin{equation}
\mathcal{L}_{\text{Re}} \propto \sum_{i=1}^{B} \frac{ \text{ReLU}(\text{Re}_{\min} - \text{Re}_i) + \text{ReLU}(\text{Re}_i - \text{Re}_{\max})}{B}
\end{equation}
where $\text{Re} = \rho V c / \mu$ with $\text{Re} \in [10^5, 10^8]$. Ramp schedule: $\lambda_{\mathrm{Re}} = 1$ for epochs $< 150$, linear ramp to $1000$ over epochs $150$--$200$.

The performance-augmented \ac{DPP} loss \cite{chenPaDGANLearningGenerate2020} is given by:
\begin{equation}
\mathcal{L}_{\text{\acs{PA-DPP}}} = -\log \det(\bm{L}), \quad \bm{L} = \bm{K} \odot (\bm{q} \bm{q}^\top)
\end{equation}
where the similarity kernel is defined as $K_{ij} = \exp\left(-\frac{\|\bm{x}_i - \bm{x}_j\|_2^2}{2\sigma_{\text{DPP}}^2}\right)$ with $\sigma_{\text{DPP}} = 0.1$, applied to normalized wing parameters $\bm{x} = (c, b, C_L) \in [0, 1]^3$. The quality score is computed as $q_i = (L/D)_i^2 / m_i^2$, where $L/D = \pi (b/c) e / C_\ell$ represents the lift-to-drag ratio and $m = C_{mw} M_{\text{root}} / g + C_{ar} b$ represents the wing mass with coefficients $C_{mw} = 4 \times 10^{-4}$ and $C_{ar} = 1$.
We use a ramp schedule: $\lambda_{\mathrm{DPP}} = 0$ for epochs $< 150$, linear ramp to $5 \times 10^{-5}$ over epochs $150$--$200$.

To enforce disentanglement for the wing model, we employ a mutual information loss. This is given by:
\begin{equation}
\mathcal{L}_{\text{\acs{MI}}} = \text{\acs{MSE}}(\bm{z}_{\text{pred}}, \bm{z}), \quad \bm{z}_{\text{pred}} = Q(\bm{x}, V, h, L_{\text{req}})
\end{equation}
where $Q$ is a Q-Network trained to invert the generator mapping, with $\lambda_{\mathrm{MI}} = 0.2$.

\subsubsection{Fuselage Model}

The total loss for the fuselage model is given by:
\begin{equation}
\mathcal{L}_{\text{fuselage}} = \mathcal{L}_{\text{recon}} + \lambda_{\mathrm{KL}} \mathcal{L}_{\text{KL}} + \lambda_{\mathrm{LV}} \mathcal{L}_{\text{LV}}
\end{equation}
    
The reconstruction loss is given by:
\begin{equation}
\mathcal{L}_{\text{recon}} = \text{\acs{MSE}}(\bm{x}_{\text{recon}}, \bm{x})
\end{equation}

The \ac{KL} divergence loss is given by:
\begin{equation}
\mathcal{L}_{\text{\acs{KL}}} = -\frac{1}{2B} \sum_{i=1}^{B} \sum_{j=1}^{d_z} \left(1 + \log(\sigma_{\text{enc},ij}^2) - \mu_{\text{enc},ij}^2 - \sigma_{\text{enc},ij}^2\right)
\end{equation}
where $\mu_{\text{enc}}, \sigma^2_{\text{enc}}$ are encoder outputs, $d_z = 4$ is the latent dimension, $B$ is the batch size, and $\lambda_{\mathrm{KL}} = 0.04$. Note that $\mu_{\text{enc}}$ and $\sigma^2_{\text{enc}}$ denote the VAE encoder outputs and should not be confused with the ALM penalty parameter $\mu_k$ (\ref{ap:alm_schemes}) or the DPP kernel length scale $\sigma_{\text{DPP}}$.

The least volume loss is used for disentanglement of the latent space. It is given by:
\begin{equation}
\mathcal{L}_{\text{LV}} = \exp\left(\frac{1}{d_z} \sum_{j=1}^{d_z} \log(\text{std}(\bm{z}_j) + \eta)\right), \quad \eta = 10^{-4}
\end{equation}
where $\text{std}(\bm{z}_j)$ is the standard deviation of the $j$-th latent dimension across the batch, and $\lambda_{\mathrm{LV}} = 0.12$. Please note that this loss is used in conjunction with \ac{SN} and is in competition with the KL divergence.

\subsubsection{Internals Model}

The total loss for the internals model is given by:
\begin{align}
\mathcal{L}_{\text{internals}} = &\lambda_{\mathrm{overlap}} \mathcal{L}_{\text{overlap}} + \lambda_{\mathrm{volume}} \mathcal{L}_{\text{volume}} + \lambda_{\mathrm{compact}} \mathcal{L}_{\text{compact}} \\
&+ \lambda_{\mathrm{LDV}} \mathcal{L}_{\text{LDV}} + \lambda_{\mathrm{AR-DPP}} \mathcal{L}_{\text{AR-DPP}} + \lambda_{\mathrm{MI}} \mathcal{L}_{\text{MI}}
\end{align}

The overlap penalty is given by:
\begin{equation}
\mathcal{L}_{\text{overlap}} = \frac{1}{B} \sum_{i=1}^{B} \frac{\|\bm{A}_i^{\text{overlap}}\|_2}{\|\bm{V}_i\|_2}
\end{equation}
where $\bm{A}_i^{\text{overlap}}$ is the overlap matrix and $\bm{V}_i$ are the box volumes for sample $i$. Ramp schedule: $\lambda_{\mathrm{overlap}} = 4$ for epochs $< 500$, linear ramp to $20$ over epochs $500$--$1500$.

The volume deficit loss is given by:
\begin{equation}
\mathcal{L}_{\text{volume}} = \left\| \frac{\bm{V}_{\text{actual}} - \bm{V}_{\text{target}}}{\bm{V}_{\text{target}}} \right\|_2
\end{equation}
Ramp schedule: $\lambda_{\mathrm{volume}} = 5$ for epochs $< 500$, linear ramp to $10$ over epochs $500$--$1500$.

The compactness loss is given by:
\begin{equation}
\mathcal{L}_{\text{compact}} = 0.6 \cdot \text{slenderness} + 0.4 \cdot \text{fill}^{-1}
\end{equation}
where slenderness is defined as $\text{slenderness} = \sqrt{A_{\text{frontal}} / \sqrt{A_{\text{side}} \cdot A_{\text{top}}}}$ and the inverse fill ratio is computed as $\text{fill}^{-1} = V_{\text{bounding}} / \sum_i V_i$.
Ramp schedule: $\lambda_{\mathrm{compact}} = 2$ for epochs $< 500$, linear ramp to $4$ over epochs $500$--$1500$.

The layout diversity via adjacency loss is given by:
\begin{align}
    \mathcal{L}_{\text{LDV}} &= \mathcal{L}_{\text{DPP}}(\bm{A}_x) + \mathcal{L}_{\text{DPP}}(\bm{A}_z), \\
    \text{where} \quad 
    & \mathcal{L}_{\text{DPP}}(\bm{A}) = -\frac{1}{B} \log \det(\bm{K}_A)
\end{align}
where $\bm{A}_x, \bm{A}_z$ are adjacency matrices in $x$ and $z$ directions, $K_{A,ij} = \exp(-\|\text{vec}(\bm{A}_i) - \text{vec}(\bm{A}_j)\|_2^2 / (2\sigma_{\text{DPP}}^2))$ with $\sigma_{\text{DPP}} = 1$, and $\lambda_{\mathrm{LDV}} = 0.5$.

The aspect ratio diversity loss is given by:
\begin{equation}
\mathcal{L}_{\text{AR-DPP}} = -\frac{1}{B} \log \det(\bm{K}_{\text{AR}}), \quad \text{AR} = \log\left[\frac{w}{\ell}, \frac{h}{\ell}\right]
\end{equation}
where $K_{\text{AR},ij} = \exp(-\|\text{vec}(\text{AR}_i) - \text{vec}(\text{AR}_j)\|_2^2 / (2\sigma_{\text{DPP}}^2))$ with $\sigma_{\text{DPP}} = 1$. Ramp schedule: $\lambda_{\mathrm{AR-DPP}} = 4$ for epochs $< 500$, linear ramp to $1$ over epochs $500$--$1500$ (decreasing). 
This loss is used to encourage boxes with varying aspect ratios.

The mutual information loss is given by:
\begin{equation}
\mathcal{L}_{\text{MI}} = -\frac{1}{B} \sum_{i=1}^{B} \log \frac{\exp(\bm{z}_i^\top \hat{\bm{z}}_i / \tau)}{\sum_{j=1}^{B} \exp(\bm{z}_i^\top \hat{\bm{z}}_j / \tau)}
\end{equation}
where $\hat{\bm{z}}_i$ is the predicted latent code from the auxiliary predictor, $\tau = 0.1$ is the temperature, and $\lambda_{\mathrm{MI}} = 0.1$.

\subsection{Architecture Details}

\subsubsection{Wing Model}

The wing model consists of two networks: a generator $S_W: (z, V, h, L_{\text{req}}) \in \mathbb{R}^5 \to (c, b, C_L) \in \mathbb{R}^3$ and a Q-Network for mutual information regularization mapping $(c, b, C_L, V, h, L_{\text{req}}) \in \mathbb{R}^6 \to z \in \mathbb{R}^2$. Both networks are implemented as 4-layer \acp{MLP} with hidden size 32 and ReLU activations. Input and output normalization is performed via registered buffers using affine scaling to the range $[-1, 1]$. Training proceeds with alternating updates, first updating the Q-Network and then the generator.

\subsubsection{Fuselage Model}

The fuselage \acs{VAE} comprises an encoder mapping $\bm{x} \in \mathbb{R}^{15} \to (\mu_{\text{enc}}, \log \sigma^2_{\text{enc}}) \in \mathbb{R}^4 \times \mathbb{R}^4$ and a decoder mapping $\bm{z} \in \mathbb{R}^4 \to \bm{x} \in \mathbb{R}^{15}$. The architecture employs a main path with layer sizes 128, 64, and 32, augmented with skip connections from dimension 15 to 32 and from dimension 4 to 15. \ac{SN} is applied to most layers to ensure smoothness in the latent-to-parameter mapping. Data normalization is performed using z-score standardization: $(\bm{x} - \mu_{\text{train}}) / \sigma_{\text{train}}$.

\subsubsection{Internals Model}

The internals model employs an autoregressive architecture with three key components: an embedding network $\bm{b}_i \in \mathbb{R}^6 \to \bm{e}_i \in \mathbb{R}^{1024}$ that encodes box parameters, a box generator $(\bm{e}_{\text{cum}}, \bm{z}, \text{seq\_len}, V_i) \in \mathbb{R}^{1029} \to \bm{b}_i \in \mathbb{R}^6$ that produces the next box, and an auxiliary predictor $(\bm{x}_i, \bm{e}_{\text{cum}}, \text{seq\_len}) \in \mathbb{R}^{1028} \to \Delta\bm{z}_i \in \mathbb{R}^4$ for mutual information regularization. The model operates autoregressively, updating the cumulative embedding as $\bm{e}_{\text{cum}} \leftarrow \bm{e}_{\text{cum}} + \bm{e}_i$ after generating each box. The final latent prediction is computed as $\hat{\bm{z}} = \sum_{i=1}^{N} \Delta\bm{z}_i$. All \acp{MLP} in this architecture use LeakyReLU activations.

\subsection{Training Strategies}

\subsubsection{Wing Model}

The wing model is trained using a two-stage learning rate schedule: $10^{-3}$ for epochs 0 to 150, then $5 \times 10^{-4}$ for subsequent epochs. Diversity enforcement is introduced at epoch 150, allowing the model to first converge on physics-based objectives. Each epoch processes 12 batches of 256 samples, totaling 3,072 samples per epoch. The Reynolds constraint and \ac{DPP} loss ramps are synchronized with the learning rate transition. Training uses random seed 42 and the Adam optimizer with default PyTorch parameters ($\beta_1 = 0.9$, $\beta_2 = 0.999$, $\epsilon = 10^{-8}$).

\subsubsection{Fuselage Model}

The fuselage model uses a constant learning rate of $2 \times 10^{-4}$ with constant loss weights throughout training (no ramping). The 551 valid designs are split 80\%/20\% into training and test sets. Test set evaluation is performed every epoch for monitoring purposes. Training employs random seed 5 and the Adam optimizer with default PyTorch parameters ($\beta_1 = 0.9$, $\beta_2 = 0.999$, $\epsilon = 10^{-8}$).

\subsubsection{Internals Model}

The internals model is trained with a constant learning rate of $5 \times 10^{-4}$ over three distinct phases. Phase one (epochs 0 to 500) focuses on initial convergence with lower penalty weights. Phase two (epochs 500 to 1500) implements progressive ramping with linear weight increases. Phase three (epochs 1500 to 2500) performs fine-tuning under maximum constraint penalties. All sequence lengths (1 to 3 boxes) are trained simultaneously, with random target volumes sampled per epoch to ensure generalization. Training uses random seeds torch=42 and numpy=5, and employs the Adam optimizer with default PyTorch parameters ($\beta_1 = 0.9$, $\beta_2 = 0.999$, $\epsilon = 10^{-8}$).

\subsection{Dataset Generation: Fuselage}

The fuselage dataset is generated via rejection sampling with extensive geometric validation. Beginning with 200,000 randomly sampled parameter sets, candidates undergo a multi-stage filtering process. Stage one performs preliminary checks on minimum and maximum dimensions, aspect ratios, and z-ordering. Stage two tests for self-intersection in the front and rear primary segments. Stage three checks for intersections with the symmetry plane. Stage four evaluates convexity, requiring all candidates to be convex. Stage five detects overlaps in bridge segments. Stage six validates smooth transitions between all adjacent segments. Following validation, accepted designs undergo post-validation scaling, where a random uniform factor from the range $[0.15, 0.4]$ is applied to all length parameters. This rigorous process yields 551 valid designs from the initial 200,000 candidates, representing a 0.28\% acceptance rate. The fuselage geometry comprises four interconnected components: front primary segment, rear primary segment, bridge, and nose.

%% file: appendix/architecture_details.tex
\section{\acs{DF}-\acs{GLUE} Architecture}
\label{app:df-architecture_details}

We first detail the data-free training architecture used in all experiments.
We then ablate core components to isolate their effect and motivate the default architecture.

\subsection{Default Architecture}

\begin{figure*}[!htb]
    \centering
    \includegraphics{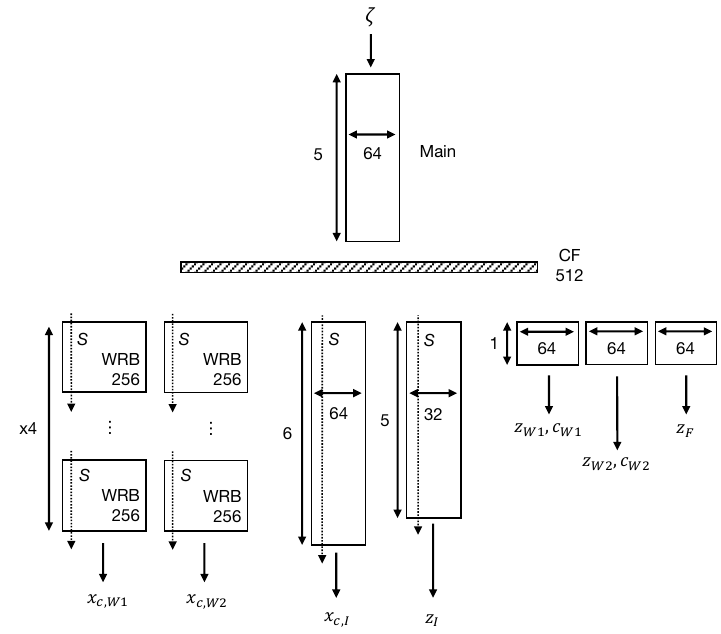}
    \caption{\acs{DF}-\acs{GLUE} architecture used throughout this work (Architecture 1). MLP-based main block with subsystem latent and placement heads. Forward pass top-to-bottom, backward pass bottom-to-top.}
    \label{fig:arch1}
\end{figure*}

Figure \ref{fig:arch1} shows the architecture for \ac{DF}-\ac{GLUE}.
A shared main block forms a 5-layer \ac{MLP} of width 64.
It maps the system-level latent $\zeta$ to a \ac{CF} vector.
Dedicated \ac{MLP} heads map \ac{CF} to each subsystem latent and condition ($z_i$, $c_i$) and to placement variables ($x_{c,i}$) where required.
The fuselage uses no placement head because it is fixed to the origin.

Latent and condition heads for the fuselage and wings 1 (front) and 2 (tail) use a single linear layer with tanh and sigmoid output activations, followed by per-dimension denormalization to the latent and condition scales of the frozen subsystem models.
For these subsystems, the shared main block carries most coordination capacity.

Internals latent and placement heads use deeper \acp{MLP} with ReLU activations and one skip connection to mitigate vanishing gradients.
This design supports gradient flow from internals objectives into the shared main block.

Wing placement heads use \acp{WRB} with ReLU activations as the basic unit.
Each \ac{WRB} forms a 3-layer \ac{MLP} with dimensions $256 \to 512 \to 256$.
A skip connection spans the full block.
The default wing placement head stacks four \acp{WRB} in sequence, which provides skip connections at every block and maintains a direct skip path from the head input to the head output.

All head outputs use tanh or sigmoid to enforce bounded variables and then apply per-output denormalization scales.

The head depth for each subsystem results from manual tuning.

\subsubsection{Weight Initialization}

For the \ac{GLUE} model, we use custom weight initialization for \texttt{nn.Sequential} modules.
We iterate through layers, identify \texttt{nn.Linear} layers, and apply initialization based on the subsequent activation function.

For each \texttt{nn.Linear} layer with weight matrix \( \mathbf{W} \in \mathbb{R}^{n_{\text{out}} \times n_{\text{in}}} \):

\paragraph{ReLU Activation} Kaiming Uniform initialization
\begin{equation}
W_{ij} \sim \mathcal{U}(-a, a), \quad \text{where} \quad a = \sqrt{\frac{6}{n_{\text{in}}}}
\end{equation}

\paragraph{Tanh Activation} LeCun-like Normal initialization with more variance
\begin{equation}
W_{ij} \sim \mathcal{N}(0, \sigma^2), \quad \text{where} \quad \sigma^2 = \frac{4}{n_{\text{in}}}
\end{equation}

\paragraph{Sigmoid Activation} LeCun-like Normal initialization with more variance
\begin{equation}
W_{ij} \sim \mathcal{N}(0, \sigma^2), \quad \text{where} \quad \sigma^2 = \frac{4}{n_{\text{in}}}
\end{equation}

where \( n_{\text{in}} \) is the fan-in (number of input features).

\paragraph{Bias Initialization}

All biases are initialized to zero:
\begin{equation}
b_i = 0 \quad \forall i
\end{equation}

This initialization only affects \texttt{nn.Linear} layers within \texttt{nn.Sequential} blocks. Skip connections and standalone linear layers defined outside sequential containers are not initialized by this function.

\subsubsection{Wing Placement Optimizer}
 
\begin{figure}[!h]
    \centering
    \includegraphics{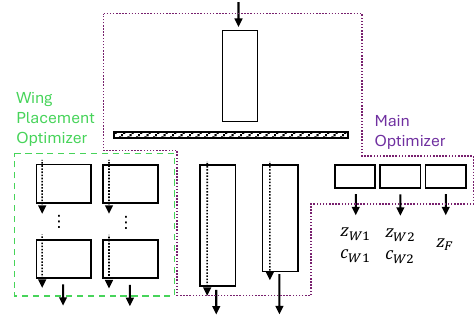}
    \caption{Dedicated optimizer for wing placement block.}
    \label{fig:inner-adam-explainer}
\end{figure}

 A dedicated optimizer for the wing placement heads as shown in Figure \ref{fig:inner-adam-explainer} mitigates mode collapse, see Section \ref{sec:mode-collapse-ablation}. In the ablation, this mechanism is denoted as inner optimizer.
Without this separation, gradients from $\Cwx$ and $\Cdi$ backpropagate into the shared main block and drive collapse that prioritizes these constraints. 
The wing placement heads predict ${x}_{c,W1}$ and ${x}_{c,W2}$ from the \ac{CF} vector produced by the main block.
This separation prevents wing placement gradients from directly shaping fuselage and wing latent outputs.

\subsection{Architecture Ablation}

With our setup, the ability of \ac{DF}-\ac{GLUE} to learn constraints and optimize performance is sensitive to its architecture.
We now ablate core components to isolate architectural factors that drive this sensitivity.
We compare three alternatives against the default, which yields four architectures in total.

\paragraph{Architecture 1}
This is the default and matches Figure \ref{fig:arch1}.

\paragraph{Architecture 2}
Matches Architecture 1 but removes the custom initialization. For Architecture 2, we use PyTorch default initialization for model weights and biases. This reduces initial weight variance and, consequently, output variance. At the first training epoch, the output designs are virtually mode-collapsed.

\begin{figure*}[htbp]
    \centering
    \includegraphics{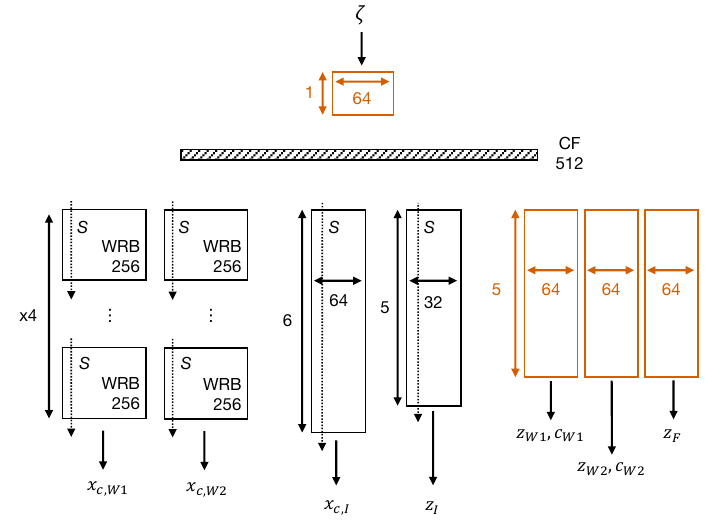}
    \caption{\acs{DF}-\acs{GLUE} Architecture 3. Compared to default Architecture 1, capacity is moved from the main block into the fuselage and wing latent heads while keeping the rest of the network unchanged.}
    \label{fig:arch3}
\end{figure*}

\paragraph{Architecture 3}
Shifts capacity from the shared main block into the fuselage and wing latent heads while keeping all other components identical to Architecture 1. Architecture 3 is shown in Figure \ref{fig:arch3}. The changes are marked in color. This ablation probes the capacity balance between the shared main block and subsystem heads.

\begin{figure*}[htbp]
    \centering
    \includegraphics{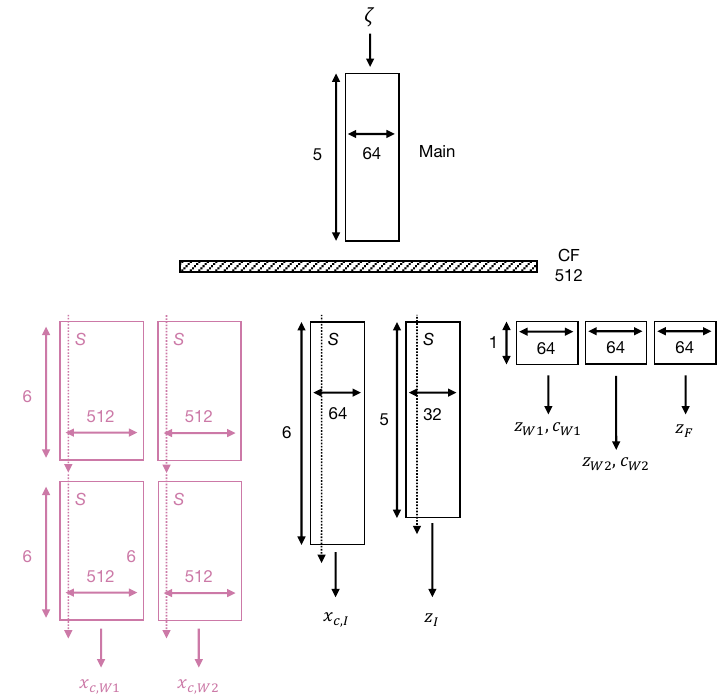}
    \caption{\acs{DF}-\acs{GLUE} Architecture 4. Replaces the Wide Residual Blocks (WRB) wing placement heads of Architecture 1 with deep sequential MLP with fewer skip connections.}
    \label{fig:arch4}
\end{figure*}

\paragraph{Architecture 4}
Modifies only the wing placement heads.
It replaces the \acp{WRB} design with a deep sequential \ac{MLP} with ReLU activations.
It adds skip connections every sixth layer, which yields substantially fewer skip connections than Architecture 1. Architecture 4 is shown in Figure \ref{fig:arch4}.

\subsubsection{Results}

\begin{figure*}[!h]
    \centering
    \includegraphics{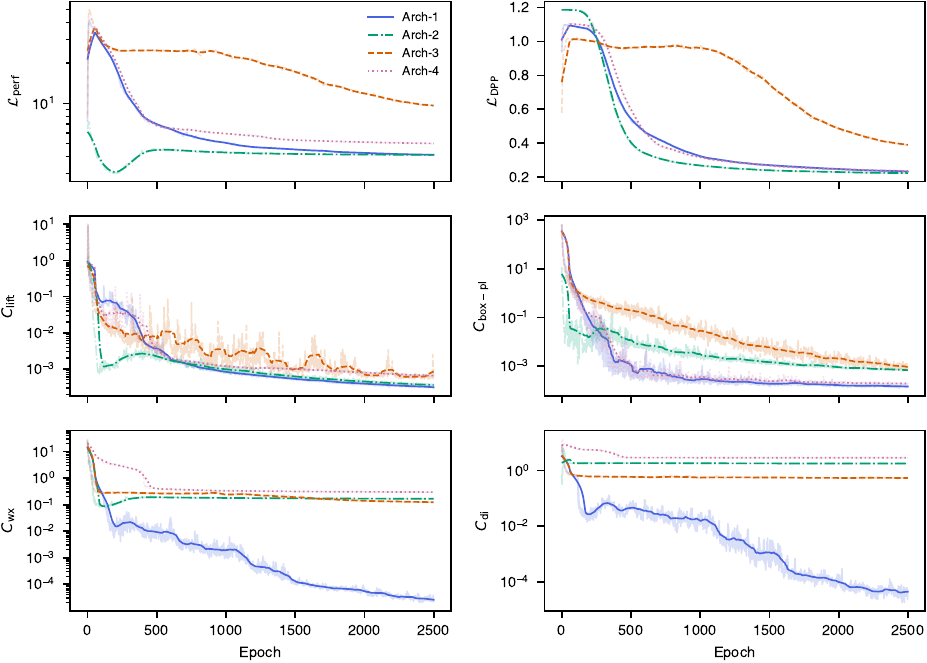}
    \hspace{1cm}
    \caption{\acs{DF}-\acs{GLUE} training losses over 2500 epochs for Architectures 1--4 (10 seeds, mean only).}
    \label{fig:arch_ablation_losses}
\end{figure*}

Figure \ref{fig:arch_ablation_losses} reports training losses over 2500 epochs with $\lambda_{\mathrm{opt}} = 0.1$ and $\lDPP = 1$ using DPP on $\mathcal{X}$.
Each architecture uses 10 seeds and the plots show mean curves only.
We omit min and max ranges for visual clarity.
Please see Eq.~\ref{eq:dfglued-loss} for the full loss formulation. 
We use $\lambda_{\mathrm{MI}} = 0$, and no subsystem-specific diversity enforcement ($\lambda_{\mathrm{DPP},z_F} = \lambda_{\mathrm{DPP},z_W} = \lambda_{\mathrm{DPP},z_I} = 0$). 
We use hypercube \ac{ALM} with $\gamma = 5\cdot 10^{-3}$. 

\begin{figure*}[!h]
    \centering
    \includegraphics{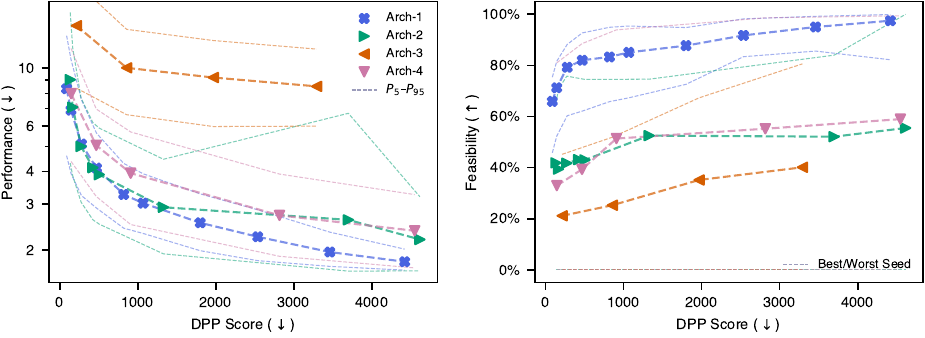}
    \caption{Feasibility and performance--diversity trade-off across Architectures 1--4 (10 seeds): Architecture 1 achieves consistently high feasibility and the best trade-off, Architecture 3 degrades both feasibility and performance most strongly, and Architecture 4 attains best-seed feasibility similar to Architecture 1 but with much poorer mean robustness.}
    \label{fig:arch_ablation_opt_vs_div}
\end{figure*}

All experiments use the dedicated wing placement optimizer in Figure \ref{fig:inner-adam-explainer}.
Figure \ref{fig:arch_ablation_opt_vs_div} summarizes feasibility and the performance--diversity trade-off across architectures. Shown are the single objective $\Omass$, diversity (\ac{DPP}), and four constraints. $\Cbb$ is omitted, as it is by far the easiest of the constraints to satisfy.

Architecture 1 (blue) is the baseline and represents successful training. In Figure \ref{fig:arch_ablation_losses}, it achieves low loss across all constraints and objectives, which is indicative of adequate learning. As seen in Figure \ref{fig:arch_ablation_opt_vs_div}, it maintains high feasibility even at high diversity while offering the best performance--diversity trade-off.

\paragraph{Architecture 2}
Architecture 2 removes custom initialization. This degrades box placement and wing placement losses relative to Architecture 1.
Consistent with this trend, Figure \ref{fig:arch_ablation_opt_vs_div} shows lower mean feasibility and a weaker performance--diversity trade-off.
Architecture 2 also shows a distinct early training trajectory in which performance improves at initialization rather than initially degrading.

\paragraph{Architecture 3}
Architecture 3 shifts capacity from the shared main block into the fuselage and wing latent heads. This degrades losses across most objectives and constraints in Figure \ref{fig:arch_ablation_losses}.
It attains the lowest mean feasibility and the weakest performance--diversity trade-off, as shown in Figure \ref{fig:arch_ablation_opt_vs_div}.

\paragraph{Architecture 4}
Architecture 4 replaces the \acp{WRB} wing placement heads with a deep sequential \ac{MLP} with fewer skip connections.
It matches Architecture 1 on $\Clift$ and $\Cboxpl$ (internals placement), but fails on wing placement constraints $\Cwx$ and $\Cdi$.
In Figure \ref{fig:arch_ablation_opt_vs_div}, its best seed reaches feasibility comparable to Architecture 1 while the mean feasibility and performance--diversity trade-off degrade. In practice, some seeds perform well, but a few runs remain at very high loss for $\Cwx$ and $\Cdi$ and, on the log scale, dominate the mean.
Thus, the high loss curves for $\Cwx$ and $\Cdi$ in Architecture 4 mainly reflect poor robustness across seeds, not uniformly poor performance. The same is also true for Architecture 2.

\subsubsection{Discussion}

\paragraph{Architecture 2}
Architecture 2 removes the custom initialization and therefore starts with substantially lower output variance.
As a result, early training exhibits low diversity in $\mathcal{Z}$, \textit{i.e.}, designs are practically mode-collapsed at training epoch 0.
For the seeds that fail to learn the wing placement constraints ($\Cwx$ and $\Cdi$), we observe that the corresponding head outputs at initialization are clustered at the saturation limits of the tanh or sigmoid nonlinearities.
In this regime, gradients through the output activations are strongly attenuated, which impairs corrective updates and can prevent the optimizer from moving the outputs back into a high-gradient region.
This explains the reduced robustness of Architecture 2. Convergence is not uniformly worse, but a subset of seeds becomes trapped by activation saturation and the resulting vanishing-gradient bottleneck. Low output variance increases sensitivity to small stochastic differences across runs, as without significant output variance the whole training batch can be in a vanishing gradient regime.

\paragraph{Architecture 3}
Architecture 3 shifts capacity from the shared main block into the fuselage and wing latent heads.
While this increases expressivity for predicting $z_F$, $z_{W,1}$, and $z_{{W,2}}$ from \ac{CF}, it simultaneously reduces the capacity of the shared mapping $\zeta \mapsto \ac{CF}$ that all other heads rely on.
In effect, \ac{CF} becomes closer to a shallow, near-linear (single nonlinearity due to one layer) lifting of the system-level latent $\zeta$, rather than a rich intermediate representation that encodes higher-order interactions between subsystems.

This change particularly disadvantages heads that must condition on detailed wing and fuselage geometry to make compatible downstream decisions, such as wing placement and internals (latent and placement).
In Architecture 1, these heads operate on \ac{CF} that can reflect ``nearly finalized'' wing and fuselage implications.
In Architecture 3, the reduced-capacity main block cannot reliably distill such information into \ac{CF}, so the placement and internals heads are effectively asked to infer compatibility decisions from a representation that is close to the system-level latent $\zeta$.
Consequently, even if the wing/fuselage latent predictions themselves are expressive, the remaining subsystems receive an impoverished conditioning signal and cannot adapt their outputs to the realized wing and fuselage designs, which degrades constraint satisfaction and the overall performance--diversity trade-off.

Overall, Architecture 3 highlights that the shared main block must have sufficient capacity. It is not merely a routing module, but the primary mechanism that converts the system-level latent into a globally informative \ac{CF} representation that supports coordination across heterogeneous heads.

More broadly, this ablation also raises a question about what information the system-level latent $\zeta$ is expected to carry under our training setup. We view this as an interesting direction for future work.
In principle, one might hope that $\zeta$ alone is sufficiently informative to support all downstream decisions (including detailed placement).
However, Architecture 3 suggests that, in practice, predicting high-quality placement and internals may require an intermediate representation (\ac{CF}) that is \emph{more information-dense and interaction-aware} than a shallow lifting of $\zeta$.
A key question is whether $\zeta$ \emph{should} be trained to be directly usable for all subsystem placement and constraint satisfaction, or whether it is preferable for $\zeta$ to remain a compact global representation that is intentionally refined into task-specific coordination features by the main block.

\paragraph{Architecture 4}
For Architecture 4, we also observe failures that are consistent with vanishing gradients.
A plausible explanation is the reduced number of skip connections in the deep sequential wing placement head.
However, our diagnostics indicate that the vanishing gradients also occur at the final output linear layer, directly before the bounded $\tanh/\sigma$ activation and subsequent denormalization.
This makes depth (without adequate skip connections) an unlikely root cause for gradient vanishing, since one might expect vanishing gradients to only appear earlier in the stack if depth alone were the primary driver.

We therefore cannot conclusively attribute the robustness degradation in Architecture 4 to the reduced skip-connection structure.
Another possibility is that the custom initialization is not applied as intended for this architecture (\textit{e.g.}, due to differences in module structure), which could change early pre-activation magnitudes and increase the likelihood of output saturation.
However, despite careful checks and repeated attempts to match initialization behavior, we were not able to make Architecture 4 converge as robustly as Architecture 1.

\subsubsection{Recommendations and Future Work}
Our main recommendation from this ablation is to avoid sigmoid and $\tanh$ at head outputs where possible, since saturation leads to vanishing gradients and can strongly reduce robustness across seeds.

More broadly, the sensitivity of \ac{DF}-\ac{GLUE}'s ability to learn constraints and optimize performance to its architecture appears to be driven mostly by classical neural-network training issues rather than fundamental roadblocks.
We hypothesize that with a better choice of output activations and sufficient skip connections to preserve gradient flow, the dominant remaining architecture choice is how to distribute capacity between the shared main block and the output heads as discussed in relation to Architecture 3.

Furthermore, future work for \ac{DF}-\ac{GLUE} should investigate state-of-the-art architectures beyond our simple \ac{MLP} (\textit{e.g.}, graph-based networks, diffusion, Transformers, ...) and output parameterizations that train robustly and generalize across different problem settings without manual tuning.

%% file: appendix/aalm_scheme.tex
\section{Adaptive Augmented Lagrangian Update Schemes}
\label{ap:alm_schemes}

\begin{algorithm}[H]
\caption[Hypercube adaptive augmented Lagrangian method]{Hypercube Adaptive Augmented Lagrangian Method}
\label{alg:hypercube_alm}
\begin{algorithmic}[1]
\Require Constraint violations $\mathbf{c}_t \in \mathbb{R}^{B \times K}$ at epoch $t$
\Require \ac{ALM} parameters $\alpha, \gamma, \varepsilon$, warmup epochs $T_{\text{warmup}}$
\State \textbf{Input:} Current state variables $\mathbf{v}_{t-1}, \boldsymbol{\mu}_{t-1}, \boldsymbol{\lambda}_{t-1} \in \mathbb{R}^{B \times K}$

\State $\mathbf{c}_t = c_{\text{cap}} \cdot \text{tanh}(\mathbf{c}_t / c_{\text{cap}})$   \Comment{Smooth cap on $\mathbf{c}_t$}

\State $\mathbf{c}^2_t \leftarrow \mathbf{c}_t \odot \mathbf{c}_t$ \Comment{Element-wise square}

\State $\mathbf{v}_t \leftarrow \alpha \mathbf{v}_{t-1} + (1-\alpha) \mathbf{c}^2_t$

\State $s \leftarrow \min(1, t / T_{\text{warmup}})$ \Comment{Warmup ramp}

\State $\Delta\boldsymbol{\mu}_t \leftarrow \gamma \cdot s \cdot \frac{|\mathbf{c}_t|}{\sqrt{\mathbf{v}_t} + \varepsilon}$ \Comment{Adaptive penalty increment}

\State $\boldsymbol{\mu}_t \leftarrow \boldsymbol{\mu}_{t-1} + \max(0, \Delta\boldsymbol{\mu}_t)$ \Comment{Non-negative penalty parameters}

\State $\boldsymbol{\lambda}_t \leftarrow \boldsymbol{\lambda}_{t-1} + \boldsymbol{\mu}_t \odot \mathbf{c}_t$ \Comment{Lagrange multiplier update}

\State \textbf{Return:} Updated weights $\boldsymbol{\lambda}_t$
\end{algorithmic}
\end{algorithm}

Here, $B$ denotes the batch size (number of hypercube volumes), $K$ represents the number of constraints, and $\odot$ denotes element-wise multiplication. The warmup epochs $T_{\text{warmup}}$ enable a gradual ramp-up of the penalty parameters $\boldsymbol{\mu}_t$ during early training. This gradual increase avoids overly aggressive constraint satisfaction and ensures better early-training stability.

The key difference between pooled and hypercube schemes lies in how they handle constraint violations across the batch. The pooled approach takes the mean across all batch elements for each constraint, while the hypercube approach accounts for potentially poor scaling of constraint violations across different designs. For example, a constraint could be poorly normalized and exhibit much larger values for bigger aircraft or at specific design points, which could skew penalty growth in the pooled approach. The hypercube scheme maintains separate tracking of distinct zones in $\mathcal{Z}$, with each zone having its own weight set for every constraint. Importantly, the hypercube approach only works because we consistently sample each batch element from the same hypercube region in $\zeta$ across training. In contrast, the pooled approach is more flexible and works with other sampling methods, such as $\zeta \sim \mathcal{N}$ or $\zeta \sim \mathcal{U}$.

\input{tables/alm_hyperparams_test}

\begin{algorithm}[H]
\caption[Pooled adaptive augmented Lagrangian method]{Pooled Adaptive Augmented Lagrangian Method}
\label{alg:pooled_alm}
\begin{algorithmic}[1]
\Require Constraint violations $\mathbf{c}_t \in \mathbb{R}^{B \times K}$ at epoch $t$
\Require \ac{ALM} parameters $\alpha, \gamma, \varepsilon$, warmup epochs $T_{\text{warmup}}$

\State \textbf{Input:} Current state variables $\mathbf{v}_{t-1}, \boldsymbol{\mu}_{t-1}, \boldsymbol{\lambda}_{t-1} \in \mathbb{R}^{1 \times K}$

\State $\overline{\mathbf{c}}_t \leftarrow \frac{1}{B} \sum_{i=1}^{B} \mathbf{c}_{t,i}$ \Comment{Pool constraints across batch}
\State $\mathbf{c}_t \leftarrow \text{repeat}(\overline{\mathbf{c}}_t, B)$ \Comment{Broadcast to all samples}

\State $\mathbf{c}_t = c_{\text{cap}} \cdot \text{tanh}(\mathbf{c}_t / c_{\text{cap}})$   \Comment{Smooth cap on $\mathbf{c}_t$}
  
\State $\mathbf{c}^2_t \leftarrow \mathbf{c}_t \odot \mathbf{c}_t$ \Comment{Element-wise square}

\State $\mathbf{v}_t \leftarrow \alpha \mathbf{v}_{t-1} + (1-\alpha) \mathbf{c}^2_t$

\State $s \leftarrow \min(1, t / T_{\text{warmup}})$ \Comment{Warmup ramp}

\State $\Delta\boldsymbol{\mu}_t \leftarrow \gamma \cdot s \cdot \frac{|\mathbf{c}_t|}{\sqrt{\mathbf{v}_t} + \varepsilon}$ \Comment{Adaptive penalty increment}

\State $\boldsymbol{\mu}_t \leftarrow \boldsymbol{\mu}_{t-1} + \max(0, \Delta\boldsymbol{\mu}_t)$ \Comment{Non-negative penalty parameters}

\State $\boldsymbol{\lambda}_t \leftarrow \boldsymbol{\lambda}_{t-1} + \boldsymbol{\mu}_t \odot \mathbf{c}_t$ \Comment{Lagrange multiplier update}

\State \textbf{Return:} Updated weights $\boldsymbol{\lambda}_t$ for constraint penalties
\end{algorithmic}
\end{algorithm}

Table \ref{tab:alm-hyperparams-text} demonstrates that the \ac{ALM} dual growth rate $\gamma$ has a direct effect on the trade-off between constraint satisfaction, optimality, and diversity. Generally, a higher growth rate $\gamma$ leads to better feasibility, as constraint satisfaction is weighted more heavily, but at the expense of worse optimality and diversity. For all \ac{GLUE} model training in this work, we use $c_{\text{cap}} = 5.0$ and $\gamma=5 \times 10^{-3}$.

%% file: tables/alm_hyperparams_test.tex
\begin{table*}[!h]
\centering
\footnotesize
\caption[ALM hyperparameter comparison for \ac{DF}-\ac{GLUE} training]{Comparison of performance, diversity, and feasibility at various hyperparameters $\gamma$ and $c_{\text{cap}}$ for \ac{DF}-\ac{GLUE} training with case 1.}
\begin{tabular}{l c c c c c}
\toprule
$\gamma$ (ALM) & $c_{\text{cap}}$ & $O_{\text{mass}}$ & Feasible & \ac{DPP}($x$, $\sigma_{\text{DPP}}{=}0.5$) & \ac{DPP}($z_F$, $\sigma_{\text{DPP}}{=}0.05$) \\
\midrule
$5\times10^{-2}$ & 5  & 4.34 & 75.6\% & 3647 & 213 \\
$5\times10^{-3}$ & 5  & 4.21 & 61.0\% & 3597 & 190 \\
$5\times10^{-4}$ & 5  & 4.07 & 42.7\% & 3558 & 187 \\
$5\times10^{-2}$ & 20 & 4.32 & 75.2\% & 3633 & 155 \\
$5\times10^{-3}$ & 20 & 4.21 & 59.7\% & 3577 & 133 \\
\bottomrule
\end{tabular}
\label{tab:alm-hyperparams-text}
\end{table*}

%% file: appendix/uav_design_problem.tex
\section{Details on UAV Design Problem}
\label{ap:uav_design_problem}

Single-objective optimization (lowest mass given conditions) is performed. The conditions, which are fixed, are:

\begin{itemize}
    \item Velocity and altitude: Affect drag and wing lift, and thus affect wing geometry
    \item Payload mass: simulates internal subsystems like radars, battery, and cargo. Essentially, we fix masses for internal boxes which the aircraft must accommodate.
    \item Payload box volumes: Additionally, we prescribe volumes for internal boxes which the aircraft must accommodate.
\end{itemize}

For more details on the specific test cases and experimental conditions, refer to \ref{ap:aircraft_cases}. Drag estimation is also implemented in the geometry layer, but due to scoping we focus only on the mass objective. Drag optimization also tends to promote high-aspect-ratio configurations that complicate visualization and analysis. Issues considering stability or dynamic aircraft performance are entirely ignored. For this reason, the lift requirement for the rear wing $L_{req,2}$ is set to $1/5 \cdot L_{req,1}$. Only level flight performance is considered. 

Experiments are run at three different sets of velocity, altitude, cargo mass, and volume. See \ref{ap:aircraft_cases} for more details.

\subsection{Dimensionality of Design Spaces}
\label{ap:dimensionality}

Provided here is a detailed breakdown of the dimensionalities for the latent space $\mathcal{Z}$, design space $\mathcal{X}$, conditions $c$, and system-level latent space $\zeta$ used in the aircraft design problem.

\input{tables/dimensionality_breakdown.tex}

Additional notes:
\begin{itemize}
    \item \textbf{Wing geometry ($x_W$)}: Each wing is parameterized by angle of attack (AoA), 64 airfoil points (x,z coordinates = 128 values), span, and chord, totaling 131 dimensions per wing. With 2 wings: $2 \times 131 = 262$ dimensions.
    \item \textbf{Internals geometry ($x_I$)}: Three bounding boxes are placed inside the fuselage. Each box requires \ac{2D} position (x,z) since y=0 for each bounding box's center, and \ac{3D} dimensions (length, width, height). Total: $3 \times (2 + 3) = 15$ dimensions. The first box is fixed in space, removing 2 degrees of freedom: $15 - 2 = 13$ dimensions.
    \item \textbf{Coupling variables ($x_C$)}: Lift requirements $l_{1,req}$ and $l_{2,req}$ are used as conditions during wing model inference rather than direct geometry instantiation, but are still counted in $\mathcal{X}$. Placement variables determine spatial relationships between subsystems.
    \item \textbf{Dimensionality reduction}: The subsystem models compress from $dim(\mathcal{X}) = 324$ to $dim(\mathcal{Z}) = 2$, a reduction factor of 14.7$\times$.
    \item \textbf{System-level latent space ($\zeta$)}: For visualization purposes (latent space sweeps), $dim(\zeta) = 2$ is used. For all other experiments, $dim(\zeta) = 4$ is used.
\end{itemize}

\subsection{Constrained Optimization Problem}

\label{ap:opt_problem_formulation}

For our \ac{UAV} design problem, we have $n_o = 1$ (mass minimization) and $n_c = 5$. Of the five constraints, three are equality constraints for wing positioning and lift, and two are inequality constraints for wing-fuselage and internals-fuselage interfacing.
The geometry layer outputs:
\begin{equation}
\label{eq:geometry_layer_output}
\mathcal{G}(\mathbf{x}) = [\Omass(\mathbf{x}), \Clift(\mathbf{x}), \Cwx(\mathbf{x}), \Cdi(\mathbf{x}), \Cbb(\mathbf{x}), \Cboxpl(\mathbf{x})]
\end{equation}

where $\Omass$ is the mass objective (to be minimized), $\Clift$ is the lift constraint, $\Cwx$ represents wing placement constraints, $\Cdi$ are dihedral constraints, $\Cbb$ is the bounding box constraint, and $\Cboxpl$ is the box placement constraint. These are detailed in Section \ref{sec:constraint_formulations}.

The formal optimization problem is:
\begin{align}
\label{eq:optimization_problem}
\min_{\mathbf{z} \in \mathcal{Z}} \quad & \Omass(\mathbf{x}(\mathbf{z})) \\
\text{subject to} \quad & \Clift(\mathbf{x}(\mathbf{z})) = 0 \label{eq:constraint_lift}\\
& \Cwx(\mathbf{x}(\mathbf{z})) = 0 \label{eq:constraint_wx}\\
& \Cdi(\mathbf{x}(\mathbf{z})) = 0 \label{eq:constraint_di}\\
& \Cbb(\mathbf{x}(\mathbf{z})) \leq 0 \label{eq:constraint_bb}\\
& \Cboxpl(\mathbf{x}(\mathbf{z})) \leq 0 \label{eq:constraint_boxpl}\\
& \mathbf{z} \in \mathcal{Z} \nonumber
\end{align}

where $\mathbf{x}(\mathbf{z}) = \{S_i(z_i)\}_{i} \cup \{x_C\}$ represents the complete design parameters obtained from subsystem latent codes via their respective generative models and the coupling variables.

\subsubsection{Heuristics-Based Objective Function}
\label{sec:objective_formulation}

The mass objective $\Omass$ (Eq. \ref{eq:optimization_problem}) is formulated as:
\begin{equation}
\Omass(\mathbf{x}) \propto m_{\text{wing}}(\mathbf{x}) + m_{\text{fuselage}}(\mathbf{x})
\end{equation}

\noindent where:
\begin{align}
m_{\text{wing}}(\mathbf{x}) &\propto \alpha_w \cdot \left( \text{span}_1^2 \cdot \text{chord}_1 + \text{span}_2^2 \cdot \text{chord}_2 \right) \\
m_{\text{fuselage}}(\mathbf{x}) &\propto \alpha_f \cdot A_{\text{surface, fus}}
\end{align}

The wing mass scales with the square of span times chord (accounting for bending moment), while fuselage mass is proportional to surface area. Here $\alpha_w$ and $\alpha_f$ are scaling constants. Note that these mass formulations are heuristic approximations rather than high-fidelity structural analyses. In the future, these heuristics could be replaced with high-fidelity differentiable surrogates or simulators for more accurate performance predictions.

\subsubsection{Constraint Formulations}
\label{sec:constraint_formulations}

\paragraph{Lift Constraint ($\Clift$)} Ensures adequate lift generation:
\begin{equation}
\Clift \propto \frac{L_{\text{required}} - L_{\text{supplied}}}{L_{\text{required}}}
\end{equation}
where $L_{\text{required}} = g(m_{\text{wing}} + m_{\text{fuselage}} + m_{\text{cargo}})$ and $L_{\text{supplied}} = L_1 + L_2$ where $L_1$ is the lift of the front wing, and $L_2$ is the lift of the rear wing.

\paragraph{Wing Placement Constraints ($\Cwx$)} Ensure wing base positions match fuselage interface:
\begin{equation}
\Cwx \propto C_{\text{wx,1}} + C_{\text{wx,2}}
\end{equation}

\noindent where:
\begin{align}
C_{\text{wx,1}} &\propto |y_{\text{w1}} - y_{\text{PS1}}| \\
C_{\text{wx,2}} &\propto \sqrt{(x_{\text{w2}} - x_{\text{PS2}})^2 + (y_{\text{w2}} - y_{\text{PS2}})^2 + (z_{\text{w2}} - z_{\text{PS2}})^2}
\end{align}

Here, PS denotes the primary segments of the fuselage parametrization (see Figure \ref{fig:fuselage_parametrization}), whose planar faces define the wing base attachment points. Wing 1 (w1) is constrained in the $y$-coordinate only (spanwise position), as the primary segment 1 $x$ and $z$ positions are fixed anchors for the entire aircraft (origin of coordinate system). Wing 2 (w2) requires full \ac{3D} position matching.

\paragraph{Dihedral Constraints ($\Cdi$)} Ensure wing orientations match fuselage interface:
\begin{align}
\Cdi \propto \sum_{i=1}^{2} |\Gamma_{\text{wing},i} - \Gamma_{\text{fuselage},i}|
\end{align}
For the sake of simplicity, the wing orientation is only defined by the dihedral angle $\Gamma_{wing,i}$. The wing's two remaining degrees of freedom are sweep (fixed) and the angle of attack, which is handled by airfoil generation (BézierGAN).

\paragraph{Bounding Box Constraints ($\Cbb$)} Ensure wing roots fit within bounding box interface:
\begin{align}
    \Cbb &\propto \sum_{i=1}^{2} 
    \left[ \max(0, a_{bb,\text{wing},i} - a_{bb,\text{fuselage},i}) \right. \nonumber \\
    &\quad + \left. \max(0, b_{bb,\text{wing},i} - b_{bb,\text{fuselage},i}) \right] \le 0
\end{align}
where $(a_{bb,i}, b_{bb,i})$ are the bounding box dimensions at wing-fuselage interface $i$. Note that $b_{bb,i}$ is effectively the extrusion length of the primary segment.

\paragraph{Box Placement Constraints ($\Cboxpl$)} Ensure internal components fit inside fuselage:
\begin{align}
\Cboxpl \propto \sum_{j=1}^{N_{\text{boxes}}} \sum_{k=1}^{8} \max\left(0, -d_{j,k} - \delta \right) \leq 0
\end{align}
where $d_{j,k} = d(\mathbf{p}_{j,k}, \partial\Omega_{\text{fuselage}})$ is the signed distance from corner $k$ of box $j$ to the fuselage surface, and $\delta$ is a safety margin.

All constraints and objectives are normalized with conditions and scaled by a constant factor to ensure they are of similar magnitude.

%% file: tables/dimensionality_breakdown.tex
\begin{table}[h]
\centering
\small
\caption[Dimensionality breakdown for aircraft design problem]{Dimensionality breakdown for aircraft design problem.}
\label{tab:dimensionality}
\begin{tabular}{llr}
\toprule
\multicolumn{3}{l}{\textbf{Subsystem Latents} $\bm{\mathcal{Z}}$}\\
& $z_W$ (Wing) & 2$\times$2 \\
& $z_I$ (Internals) & 4 \\
& $z_F$ (Fuselage) & 4 \\
& $x_C$ (Coupling vars) & \\
& \qquad Wing lift requirement & 2$\times$1 \\
& \qquad Wing 1 placement & 2 \\
& \qquad Wing 2 placement & 4 \\
& \qquad Internals placement & 2 \\
\cmidrule{2-3}
& \textbf{Total} $\mathbf{dim(\mathcal{Z})}$ & \textbf{22} \\
\midrule
\multicolumn{3}{l}{\textbf{\ac{GLUE} Model Conditions} $\bm{c}$}\\
& Box volumes & 3 \\
& Cargo mass & 1 \\
& Altitude & 1 \\
& Velocity & 1 \\
\cmidrule{2-3}
& \textbf{Total} $\mathbf{dim(c)}$ & \textbf{6} \\
\midrule
\multicolumn{3}{l}{\textbf{Design Space} $\bm{\mathcal{X}}$}\\
& $x_F$ (Fuselage) & 39 \\
& $x_W$ (Per wing) & 2$\times$131 \\
& $x_I$ (Internals) & 13 \\
& $x_C$ (Coupling) & 10 \\
\cmidrule{2-3}
& \textbf{Total} $\mathbf{dim(\mathcal{X})}$ & \textbf{324} \\
\midrule
\textbf{System-level Latent} $\bm{\zeta}$ && 2-4\\
\bottomrule
\end{tabular}
\end{table}

%% file: appendix/dpp_loss.tex
\section{DPP Loss Implementation}
\label{ap:dpp_loss}

This appendix details the \ac{DPP} loss implementation used here, both for training of models and diversity measurements.
The \ac{DPP} loss encourages diversity in generated aircraft designs by penalizing similarity between samples in a batch, building on the foundational work of Elfeki \textit{et al.} \cite{GDPPLearningDiverseGenerations2019}.

\subsection{Experimental Details}

Table \ref{tab:dpp_experiment_settings} summarizes the settings used for the different experiments.
\input{tables/dpp_experiment_settings.tex}

\ac{DPP} scoring also depends on batch size. For benchmarking (Fig.~\ref{fig:opt_vs_diversity_case_1_b} and Table~\ref{tab:opt_feas_div_comparison}), we fix $B{=}35$ samples per evaluation and repeat $R{=}20$ times with different random sub-batches drawn from the full sample set, then average the results. This ensures statistically consistent, method-agnostic comparison across large sample sets.

The mode collapse ablation (Table~\ref{tab:mode-collapse}) uses a different protocol: \ac{DPP} is evaluated once ($R{=}1$) on the full batch of $N{=}1{,}000$ samples of the fuselage latent $z_F$ (4-D), using raw (unnormalized) values. This is sufficient for detecting mode collapse versus diversity, and the scores are not intended to be compared with benchmarking results.

Features in $\mathcal{X}$ and $\mathcal{X}_\text{vis}$ are min-max normalized to $[0,1]$ using fixed data-range bounds prior to \ac{DPP} evaluation. The mode collapse experiment uses raw $z_F$ values (no normalization).

\subsection{Formulation}

The \ac{DPP} loss is based on the log-determinant of a similarity kernel matrix:

\begin{equation}
\mathcal{L}_{\text{DPP}} = -2\log|\det(\mathbf{K} + \epsilon\mathbf{I})|
\end{equation}

where $\mathbf{K}$ is the similarity kernel matrix, $\epsilon$ is a regularization parameter, and $\mathbf{I}$ is the identity matrix.

The similarity kernel is computed using a Gaussian radial basis function:
\begin{equation}
K_{ij} = \exp\left(-\frac{d^2(\mathbf{x}_i, \mathbf{x}_j)}{2\sigma_{\text{DPP}}^2}\right)
\end{equation}

where $d(\mathbf{x}_i, \mathbf{x}_j)$ is the Euclidean distance between samples $i$ and $j$, and $\sigma_{\text{DPP}}$ is the characteristic length scale.

\begin{algorithm}[H]
\caption{DPP Loss Computation for \ac{GLUE} Model Training}
\label{alg:dpp_loss_computation}
\begin{algorithmic}[1]
\State \textbf{Input:} $\mathbf{X} \in \mathbb{R}^{B \times D}$ (batch of $B$ samples with $D$ features), $\sigma_{\text{DPP}}$, $\epsilon$
\State \textbf{Output:} $\mathcal{L}_{\text{DPP}}$

\State Compute pairwise squared distances: $\mathbf{D}^2 = \text{cdist}(\mathbf{X}, \mathbf{X})^2$
\State Compute similarity kernel: $\mathbf{K} = \exp(-\mathbf{D}^2 / (2\sigma_{\text{DPP}}^2))$
\State Add regularization: $\mathbf{L} = \mathbf{K} + \epsilon\mathbf{I}$
\State Compute log-determinant: $\text{sign}, \log|\det(\mathbf{L})| = \text{slogdet}(\mathbf{L})$
\State Return: $\mathcal{L}_{\text{DPP}} = -2 \cdot \log|\det(\mathbf{L})|$
\end{algorithmic}
\end{algorithm}

Algorithm \ref{alg:dpp_loss_computation} shows the implementation used for training of \ac{DF}-\ac{GLUE} and benchmarking. It uses \texttt{torch.linalg.slogdet} for numerical stability to avoid overflow and underflow issues. The parameter $\sigma_{\text{DPP}}$ can be varied to control the kernel length scale, where higher values of $\sigma_{\text{DPP}}$ mean that samples farther away from each other will have greater influence on the diversity loss. The regularization parameter $\epsilon = 1 \times 10^{-2}$ prevents numerical issues with singular matrices. 

In addition, for the training of some subsystem models, we use a slightly modified version of the \ac{DPP} loss. The wing generator uses a performance-augmented \ac{DPP} loss that incorporates aerodynamic quality metrics, following the approach of Chen and Ahmed \cite{chenPaDGANLearningGenerate2020}:

\begin{equation}
\mathcal{L}_{\text{PA-DPP}} = -\log\det(\mathbf{L})
\end{equation}

where $\mathbf{L} = \mathbf{K} \odot (\mathbf{q} \mathbf{q}^T)$ and $q_i = \frac{(L/D)_i^2}{m_i^2}$ is the quality score for sample $i$.

The internals module employs three specialized \ac{DPP} losses:

\begin{itemize}
\item {Standard DPP:} Z-score normalized diversity for box arrangements
\item {Aspect Ratio DPP:} Diversity in box aspect ratios (width/length, height/length)
\item {Adjacency DPP:} Diversity in spatial adjacency patterns between boxes
\end{itemize}

These variants use batch size normalization and robust Cholesky decomposition with eigenvalue fallback for numerical stability.

%% file: tables/dpp_experiment_settings.tex
\begin{table}[h]
    \centering
    \footnotesize
    \caption{DPP settings for the different experiments. The domain indicates which feature space the DPP loss is applied to. $\mathcal{X}_\text{vis}$ = 6 visually important features in $\mathcal{X}$ (rear dihedral, rear wingbase $x$/$z$, nose-tip offset, chord$_1$, chord$_2$). All features are min-max normalized to $[0,1]$ before kernel evaluation, except $z_F$ which uses raw values (see $^\dagger$).}
    \label{tab:dpp_experiment_settings}
    \begin{tabular}{lccc}
    \toprule
    \textbf{Experiment} & \textbf{Kernel $\sigma_{\text{DPP}}$} & \textbf{Domain} & \textbf{Dim.} \\
    \midrule
    Opt-Feas-Div (Fig. \ref{fig:opt_vs_diversity_case_1_b}) & 0.5 & $\mathcal{X}$ & 39 \\
    Opt-Feas-Compute (Fig. \ref{fig:opt_vs_compute_case_1_b}) & - & - & - \\
    SN Ablation (Fig. \ref{fig:smoothness-fuselage-model}) & 0.5 & $\mathcal{X}_\text{vis}$ & 6 \\
    Mode Collapse Abl. (Tab. \ref{tab:mode-collapse}) & 0.05 & $z_F$$^\dagger$ & 4 \\
    \bottomrule
    \end{tabular}%
    \smallskip\\
    \raggedright $^\dagger$ Full batch ($N{=}1{,}000$), single evaluation ($R{=}1$), raw (unnormalized) $z_F$ values. All other experiments use $B{=}35$ random sub-batches, $R{=}20$ evaluations.
\end{table}

%% file: appendix/aircraft_cases.tex
\section{Overview of Test Cases for Aircraft Design Problem}
\label{ap:aircraft_cases}
\input{tables/case_comparison.tex}

%% file: tables/case_comparison.tex
\begin{table}[!h]
    \centering
    \footnotesize
    \caption[Comparison of mission and cargo requirements]{Comparison of mission and cargo requirements for the three optimization cases.}
    \label{tab:case_comparison}
    \begin{tabular}{lccc}
        & \textbf{Case 1} & \textbf{Case 2} & \textbf{Case 3} \\
        \midrule
         Velocity (m/s) & 35 & 50 & 20 \\
         Altitude (m) & 2000 & 2000 & 1000 \\
        \midrule
         Battery Mass (kg) & 100 & 150 & 30 \\
         Electronics Mass (kg) & 20 & 20 & 10 \\
         Payload Mass (kg) & 30 & 30 & 20 \\
        \textbf{Total Mass} & \textbf{150} & \textbf{200} & \textbf{60} \\
        \midrule
         Battery Volume (L) & 45 & 75 & 15 \\
         Electronics Volume (L) & 15 & 15 & 5 \\
         Payload Volume (L) & 30 & 60 & 30 \\
        \textbf{Total Volume} & \textbf{90} & \textbf{150} & \textbf{50} \\
        \bottomrule
    \end{tabular}
\end{table}

%% file: appendix/benchmarking_all_cases.tex
\section{Cross-Case Benchmarking}
\label{ap:benchmarking_all_cases}

\input{tables/opt_feas_div_comparison.tex}

Table \ref{tab:opt_feas_div_comparison} extends the analysis across three different aircraft design cases. Diversity measurements and sampling are identical to the analysis of case 1 above (DPP: $\sigma_{\text{DPP}} = 0.5$, $\mathcal{X}$, 39-D; see \ref{ap:dpp_loss}).

For case 1, the results are excerpts from Figure \ref{fig:opt_vs_diversity_case_1_b}.
Depending on hyperparameters, \ac{DF}-\ac{GLUE} can either reliably converge to a highly optimal solution or exhibit the highest diversity among all methods.
In terms of constraint satisfaction, the \ac{DF}-\ac{GLUE} model outperforms all data-driven \ac{GLUE} models and is nearly competitive with the optimization algorithms (\ac{ALM}-\ac{GD}, \ac{TuRBO}-\ac{iGD}).

The same trend holds for cases 2 and 3.
\ac{DF}-\ac{GLUE} consistently achieves higher feasibility than all data-driven approaches.
While it does not quite match the precision with which the optimization algorithms can satisfy constraints, it remains close and clearly superior to purely \ac{DD}-\ac{GLUE}.

Regarding performance, the data-driven \ac{GLUE} models tend to mimic the behavior of \ac{ALM}-\ac{GD} across all cases, suggesting they effectively learn similar design trade-offs.
For case 3, \ac{TuRBO}-\ac{iGD} performs particularly poorly in terms of optimality.
This further confirms our finding that \ac{TuRBO}-\ac{iGD} struggles with high-dimensional, constrained settings and is not well suited for our problem.

Note that the best objective of 0.87 reported for \ac{ALM}-\ac{GD} in case 3 appears anomalous. As of the time of writing, we are unsure what causes this.

%% file: tables/opt_feas_div_comparison.tex
\begin{table*}[t]
    \centering
    \caption[Comparison of optimality, feasibility, and diversity across training conditions]{Comparison of optimality, feasibility, and diversity across conditions (see \ref{ap:aircraft_cases}). DPP on $\mathcal{X}$ (39-D) with $\sigma_{\text{DPP}} = 0.5$.}
    \label{tab:opt_feas_div_comparison}
    \vspace{0.1cm}
    \footnotesize
    \begin{tabular}{clcccccccrr}
        \toprule
        &\textbf{Experiment} & \textbf{Samples} & \textbf{Seeds} & \textbf{Feas. ($\uparrow$)} & \textbf{Worst/Best Feas. ($\uparrow$)} & \textbf{Best Obj. ($\downarrow$)} & \textbf{Mean Obj. ($\downarrow$)} & \textbf{DPP ($\downarrow$)} \\
        \midrule
        \multirow{8}{*}{\rotatebox{90}{Case 1}} & \ac{TuRBO}-\ac{iGD} Top 1\% & 400 & 200 & 94.5\% & - & 1.89 & 4.23 & 1544 \\
        & \ac{ALM}-\ac{GD} 100\% & 70860 & 14180 & \textbf{97.2\%} & - & 1.63 & 2.93 & 1424 \\
        & \acs{cVAE} & 10000 & 10 & 48.1\% & (28.6\% / 54.8\%) & 1.70 & 2.83 & 1689 \\
        & \acs{MDD-GAN} 80\% & 10000 & 10 & 71.4\% & (64.5\% / 79.6\%) & \textbf{1.61} & 2.31 & 2301 \\
        & \acs{OT-GAN} & 10000 & 10 & 50.7\% & (38.6\% / 54.9\%) & 1.68 & 2.77 & 1886 \\
        & \acs{DDPM} & 10000 & 10 & 78.5\% & (75.9\% / 81.1\%) & 1.65 & 2.75 & 1460 \\
        & \ac{DF}-\ac{GLUE} $\lambda_{\text{perf}}$ 0.05, $\lambda_{\text{DPP}}$ 0.25 & 10000 & 10 & 85.9\% & (70.7\% / 94.6\%) & 1.80 & 3.33 & \textbf{986} \\
        & \ac{DF}-\ac{GLUE} $\lambda_{\text{perf}}$ 0.05, $\lambda_{\text{DPP}}$ 0.0 & 10000 & 10 & 96.5\% & (73.5\% / 99.7\%) & 1.63 & \textbf{1.90} & 4543 \\
        \midrule
        \multirow{8}{*}{\rotatebox{90}{Case 2}} & \ac{TuRBO}-\ac{iGD} Top 1\% & 191 & 200 & - & - & 0.98 & 1.87 & \textbf{658} \\
        & \ac{ALM}-\ac{GD} 100\% & 47960 & 9592 & \textbf{92.6\%} & - & 0.74 & 1.47 & 1375 \\
        & \acs{cVAE} & 10000 & 10 & 37.1\% & (28.6\% / 42.7\%) & 0.79 & 1.43 & 1568 \\
        & \acs{MDD-GAN} 100\% & 10000 & 10 & 56.2\% & (48.8\% / 62.0\%) & 0.76 & 1.41 & 1551 \\
        & \acs{OT-GAN} & 10000 & 10 & 43.6\% & (37.1\% / 51.6\%) & 0.78 & \textbf{1.35} & 1740 \\
        & \acs{DDPM} & 10000 & 10 & 56.9\% & (53.9\% / 59.8\%) & 0.76 & 1.38 & 1439 \\
        & \ac{DF}-\ac{GLUE} $\lambda_{\text{perf}}$ 0.1, $\lambda_{\text{DPP}}$ 0.2 & 10000 & 10 & 78.0\% & (0.0\% / 98.6\%) & \textbf{0.67} & 1.37 & 1589 \\
        \midrule
        \multirow{8}{*}{\rotatebox{90}{Case 3}} &  \ac{TuRBO}-\ac{iGD} Top 1\% & 400 & 200 & 93.3\% & - & 16.32 & 35.98 & \textbf{1352} \\
        & \ac{ALM}-\ac{GD} 100\% & 64185 & 12837 & \textbf{94.0\%} & - & \textbf{0.87} & 15.63 & 2609 \\
        & \acs{cVAE} & 10000 & 10 & 41.3\% & (32.5\% / 51.2\%) & 11.97 & 15.41 & 2955 \\
        & \acs{MDD-GAN} & 10000 & 10 & 62.9\% & (45.5\% / 70.8\%) & 11.74 & 14.75 & 3116 \\
        & \acs{OT-GAN} & 10000 & 10 & 50.9\% & (44.5\% / 59.1\%) & 11.85 & 15.51 & 2916 \\
        & \acs{DDPM} & 10000 & 10 & 54.1\% & (47.8\% / 57.8\%) & 11.94 & 14.12 & 2728 \\
        & \ac{DF}-\ac{GLUE} $\lambda_{\text{perf}}$ 0.1, $\lambda_{\text{DPP}}$ 0.75 & 10000 & 10 & 80.3\% & (69.4\% / 87.7\%) & 11.85 & \textbf{13.18} & 3158 \\
        \bottomrule
    \end{tabular}%
    
\end{table*}

%% file: appendix/qualitative_comparison.tex
\section{Qualitative Comparison of Methods}
\label{ap:qualitative_comparison}

\input{figures/visual_designs_all_models/visual_designs_all_models.tex}

Provided here is a visual comparison of samples obtained using optimization algorithms, data-driven \ac{GLUE} models, and the data-free \ac{GLUE} model.

\begin{itemize}
    \item Bayesian optimization (\ac{TuRBO}-\ac{iGD}): Exhibits great diversity across configurations, but designs are generally low-performance (large aircraft mass).
    \item \ac{ALM}-\ac{GD}: Finds better designs than Bayesian Optimization, but can struggle with local minima, as shown in several configurations. For example, when the internals model places component bounding boxes on top of each other rather than in a single row (which is beneficial for minimum fuselage diameter and thus weight), the fuselage cannot shrink (See second from the right in Figure \ref{fig:visual_designs_all_models_appendix}). 
    In this case, there is a conflict between the box placement constraint (Eq. \ref{eq:constraint_boxpl}, requiring boxes inside the fuselage) and the mass objective (Eq. \ref{eq:optimization_problem}, favoring smaller fuselage) which cannot be resolved.
    In other words, the optimizer is prone to get trapped in local minima.
    \item Data-driven \ac{GLUE} Models: Since these models (\acs{cVAE}, \acs{MDD-GAN}, \acs{OT-GAN}, \acs{DDPM}) are trained on data acquired with \ac{ALM}-\ac{GD}, they tend to replicate similar designs to \ac{ALM}-\ac{GD}.
    \item Data-Free \ac{GLUE} Model: Exhibits low diversity, with designs consistently in high-performance regions of the design space. This is especially pronounced when trained without diversity losses, as shown here. In other words, the \ac{DF}-\ac{GLUE} consistently converges to the best observed optimum if diversity is not explicitly incentivized. 
    \ac{DF}-\ac{GLUE} does not produce \textit{inherently} non-diverse designs. In fact, it can outperform optimization algorithms (I.) and data-driven  models (II.) in terms of diversity if trained with large diversity loss weight $\lambda_{\text{DPP}}$.
\end{itemize}

%% file: figures/visual_designs_all_models/visual_designs_all_models.tex
\begin{figure*}[!p]
    \centering
    \vspace{-10pt}
    {I. TuRBO}\\[-7.5pt]
    \includegraphics[width=0.90\textwidth]{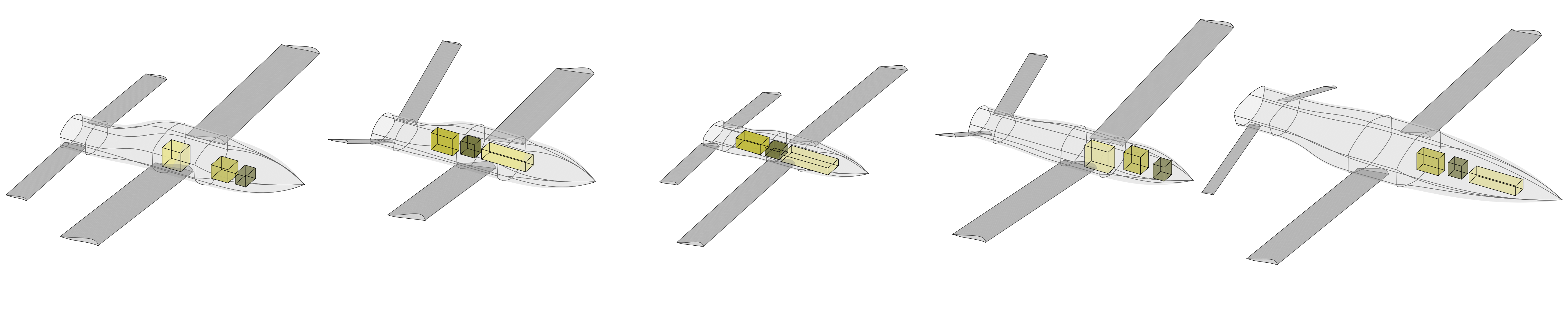}\\[-8pt]
    
    {I. ALM-GD}\\[-7.5pt]
    \includegraphics[width=0.90\textwidth]{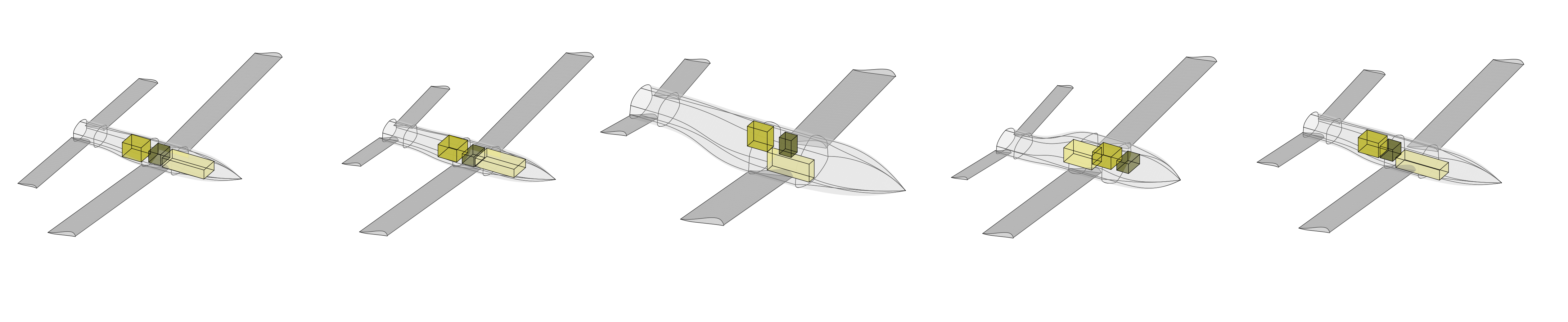}\\[-8pt]
    
    {II. cVAE}\\[-7.5pt]
    \includegraphics[width=0.90\textwidth]{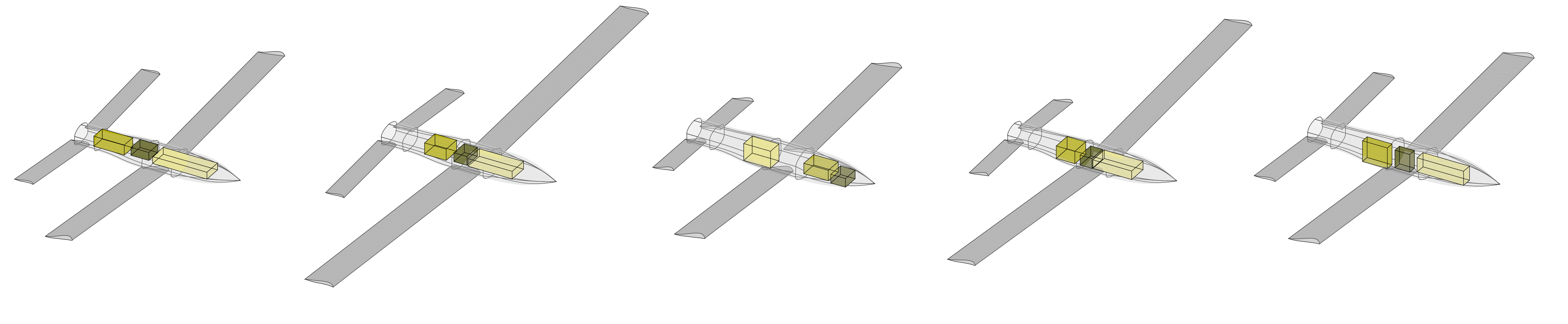}\\[-8pt]
    
    {II. OT-GAN}\\[-7.5pt]
    \includegraphics[width=0.90\textwidth]{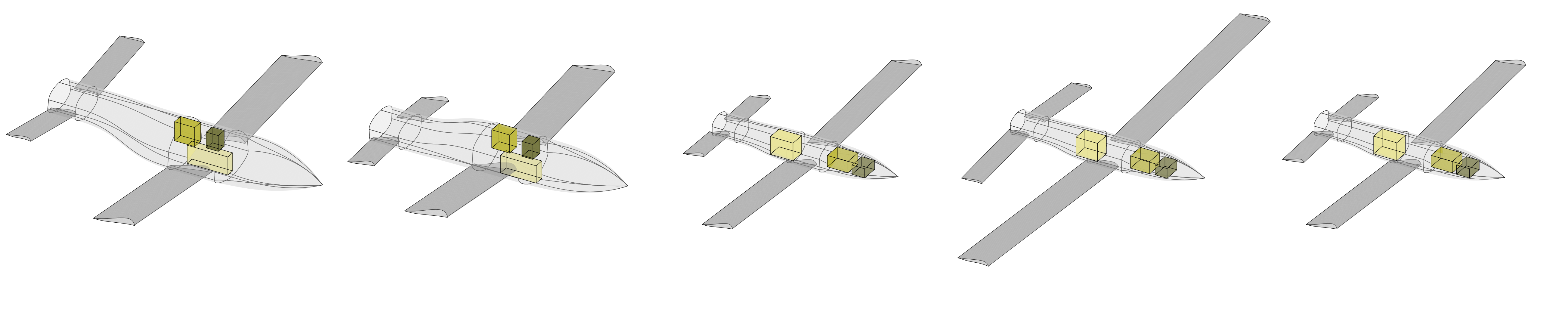}\\[-8pt]
    
    {II. MDD-GAN}\\[-7.5pt]
    \includegraphics[width=0.90\textwidth]{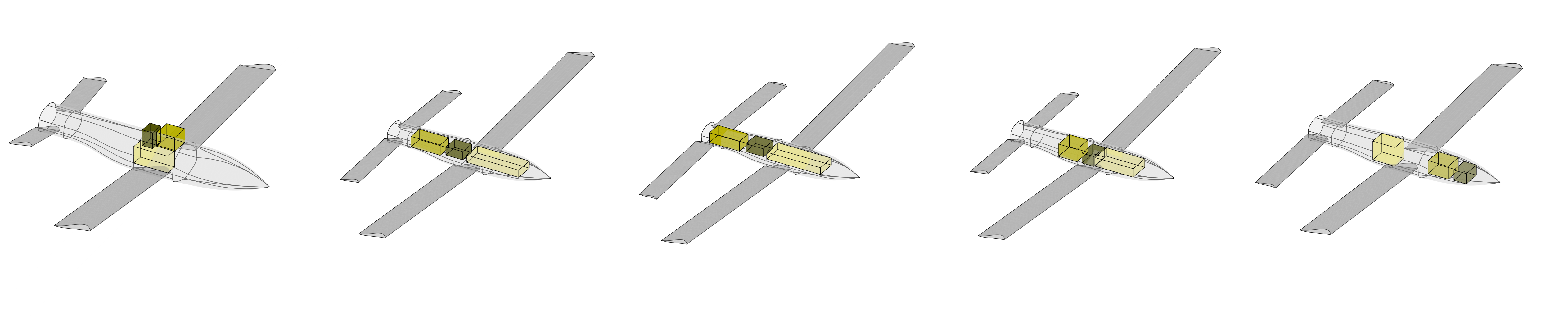}\\[-8pt]
    
    {II. DDPM}\\[-7.5pt]
    \includegraphics[width=0.90\textwidth]{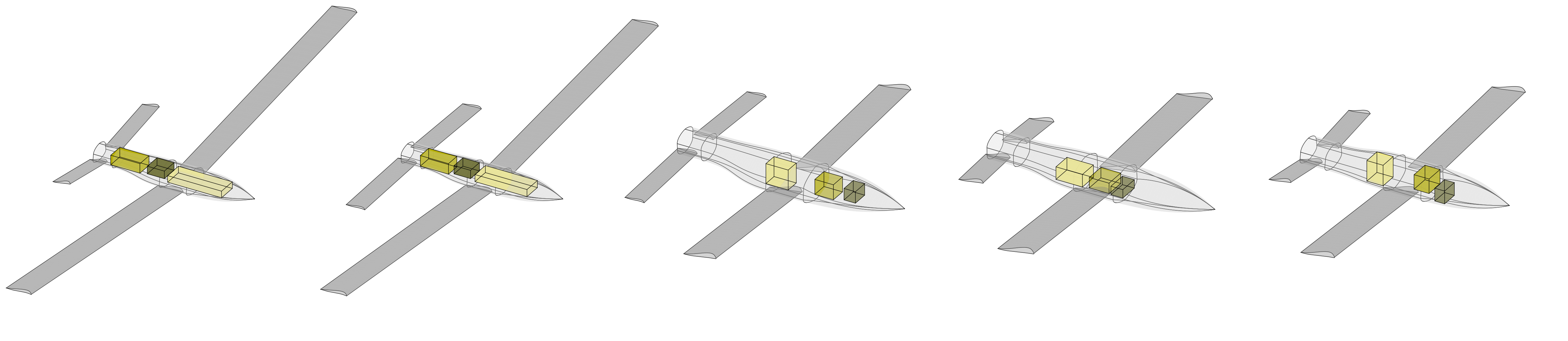}\\[-8pt]
    
    {III. \ac{DF}-\ac{GLUE} ($\lambda_{\text{DPP}}=0$)}\\[-7.5pt]
    \includegraphics[width=0.90\textwidth]{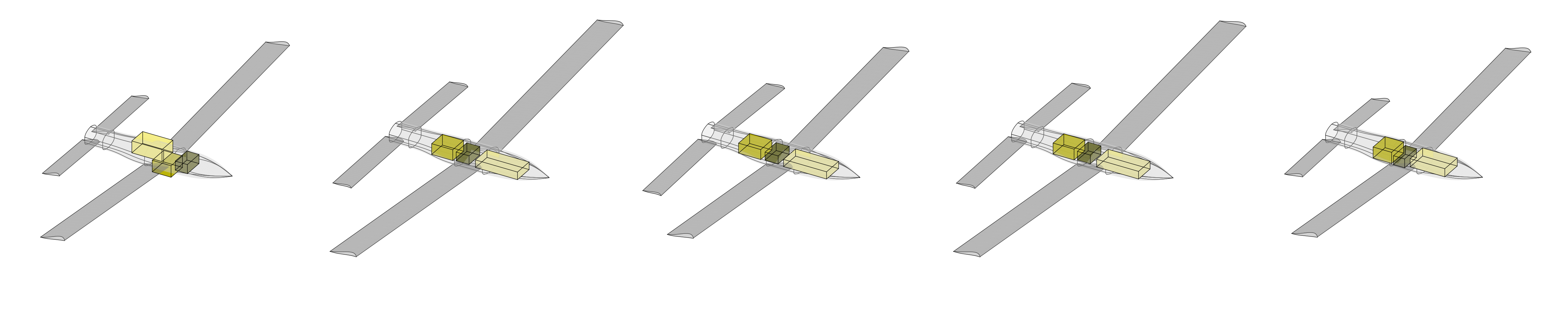}
    \vspace{-20pt}

    \caption[Exemplar aircraft designs from all evaluated methods]{Exemplar aircraft designs generated by all evaluated methods.}
    \label{fig:visual_designs_all_models_appendix}
\end{figure*}

%% file: appendix/opt_algos.tex
\section{Details on Optimization Algorithms}
\label{ap:opt_algos}

Table \ref{tab:optimizer_hyperparameters} shows detailed hyperparameters for the optimization algorithms. For both methods, these are hand-tuned. 
\input{tables/optimizer_hyperparameters.tex}

\subsection{TuRBO-iGD}

Bayesian optimization offers resilience against local minima and provides a valuable benchmark for global optimality. However, it suffers from poor scaling in high dimensions \cite{erikssonScalableGlobalOptimization2020}. Our search space involves subsystem latent codes and coupling variables with $\dim(\mathcal{Z}) = 22$. While specialized methods exist for high-dimensional BO \cite{lethamReExaminingLinearEmbeddings, erikssonHighDimensionalBayesianOptimization2021}, they typically rely on pruning or embedding to compress the space. Since we already optimize in a compressed latent space, we estimate the marginal gain from these methods to be low. We therefore use \ac{TuRBO} \cite{erikssonScalableGlobalOptimization2020}, which builds trust regions with lightweight \acp{GP}.

We find that the number of constraints poses an even greater challenge than dimensionality. 
We attribute this to the feasible hypervolume shrinking multiplicatively with each additional constraint, especially for equality constraints with tight tolerances. For our five-constraint problem, standard \ac{TuRBO} is effectively intractable. We address this by using an inner gradient descent (\ac{iGD}) step for equality constraints with known convex structure ($C_{wx}$, $C_{di}$, $C_{lift}$), yielding {\ac{TuRBO}-\ac{iGD}}. This speeds up convergence because \ac{TuRBO} only manages the feasibility of the remaining constraints, and the search dimensionality is reduced (22 to 15) as some variables are handled by \ac{iGD}.

\begin{figure}[htbp]
    \centering
    \includegraphics[width=0.85\columnwidth]{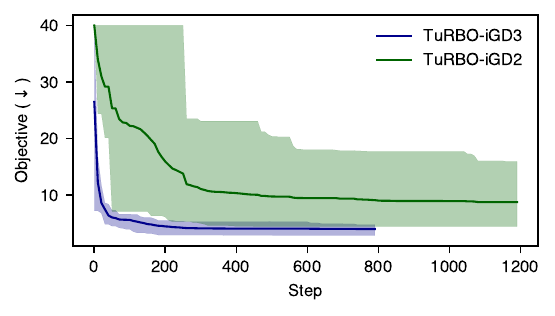}
    \caption[Effect of inner gradient descent on convergence for TuRBO-iGD]{Effect of inner gradient descent on convergence: \ac{TuRBO}-\ac{iGD}3 converges substantially faster than \ac{TuRBO}-\ac{iGD}2. Pure \ac{TuRBO} rarely reaches feasibility.}
    \label{fig:turbo_igd_constraints}
\end{figure}

Figure \ref{fig:turbo_igd_constraints} illustrates the impact of this hybrid approach. Moving from two equality constraints handled by \ac{iGD} (\ac{TuRBO}-\ac{iGD}2) to all three (\ac{TuRBO}-\ac{iGD}3) markedly improves convergence. Conversely, pure \ac{TuRBO} without inner gradient descent rarely reaches feasibility.

To implement \ac{TuRBO}-\ac{iGD}, we decouple inequality constraints $C_{\text{box-pl}}$ and $C_{\text{bb}}$, which are handled in the outer loop (\ac{TuRBO}) from equality constraints ($C_{\text{lift}}$, $C_{\text{wx}}$, $C_{\text{di}}$) in the inner loop (\ac{iGD}).

\begin{itemize}
    \item \textit{Outer Variables ($\mathbf{x}$, TuRBO):} Shape latents ($z_W, z_F, z_I$) and internals placement parameters $x_I$.
    \item \textit{Inner Variables ($\mathbf{y}$, iGD):} Wing placement parameters $x_{W,pl}$ (position and dihedral angle) and lift requirements $L_{req,1}, L_{req,2}$.
\end{itemize}

Wing coordinates are reset to a fixed location before every iGD step regardless of fuselage shape. The solver must then position them to valid attachment points.
We perform 8,000 iterations without auto-stopping, relying on trust region restarts to escape local optima (as implemented by default in the \ac{TuRBO} library).

The objective function $f$ minimized by TuRBO (line 3 of Algorithm \ref{alg:turbo_igd}) combines the mass $O_{\text{mass}}$ with soft penalties for the inequality constraints $C_{\text{box-pl}}$ and $C_{\text{bb}}$, which are not handled by the inner loop. 
We define $f = O_{\text{mass}} \cdot \psi(C_{\text{box-pl}}) \cdot \psi(C_{\text{bb}})$, where $\psi(C) = 1 + a_1 \tanh(C) + a_2 C$ imposes a steep penalty for violation. 
We use hand-tuned hyperparameters $a_1 = 5$ and $a_2 = 0.1$ in our experiments.

\begin{algorithm}[htbp]
    \caption{TuRBO-iGD Algorithm}
    \label{alg:turbo_igd}
    \begin{algorithmic}
    \State Initialize TuRBO state and hyperparameters
    \While{iterations $<$ max iterations}
        \State $\mathbf{x} \gets \text{TuRBO}(\text{GPs})$ \Comment{Sample candidate}
        \State Initialize inner variables $\mathbf{y}$
        \Repeat \Comment{Enter inner loop}
            \State $O_p, C_k \gets \mathcal{G}(\mathbf{x}, \mathbf{y})$ \Comment{Evaluate geometry}
            \State Update $\mathbf{y}$ via gradient descent on $C_{\text{lift}}, C_{\text{di}}, C_{\text{wx}}$
        \Until{$C_{\text{lift}}, C_{\text{di}}, C_{\text{wx}} \le \text{tol}$}
        \State $f \gets O_{\text{mass}} \cdot \psi(C_{\text{box-pl}}) \cdot \psi(C_{\text{bb}})$ \Comment{Compute penalized objective}
        \State Update GPs with $(\mathbf{x}, f)$
    \EndWhile
    \end{algorithmic}
\end{algorithm}

%% file: tables/optimizer_hyperparameters.tex
\begin{table*}[htbp]
    \centering
    \small
    \caption[Hyperparameters for optimization algorithms]{Hyperparameters for optimization algorithms.}
    \label{tab:optimizer_hyperparameters}
    \begin{tabular}{lc}
    \toprule
    \multicolumn{2}{c}{\textbf{\ac{ALM}-GD}} \\
    \midrule
    \textbf{Parameter} & \textbf{Value} \\
    \midrule
    Batch Size                      & 5             \\
    Max Iterations                  & 10\,000       \\
    Initial Sampling                & Sobol         \\
    \midrule
    Primal \ac{LR}                       & 2.0e-3        \\
    \ac{LR} Decay                        & 0.99          \\
    Dual \ac{LR} ($\gamma$)              & 1.0e-2        \\
    Constraint Cap                  & 2.0           \\
    Gradient Clipping               & Disabled      \\
    \midrule
    Patience (epochs)               & 50            \\
    Min. Improvement                & 0.02          \\
    Constraint Tolerance            & 1.0e-4        \\
    \bottomrule
    \end{tabular}
    \hspace{1cm}
    \begin{tabular}{lc}
    \toprule
    \multicolumn{2}{c}{\textbf{BayesOpt (TuRBO-iGD)}} \\
    \midrule
    \textbf{Parameter} & \textbf{Value} \\
    \midrule
    Trust Regions ($m$)             & 5             \\
    Max Iterations                  & 8\,000       \\
    Initial Sampling                & LHS           \\
    Init. Points per Region         & 10            \\
    Acquisition Batch ($q$)         & 1             \\
    \midrule
    Trust Region Length Min         & $0.5^7$       \\
    Trust Region Length Max         & 1.6           \\
    \midrule
    \textit{Inner GD:}  &               \\
    \quad Max Steps                 & 8             \\
    \quad Convergence Tol.          & 1.0e-4        \\
    \bottomrule
    \end{tabular}
\end{table*}

%% file: appendix/ddjm.tex
\section[Details on Data-Driven GLUE Models]{Details on Data-Driven \ac{GLUE} Models}
\label{ap:ddjm}

These models are set up to condition on velocity, altitude, internals mass, and internals volumes, but for the benchmarks and ablations in this work we train them at fixed (constant) conditions.

\subsection{{OT-GAN}} The \acs{OT-GAN} uses an entropic regularized optimal transport distance loss \cite{genevayLearningGenerativeModels2017} implemented with the geomloss package \cite{feydyInterpolatingOptimalTransport}:
\begin{equation}
\mathcal{L}_{\text{OT}} = W_{\varepsilon}(G_{\theta}(\mathbf{z}), \mathbf{x}_{\text{real}})
\end{equation}
where the blur parameter $\varepsilon(t)$ is annealed from $0.1 \to 0.02$ using a cosine schedule. Thanks to this, there is no adversarial training, only a Sinkhorn loss. The model uses only positive (feasible) training samples, with no negative data. We employ the 2-Wasserstein distance ($p=2$).  No \ac{DPP} loss is used since training with optimal transport relieves \acsp{GAN}' typical propensity to collapse to a single mode.

\subsection{{MDD-GAN}} The \acs{MDD-GAN} \cite{regenwetterConstrainingGenerativeModels2024} distinguishes fake, feasible, and infeasible samples:
\begin{align}
\mathcal{L}_D &= \mathcal{L}_{\text{CE}}(\mathbf{D}(G(\mathbf{z})), 0) \nonumber \\
& + \mathcal{L}_{\text{CE}}(\mathbf{D}(\mathbf{x}_{\text{valid}}), 1) \nonumber \\
& + \mathcal{L}_{\text{CE}}(\mathbf{D}(\mathbf{x}_{\text{invalid}}), 2) \\
\mathcal{L}_G &= \mathcal{L}_{\text{CE}}(\mathbf{D}(G(\mathbf{z})), 1) + \lambda_{\text{div}} \cdot \mathcal{L}_{\text{DPP}}(G(\mathbf{z}))
\end{align}
The model uses a three-way classification scheme, assigning label 0 to fake samples, label 1 to feasible samples, and label 2 to infeasible samples. To prevent mode collapse, we incorporate a \ac{DPP} loss with weight $\lambda_{\text{div}} = 0.05$. This approach requires both positive and negative training samples.

\subsection{{cVAE}} \acs{cVAE} with binary feasibility labels:
\begin{equation}
\mathcal{L}_{\text{cVAE}} = \mathcal{L}_{\text{recon}} + \beta_{\text{KL}} \cdot \mathcal{L}_{\text{KL}}
\end{equation}
where $\mathcal{L}_{\text{recon}} = \mathbb{E}[\|\mathbf{x} - \hat{\mathbf{x}}\|_2^2]$ and $\mathcal{L}_{\text{KL}} = \mathbb{E}[-\frac{1}{2} \sum_{i}(1 + \log\sigma_{\text{enc},i}^2 - \mu_{\text{enc},i}^2 - \sigma_{\text{enc},i}^2)]$ with $\beta_{\text{KL}} = 0.3$. We augment the conditioning variables with a binary feasibility label such that $\mathbf{c}_{\text{aug}} = [\mathbf{c}, y]$ where $y \in \{0,1\}$. The model employs the reparameterization trick $\mathbf{z} = \boldsymbol{\mu}_{\text{enc}} + \boldsymbol{\sigma}_{\text{enc}} \odot \boldsymbol{\epsilon}$ where $\boldsymbol{\epsilon} \sim \mathcal{N}(\mathbf{0}, \mathbf{I})$ to enable backpropagation through the sampling process. At inference time, we only generate samples with the positive feasibility label ($y=1$). Similar to MDD-GAN, this model requires both positive and negative training samples.

\subsection{{DDPM}} The \acs{DDPM} uses denoising diffusion with cosine noise schedule \cite{hoDenoisingDiffusionProbabilistic2020}:
\begin{equation}
\mathcal{L}_{\text{DDPM}} = \mathbb{E}_{t, \mathbf{x}_0, \boldsymbol{\epsilon}}\left[\|\boldsymbol{\epsilon} - \boldsymbol{\epsilon}_{\theta}(\mathbf{x}_t, \mathbf{c}, t)\|_2^2\right]
\end{equation}
where $\mathbf{x}_t = \sqrt{\bar{\alpha}_t}\mathbf{x}_0 + \sqrt{1-\bar{\alpha}_t}\boldsymbol{\epsilon}$ with cosine schedule $\bar{\alpha}(t) = \cos^2\left(\frac{t/T + s}{1 + s} \cdot \frac{\pi}{2}\right)$ and $s = 0.008$. The model uses a 128-dimensional sinusoidal time embedding to encode the diffusion timestep. We apply \ac{EMA} with rate $\gamma = 0.999$ to the model weights, which provides more stable sampling behavior. At inference time, the model applies 1000 denoising steps to generate samples. Unlike \acs{MDD-GAN} and \acs{cVAE}, this model uses only positive (feasible) training samples.

\subsection{Model Capacity}

All \ac{DD}-\ac{GLUE} models are based on fully connected architectures and are set up to use either deep \ac{MLP} blocks (with skip connections every 6 layers) 
or wide residual blocks (WRBs) with configurable width multipliers. 
The latter are similar to those used in the \acs{DF}-\acs{GLUE} architecture, which is described in more detail in \ref{app:df-architecture_details}.
Based on hyperparameter optimizations, we find that the \acs{DDPM} benefits from a wide residual block architecture.
The \acs{DDPM} thus uses 6 WRBs, which was found to be best during hyperparameter optimization.
The \acs{cVAE}, \acs{OT-GAN}, and \acs{MDD-GAN} use generic deep \ac{MLP} blocks with skip connections every 6 layers.
Table \ref{tab:hpo_architecture_comparison} summarizes the feasibility scores (mean of two worst across four seeds per trial) obtained for each model and architecture choice during hyperparameter optimization.

\begin{table*}[t]
    \centering
    \footnotesize
    \caption{Best robust feasibility scores (feasibility assessed as mean of 2 worst across 4 seeds per trial) obtained via HPO for different generative models and architectures.}
    \label{tab:hpo_architecture_comparison}
    \renewcommand{\arraystretch}{1.2}{
        \begin{tabular}{llccc}
        \toprule
        Model  & Architecture      & HPO Trials & Seeds / Trial & Best Feasibility ($\uparrow$) \\
        \hline
        \acs{cVAE}   & Deep \ac{MLP}          & 40         & 4             & 48.63 \% \\
        \acs{cVAE}   & Wide ResBlocks        & 50         & 4             & \textbf{51.95 \%} \\
        \hline
        \acs{DDPM}   & Deep \ac{MLP}          & 50         & 4             & 70.70 \% \\
        \acs{DDPM}   & Wide ResBlocks        & 50         & 4             & \textbf{77.34 \%} \\
        \hline
        \acs{MDD-GAN} & Deep \ac{MLP}        & 60         & 4             & \textbf{69.14 \%} \\
        \acs{MDD-GAN} & Wide ResBlocks      & 50         & 4             & 66.02 \% \\
        \hline
        \acs{OT-GAN}  & Deep \ac{MLP}        & 50         & 4             & \textbf{52.34 \%} \\
        \acs{OT-GAN}  & Wide ResBlocks      & 50         & 4             & 49.22 \% \\
        \bottomrule
        \end{tabular}
    }
\end{table*}

\subsection{Hyperparameters}
\input{tables/sota_hyperparameters.tex}

Hyperparameter Optimization (HPO) is carried out for all data-driven \ac{GLUE} models using the ax library \cite{olson2025ax} to ensure the relative difference in quality between the data-free and data-driven approach is not due to badly tuned hyperparameters.
Table \ref{tab:sota_hyperparameters} shows the hyperparameters for the data-driven \ac{GLUE} models. Hyperparameters optimized using HPO are shown in bold. The augmentation sampling factor $N$ is capped at $N \le 20$ to limit \ac{GPU} memory use.

As an alternative augmentation strategy, we considered storing pre-convergence samples along the \ac{ALM}-\ac{GD} trajectory. With this method, however, tiny steps near convergence produce many near-duplicates and thereby collapse diversity.

%% file: tables/sota_hyperparameters.tex
\begin{table*}[htbp]
    \centering
    \footnotesize
    \renewcommand{\arraystretch}{1.1}
    \caption[Hyperparameters for data-driven generative models]{Hyperparameters for data-driven generative models. Shown in bold are those hyperparameters optimized using \ac{HPO}.}
    \label{tab:sota_hyperparameters}
    \begin{tabular}{lcccc}
    \toprule
    \textbf{Parameter} & \textbf{OT-\acs{GAN}} & \textbf{\acs{MDD-GAN}} & \textbf{\acs{cVAE}} & \textbf{\acs{DDPM}} \\
    \midrule
    Architecture                   & Deep MLP & Deep MLP & Deep MLP & Wide Residual \\
    Latent Dimension ($d_z$)       & 4        & 4        & 4        & ---      \\
    Hidden Dimension               & 620      & 620      & 280      & 512      \\
    Hidden Layers (G/Enc)          & 12       & 12       & 8        & 12       \\
    Hidden Layers (D/Dec)          & ---      & 12       & 8        & ---      \\
    Model Capacity (Params)         & 7M   & 14.8M    & 10.4M    & 54M + 54M EMA     \\
    \quad \textit{\footnotesize Breakdown}  & ---  & \textit{\footnotesize G: 7M, D: 7.8M}  & \textit{\footnotesize E: 5.8M, D: 4.6M}  & ---  \\
    Learning Rate (G/Model)        & \textbf{1.20e-4}  & \textbf{5.50e-5}  & \textbf{2.00e-4}  & \textbf{4.50e-5}  \\
    Learning Rate (D)              & ---      & \textbf{9.20e-5}  & ---      & ---      \\
    Batch Size                     & 1024     & 1024     & 1024     & 1024     \\
    Epochs                         & 500      & 500      & 500      & 500      \\
    Noise Mode                     & Uniform  & Uniform  & Gaussian & Gaussian \\
    Latent Scale                   & 3.5      & 3.5      & ---      & ---      \\
    Augmentation Noise Scale $\sigma_{aug}$ & \textbf{5.80e-2}  & \textbf{1.00e-2}  & \textbf{2.00e-2}  & \textbf{1.48e-1}  \\
    Augmentation Sampling Factor $N$ & \textbf{5} & \textbf{8} & \textbf{5} & \textbf{10} \\
    \ac{KL} Weight                      & ---      & ---      & \textbf{0.3}      & ---      \\
    Diversity Weight ($\lambda_{\text{div}}$) & --- & \textbf{0.05} & --- & --- \\
    Time Embedding Dimension       & ---      & ---      & ---      & 128      \\
    Denoising Steps                & ---      & ---      & ---      & 1000     \\
    \ac{EMA} Rate                       & ---      & ---      & ---      & 0.999    \\
    Optimal Transport Blur ($\varepsilon$)        & 0.1 $\to$ 0.02 & --- & --- & --- \\
    Discriminator Outputs          & ---      & 3        & ---      & ---      \\
    \bottomrule
    \end{tabular}
\end{table*}

%% file: appendix/geometry_layer.tex
\section{Differentiable Geometry Layer}
\label{ap:geometry_layer}

This appendix provides technical details for the batched parametric geometry layer used in \ac{UAV} design generation. 
It represents complete aircraft configurations through a modular, differentiable parametric representation comprising fuselage segments (cubic B\'ezier curves and bicubic B\'ezier surfaces, see Figure \ref{fig:fuselage_parametrization}), wings (scaled and extruded airfoil profiles), and internal payload boxes. 
The layer computes key metrics including frontal area, surface (wetted) area, and volume for fuselage drag and mass estimates, as well as wing induced drag. 
The mass estimate heuristically combines wing moments and fuselage surface area, while fuselage drag follows OpenVSP's Hoerner drag \cite{Hoerner1965Drag,kimOPENVSPBASEDAERODYNAMIC}. 
Additionally, the layer performs validity checks for self-intersection, convexity, wing-fuselage interface quality, and containment of internal components ($\Cboxpl$). 
These evaluations produce the outputs described in Equation \ref{eq:geometry_layer_output}. 
The implementation enables \textit{batch processing} while maintaining full differentiability for gradient-based optimization by treating discrete indexing operations as constants during backpropagation and using numerically stable formulations in PyTorch \cite{paszkePyTorchImperativeStyle2019}.

\subsection{Overview}

\begin{figure*}[h!]
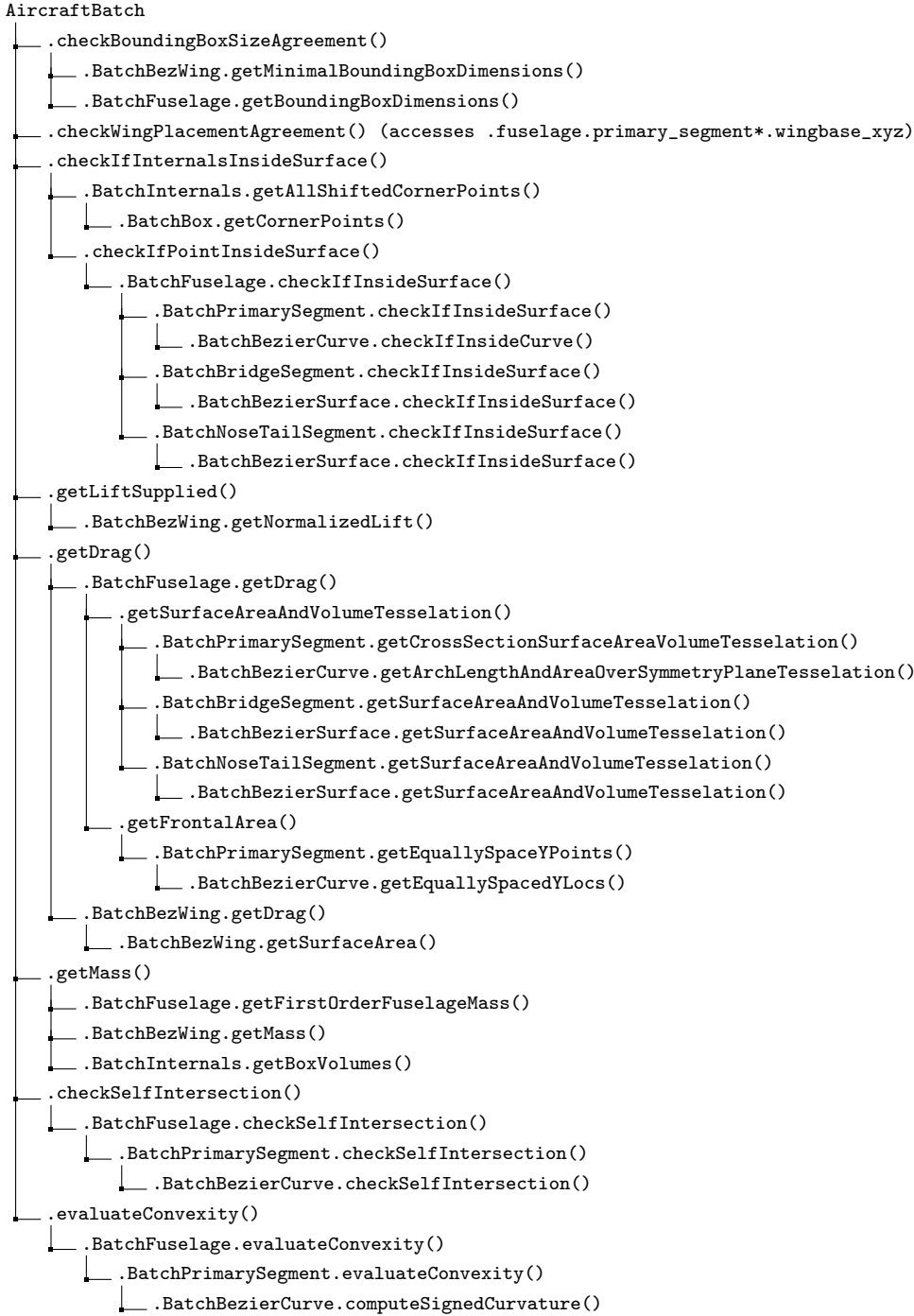

    \footnotesize
    \dirtree{%
    .1 AircraftBatch.
    .2 .checkBoundingBoxSizeAgreement().
    .3 .BatchBezWing.getMinimalBoundingBoxDimensions().
    .3 .BatchFuselage.getBoundingBoxDimensions().
    .2 .checkWingPlacementAgreement() (accesses .fuselage.primary\_segment*.wingbase\_xyz).
    .2 .checkIfInternalsInsideSurface().
    .3 .BatchInternals.getAllShiftedCornerPoints().
    .4 .BatchBox.getCornerPoints().
    .3 .checkIfPointInsideSurface().
    .4 .BatchFuselage.checkIfInsideSurface().
    .5 .BatchPrimarySegment.checkIfInsideSurface().
    .6 .BatchBezierCurve.checkIfInsideCurve().
    .5 .BatchBridgeSegment.checkIfInsideSurface().
    .6 .BatchBezierSurface.checkIfInsideSurface().
    .5 .BatchNoseTailSegment.checkIfInsideSurface().
    .6 .BatchBezierSurface.checkIfInsideSurface().
    .2 .getLiftSupplied().
    .3 .BatchBezWing.getNormalizedLift().
    .2 .getDrag().
    .3 .BatchFuselage.getDrag().
    .4 .getSurfaceAreaAndVolumeTesselation().
    .5 .BatchPrimarySegment.getCrossSectionSurfaceAreaVolumeTesselation().
    .6 .BatchBezierCurve.getArchLengthAndAreaOverSymmetryPlaneTesselation().
    .5 .BatchBridgeSegment.getSurfaceAreaAndVolumeTesselation().
    .6 .BatchBezierSurface.getSurfaceAreaAndVolumeTesselation().
    .5 .BatchNoseTailSegment.getSurfaceAreaAndVolumeTesselation().
    .6 .BatchBezierSurface.getSurfaceAreaAndVolumeTesselation().
    .4 .getFrontalArea().
    .5 .BatchPrimarySegment.getEquallySpaceYPoints().
    .6 .BatchBezierCurve.getEquallySpacedYLocs().
    .3 .BatchBezWing.getDrag().
    .4 .BatchBezWing.getSurfaceArea().
    .2 .getMass().
    .3 .BatchFuselage.getFirstOrderFuselageMass().
    .3 .BatchBezWing.getMass().
    .3 .BatchInternals.getBoxVolumes().
    .2 .checkSelfIntersection().
    .3 .BatchFuselage.checkSelfIntersection().
    .4 .BatchPrimarySegment.checkSelfIntersection().
    .5 .BatchBezierCurve.checkSelfIntersection().
    .2 .evaluateConvexity().
    .3 .BatchFuselage.evaluateConvexity().
    .4 .BatchPrimarySegment.evaluateConvexity().
    .5 .BatchBezierCurve.computeSignedCurvature().
    }
    \caption[Differentiable geometry layer functionality overview]{Differentiable geometry layer functionality overview.}
    \label{fig:geometry_layer_overview}
\end{figure*}

Figure \ref{fig:geometry_layer_overview} shows the high-level overview of the differentiable geometry layer's evaluation functionality. Some of these functions are used in the constraint checks and the objective function for \ac{DF}-\ac{GLUE} training and optimization algorithms. Others are used to train our subsystem models.

\subsection{Fuselage Surface Area and Volume Computation}

Computing geometric properties of parametric surfaces requires numerical integration. We employ adaptive tessellation where bicubic B\'ezier patches are uniformly sampled on an $(n \times n)$ grid in parameter space $(u,v) \in [0,1]^2$. Each resulting quadrilateral cell's area is approximated via the cross product of its edge vectors. Volume is computed by projecting these elements onto the $xz$-plane and integrating, exploiting the fuselage's bilateral symmetry about the $xz$-plane.

The bicubic B\'ezier surface is defined as:
\begin{equation}
\mathbf{B}(u,v) = \sum_{i=0}^{3} \sum_{j=0}^{3} B_i^3(u) B_j^3(v) \mathbf{P}_{ij}
\end{equation}
where $B_i^3(t)$ are cubic Bernstein polynomials and $\mathbf{P}_{ij} \in \mathbb{R}^{B \times 3}$ are control points for batch size $B$. In matrix form: $\mathbf{B}(u,v) = \mathbf{u}^T M \mathbf{P} M^T \mathbf{v}$, where $M$ is the B\'ezier basis matrix.

All operations are fully differentiable, utilizing only standard tensor operations (cross products, norms, summations). The method is vectorized across the batch dimension, enabling efficient parallel processing.

\begin{algorithm}[htbp]
\caption[Batched surface area and volume computation]{Batched Surface Area and Volume Computation}
\label{alg:batched_surface_area_volume}
\begin{algorithmic}[1]
\Require B\'ezier control points $\mathbf{P} \in \mathbb{R}^{B \times 4 \times 4 \times 3}$, tessellation resolution $n$
\Ensure Surface area $S \in \mathbb{R}^B$, volume $V \in \mathbb{R}^B$
\State $\mathbf{u}, \mathbf{v} \gets \mathrm{linspace}(0, 1, n)$ \Comment{Parameter grid}
\State Initialize $S \gets \mathbf{0}_B$, $V \gets \mathbf{0}_B$
\For{$i = 0$ to $n-1$}
    \For{$j = 0$ to $n-1$}
        \State $\mathbf{A} \gets \mathrm{BezierSurface}(\mathbf{P}, u_i, v_j)$ \Comment{Shape: $[B, 3]$}
        \State $\mathbf{B} \gets \mathrm{BezierSurface}(\mathbf{P}, u_{i+1}, v_j)$
        \State $\mathbf{C} \gets \mathrm{BezierSurface}(\mathbf{P}, u_i, v_{j+1})$
        \State $\mathbf{d}_1 \gets \mathbf{B} - \mathbf{A}$, \quad $\mathbf{d}_2 \gets \mathbf{C} - \mathbf{A}$ \Comment{Edge vectors}
        \State $\mathbf{n} \gets \mathbf{d}_1 \times \mathbf{d}_2$ \Comment{Cross product in $\mathbb{R}^3$}
        \State $S \mathrel{+}= \|\mathbf{n}\|_2$ \Comment{Accumulate parallelogram area}
        \State $V \mathrel{+}= |\mathbf{n}_y| \cdot \frac{|\mathbf{B}_y + \mathbf{C}_y|}{2}$ \Comment{Volume via $y$-projection}
    \EndFor
\EndFor
\State \Return $S, V$
\end{algorithmic}
\end{algorithm}

\textbf{Implementation notes:} Typical resolution is $n=10$, yielding $(10-1)^2 = 81$ parallelogram elements per surface patch. The volume calculation projects each surface element onto the $xz$-plane by taking the $y$-component of the normal vector, then multiplies by the average $y$-coordinate of the element vertices. We only list the algorithm for bicubic area and volume computation here; for the primary segments, which are extruded cubic B\'ezier curves, we apply essentially the same approach but in 1D. Instead of \ac{2D} patches, we tessellate 1D line segments along the curve and compute cross-sectional properties accordingly.

\subsection{Fuselage Frontal Area Computation}

Frontal area (cross-sectional area in the $yz$-plane) is critical for drag estimation but cannot be computed via simple tessellation. The fuselage comprises multiple overlapping segments (primary segments, bridge, nose), and we require the \textit{outer envelope} formed by their union. Our approach: (1) determine vertical extent $[z_{\min}, z_{\max}]$ from segment control points, (2) generate $N$ equally-spaced horizontal slices at $z$-coordinates $\mathbf{z}_e$, (3) interpolate $y$-coordinates along each segment's B\'ezier curves at these $z$-values, (4) take the maximum $y$ across all segments to obtain the outer envelope, and (5) integrate using the trapezoidal rule. For simplification, the current implementation approximates this envelope using only the primary fuselage segments.

The critical challenge is \textit{differentiable interpolation}. Given query points $\mathbf{z}_e$ not necessarily on the B\'ezier curve's natural parameterization $t \in [0,1]$, we must invert the relationship $z = B_z(t)$ to find corresponding $y = B_y(t)$ values. We employ a two-stage approach: dense sampling of the B\'ezier curve followed by piecewise linear interpolation with careful gradient handling.

\begin{algorithm}[htbp]
\caption[Batched frontal area computation]{Batched Frontal Area Computation}
\label{alg:batched_frontal_area}
\begin{algorithmic}[1]
\Require Primary segment B\'ezier curves $\{\mathbf{C}_k\}_{k=1}^{K}$, resolution $N$
\Ensure Frontal area $A_f \in \mathbb{R}^B$
\State $z_{\max} \gets \max_{k}\left(\mathbf{C}_k^{\mathrm{roof}}, \mathbf{C}_k^{\mathrm{bb\text{-}top}}\right)$ \Comment{Per-batch maximum $[B]$}
\State $z_{\min} \gets \min_{k}\left(\mathbf{C}_k^{\mathrm{floor}}, \mathbf{C}_k^{\mathrm{bb\text{-}bottom}}\right)$ \Comment{Per-batch minimum $[B]$}
\State $\mathbf{w} \gets \mathrm{linspace}(0,1,N)$ 
\State $\mathbf{z}_e \gets z_{\min} \odot (1-\mathbf{w}) + z_{\max} \odot \mathbf{w}$ \Comment{Broadcast to $[B, N]$}
\For{each segment $k \in \{1, \ldots, K\}$}
    \State $\mathbf{y}_k \gets \mathrm{InterpolateBezier}(\mathbf{C}_k, \mathbf{z}_e)$ \Comment{See Algorithm~\ref{alg:bezier_interp}}
\EndFor
\State $\mathbf{y}_{\mathrm{outer}} \gets \max_{k}(\mathbf{y}_k)$ \Comment{Outer envelope $[B, N]$}
\State $\Delta z \gets \mathbf{z}_e[:, 1:] - \mathbf{z}_e[:, :-1]$ \Comment{Slice heights $[B, N-1]$}
\State $\mathbf{y}_{\mathrm{avg}} \gets \frac{\mathbf{y}_{\mathrm{outer}}[:, 1:] + \mathbf{y}_{\mathrm{outer}}[:, :-1]}{2}$ \Comment{Average widths}
\State $A_f \gets 2 \sum_{i=1}^{N-1} \Delta z_i \cdot \mathbf{y}_{\mathrm{avg}, i}$ \Comment{Factor 2 for symmetry}
\Return $A_f$
\end{algorithmic}
\end{algorithm}

\textbf{Gradient handling:} The \texttt{searchsorted} operation is discrete. For backpropagation, the returned indices are treated as constants (zero gradient); gradients flow only through the values gathered and used in the subsequent linear interpolation. Small epsilon clamping ($\epsilon = 10^{-12}$) in the denominator prevents division by zero for perfectly horizontal curve segments while maintaining numerical stability. Typical parameters: $r=50$, $N=20$.

\begin{algorithm}[H]
\caption[Differentiable B\'ezier curve interpolation]{Differentiable B\'ezier Curve Interpolation at Specified $z$-Coordinates}
\label{alg:bezier_interp}
\begin{algorithmic}[1]
\Require B\'ezier curve $\mathbf{C}$, query $z$-coordinates $\mathbf{z}_e \in \mathbb{R}^{B \times N}$, sampling resolution $r$
\Ensure Interpolated $y$-coordinates $\mathbf{y}_e \in \mathbb{R}^{B \times N}$
\State $\mathbf{t} \gets \mathrm{linspace}(0, 1, r)$ \Comment{Dense parameter sampling}
\State $\mathbf{y}_c, \mathbf{z}_c \gets \mathrm{BezierCurve}(\mathbf{C}, \mathbf{t})$ \Comment{Evaluate curve, shapes $[B, r]$}
\State $\mathrm{idx}_{\mathrm{sort}} \gets \mathrm{argsort}(\mathbf{z}_c, \dim=1)$ \Comment{Sort by $z$-coordinate}
\State $\mathbf{z}_c \gets \mathbf{z}_c[\mathrm{idx}_{\mathrm{sort}}]$, \quad $\mathbf{y}_c \gets \mathbf{y}_c[\mathrm{idx}_{\mathrm{sort}}]$ \Comment{Monotonic in $z$}
\State $\mathrm{idx}_r \gets \mathrm{searchsorted}(\mathbf{z}_c, \mathbf{z}_e)$ \Comment{Binary search $[B,n]$}
\State $\mathrm{idx}_l \gets \mathrm{clamp}(\mathrm{idx}_r - 1, 0, r-2)$ \Comment{Left bracket indices}
\State $\mathrm{idx}_r \gets \mathrm{clamp}(\mathrm{idx}_r, 1, r-1)$ \Comment{Right bracket indices}
\State $z_l, z_r \gets \mathbf{z}_c[\mathrm{idx}_l], \mathbf{z}_c[\mathrm{idx}_r]$ \Comment{Gather bracketing values}
\State $y_l, y_r \gets \mathbf{y}_c[\mathrm{idx}_l], \mathbf{y}_c[\mathrm{idx}_r]$
\State $t_{\mathrm{lerp}} \gets \frac{\mathbf{z}_e - z_l}{\max(z_r - z_l, \epsilon)}$ \Comment{$\epsilon = 10^{-12}$ prevents NaN gradients}
\State $\mathbf{y}_e \gets (1-t_{\mathrm{lerp}}) \odot y_l + t_{\mathrm{lerp}} \odot y_r$ \Comment{Linear interpolation}
\State $\mathbf{y}_e \gets \mathbf{y}_e \odot \mathbb{1}[0 \leq t_{\mathrm{lerp}} \leq 1]$ \Comment{Zero out-of-bounds queries}
\Return $\mathbf{y}_e$
\end{algorithmic}
\end{algorithm}

\subsection{Ray-Casting Inside/Outside Test}

Determining whether payload boxes violate the fuselage envelope requires testing potentially thousands of points simultaneously (for more complex subsystems with intricate geometries and many components; currently we use rectangular boxes with 8 corner points each to test). We employ a vectorized ray-casting algorithm: for each query point $\mathbf{p}$, we cast a ray from a known interior origin $\mathbf{o}$ to $\mathbf{p}$ and compute the first intersection with the fuselage surface. If an intersection exists with parameter $t \in (0,1]$, the signed distance indicates how far the point penetrates the surface.

The geometric problem reduces to solving the parametric intersection system:
\begin{equation}
\mathbf{o} + t \mathbf{d} = \mathbf{S}(u,v)
\end{equation}
where $\mathbf{d} = \mathbf{p} - \mathbf{o}$ is the ray direction, $\mathbf{S}(u,v)$ is a B\'ezier surface patch, and we seek $(t, u, v)$. For efficiency, we tessellate the surface into linear segments (reducing bicubic patches to piecewise bilinear approximations), yielding a series of $2 \times 2$ linear systems that can be solved in parallel.

For linearly extruded surfaces like primary segments, ray-casting is simplified to \ac{2D} intersection tests in the yz-plane (setting $x_{\mathbf{o}} = x_{\mathbf{p}}$). For general bicubic surfaces (nose, bridge), we cast rays with a zero x-direction component but perform full \ac{3D} Möller-Trumbore intersection against the tessellated surface triangles.

\begin{algorithm}[H]
\caption[Batched ray-casting inside/outside test]{Batched Ray-Casting Inside/Outside Test (\ac{2D} Simplified View)}
\begin{algorithmic}[1]
\Require Query points $\mathbf{P} \in \mathbb{R}^{B \times N \times 3}$, B\'ezier surface $\mathbf{S}$, resolution $n$
\Ensure Signed distances $\mathbf{d} \in \mathbb{R}^{B \times N}$ (positive = outside, 0 = on surface)
\State $\mathbf{O} \gets \mathrm{GetInteriorOrigin}(\mathbf{P})$ \Comment{Known-inside point $[B,N,3]$}
\State $\mathbf{a} \gets \mathbf{P}_{yz} - \mathbf{O}_{yz}$ \Comment{Ray direction (YZ-plane only) $[B,N,2]$}
\State $\mathbf{u} \gets \mathrm{linspace}(0,1,n)$ \Comment{Tessellation parameters}
\State $\mathbf{A}, \mathbf{B} \gets \mathrm{TessellateSurface}(\mathbf{S}, \mathbf{u})$ \Comment{$[B, n\!-\!1, 3]$ endpoints}
\State $\mathbf{c} \gets \mathbf{B}_{yz} - \mathbf{A}_{yz}$ \Comment{Segment directions $[B, n\!-\!1, 2]$}
\State $\mathbf{b} \gets \mathbf{O}_{yz}^{(1)} - \mathbf{A}_{yz}^{(2)}$ \Comment{Broadcast to $[B, n\!-\!1, N, 2]$}
\State \Comment{\textit{Solve:} $\mathbf{O} + t\mathbf{a} = \mathbf{A} + s\mathbf{c}$ for $(t,s)$ via Cramer's rule}
\State $\mathrm{den} \gets \mathbf{a}_y^{(1)} \mathbf{c}_x^{(2)} - \mathbf{a}_x^{(1)} \mathbf{c}_y^{(2)}$ \Comment{Determinant $[B, n\!-\!1, N]$}
\State $\mathrm{num} \gets \mathbf{a}_y^{(1)} \mathbf{b}_x - \mathbf{b}_y \mathbf{a}_x^{(1)}$ 
\State $s \gets \begin{cases} \mathrm{num}/(\mathrm{den} + \epsilon) & |\mathrm{den}| > \epsilon \\ -1 & \text{otherwise} \end{cases}$ \Comment{$\epsilon = 10^{-6}$}
\State $\mathrm{valid}_s \gets (0 \leq s \leq 1) \land (|\mathrm{den}| > \epsilon)$ \Comment{On segment?}
\State $t \gets \begin{cases} (\mathbf{c}_x^{(2)} s - \mathbf{b}_x)/(\mathbf{a}_x^{(1)} + \epsilon) & \mathrm{valid}_s \\ 0 & \text{otherwise} \end{cases}$
\State $\mathrm{valid} \gets \mathrm{valid}_s \land (t > 0)$ \Comment{Ray goes forward}
\State $\mathrm{idx}_{\mathrm{first}} \gets \mathrm{argmax}_{\dim=1}(\mathrm{valid})$ \Comment{First valid intersection per ray}
\State $t_{\mathrm{hit}} \gets t[\mathrm{idx}_{\mathrm{first}}]$ \Comment{Gather intersection parameters}
\State $\mathrm{has\_hit} \gets \mathrm{any}_{\dim=1}(\mathrm{valid})$
\State $\mathbf{d} \gets \|\mathbf{P} - \mathbf{O}\|_2 \odot (1 - t_{\mathrm{hit}}) \odot \mathrm{has\_hit}$ \Comment{Signed distance}
\Return $\mathbf{d}$
\end{algorithmic}
\end{algorithm}

\textbf{Notation:} Superscripts $(1), (2)$ denote broadcasting dimensions for clarity. The {\small\texttt{GetInteriorOrigin}} function computes a guaranteed interior point by linearly interpolating between known segment boundaries at the query point's $x$-coordinate.

\textbf{Gradient flow:} The \texttt{argmax} on boolean masks is discrete. For backpropagation, the returned index is treated as a constant; gradients flow through the $t$ value gathered at that index. Division-by-zero protection via small $\epsilon$ ensures bounded gradients.

\subsection{Differentiable Box Overlap Detection}

Training the autoregressive internal box generator requires a differentiable penalty for overlapping boxes. We compute pairwise overlapping volumes using the ReLU activation to smoothly handle non-overlapping cases (where overlap = 0). For axis-aligned boxes, two boxes overlap if and only if they overlap along all three axes. The overlap extent along each axis is given by:
\begin{equation}
\ell_\alpha = \max\left(0, \frac{d_{i,\alpha} + d_{j,\alpha}}{2} + \delta - |c_{i,\alpha} - c_{j,\alpha}|\right)
\end{equation}
where $d_{i,\alpha}$ is the dimension of box $i$ along axis $\alpha$, $c_{i,\alpha}$ is its center coordinate, and $\delta$ is a safety margin encouraging spacing. The total overlap volume is $\mathcal{O}_{ij} = \ell_x \ell_y \ell_z$.

\begin{algorithm}[H]
\caption[Differentiable box overlap computation]{Differentiable Box Overlap Computation}
\begin{algorithmic}[1]
\Require Box centers $\mathbf{c} \in \mathbb{R}^{B \times K \times 3}$, dimensions $\mathbf{d} \in \mathbb{R}^{B \times K \times 3}$, margin $\delta$
\Ensure Overlap matrix $\mathbf{O} \in \mathbb{R}^{B \times K \times K}$
\State $\mathbf{c}_i \gets \mathbf{c}[:, :, \mathrm{None}, :]$ \Comment{Broadcast to $[B,K,1,3]$}
\State $\mathbf{c}_j \gets \mathbf{c}[:, \mathrm{None}, :, :]$ \Comment{Broadcast to $[B,1,K,3]$}
\State $\mathbf{d}_i, \mathbf{d}_j \gets$ \textit{analogous broadcasting}
\For{$\alpha \in \{x, y, z\}$ (index 0, 1, 2)}
    \State $\ell_\alpha \gets \mathrm{ReLU}\left(\frac{\mathbf{d}_{i,\alpha} + \mathbf{d}_{j,\alpha}}{2} + \delta - |\mathbf{c}_{i,\alpha} - \mathbf{c}_{j,\alpha}|\right)$
\EndFor
\State $\mathbf{O} \gets \ell_x \odot \ell_y \odot \ell_z$ \Comment{Element-wise product $[B,K,K]$}
\State Set diagonal: $\mathbf{O}[:, i, i] \gets 0$ for all $i$ \Comment{Remove self-overlap}
\Return $\mathbf{O}$
\end{algorithmic}
\end{algorithm}

\subsection{Payload Arrangement Convex Hull View Area}

For volumetric efficiency scoring used in training of the box arrangement autoregressive model, we compute the convex hull of all internal box corners projected onto \ac{2D} views (frontal: $yz$, side: $xz$, top: $xy$). This provides a tight envelope of the usable internal volume. The challenge is implementing a \ac{2D} convex hull algorithm that is fully differentiable.

We employ a brute-force $O(n^3)$ approach based on the following geometric property: an edge $(i,j)$ is on the convex hull if and only if all other points $k$ lie on the same side of the line through $i$ and $j$. This is tested via the cross product sign. Points identified as being on the hull are then sorted counter-clockwise by angle from their centroid, and area is computed via the shoelace formula.

\begin{algorithm}[htbp]
\caption[Differentiable 2D convex hull area]{Differentiable \ac{2D} Convex Hull Area}
\begin{algorithmic}[1]
\Require \ac{2D} points $\mathbf{p} \in \mathbb{R}^{B \times N \times 2}$ (pre-sorted lexicographically, deduplicated)
\Ensure Convex hull area $A \in \mathbb{R}^B$
\State $\mathbf{p}_i \gets \mathbf{p}[:, :, \mathrm{None}, \mathrm{None}, :]$ \Comment{Broadcast to $[B,N,1,1,2]$}
\State $\mathbf{p}_j \gets \mathbf{p}[:, \mathrm{None}, :, \mathrm{None}, :]$ \Comment{Broadcast to $[B,1,N,1,2]$}
\State $\mathbf{p}_k \gets \mathbf{p}[:, \mathrm{None}, \mathrm{None}, :, :]$ \Comment{Broadcast to $[B,1,1,N,2]$}
\State $\mathbf{v}_{ij} \gets \mathbf{p}_j - \mathbf{p}_i$, \quad $\mathbf{v}_{ik} \gets \mathbf{p}_k - \mathbf{p}_i$ 
\State $\mathrm{cross} \gets \mathbf{v}_{ij,x} \mathbf{v}_{ik,y} - \mathbf{v}_{ij,y} \mathbf{v}_{ik,x}$ \Comment{\ac{2D} cross product $[B,N,N,N]$}
\State $\mathrm{all\_geq} \gets \mathrm{all}_{\dim=3}(\mathrm{cross} \geq 0)$ \Comment{All points right of edge}
\State $\mathrm{all\_leq} \gets \mathrm{all}_{\dim=3}(\mathrm{cross} \leq 0)$ \Comment{All points left of edge}
\State $\mathrm{edge}_{ij} \gets \mathrm{all\_geq} \lor \mathrm{all\_leq}$ \Comment{Valid hull edge $[B,N,N]$}
\State $\mathrm{edge}[:, i, i] \gets \mathrm{False}$ for all $i$ \Comment{Remove self-edges}
\State $\mathrm{on\_hull}_i \gets \mathrm{any}_{\dim=2}(\mathrm{edge}_{ij})$ \Comment{Point on hull if any edge $[B,N]$}
\State $n_h \gets \sum_i \mathrm{on\_hull}_i$ \Comment{Hull size per batch element}
\State $\mathbf{c} \gets \frac{1}{\max(n_h, 1)} \sum_{i: \mathrm{on\_hull}_i} \mathbf{p}_i$ \Comment{Centroid $[B,2]$}
\State $\theta \gets \mathrm{atan2}(\mathbf{p}_y - \mathbf{c}_y, \mathbf{p}_x - \mathbf{c}_x)$ \Comment{Angles from centroid}
\State $\theta[\neg \mathrm{on\_hull}] \gets +\infty$ \Comment{Sort non-hull points to end}
\State $\mathrm{idx}_{\mathrm{sort}} \gets \mathrm{argsort}(\theta, \dim=1)$ \Comment{Counter-clockwise order}
\State $\mathbf{p}_{\mathrm{hull}} \gets \mathbf{p}[\mathrm{idx}_{\mathrm{sort}}]$ \Comment{Sorted hull points (padded)}
\State \Comment{\textit{Shoelace formula with wrapping:}}
\State $\mathbf{x}, \mathbf{y} \gets \mathbf{p}_{\mathrm{hull},x}, \mathbf{p}_{\mathrm{hull},y}$
\State $A \gets \frac{1}{2} \left| \sum_{i=0}^{n_h-1} (x_i y_{(i+1) \bmod n_h} - x_{(i+1) \bmod n_h} y_i) \right|$ 
\Return $A$
\end{algorithmic}
\end{algorithm}

\textbf{Complexity and gradient handling:} The \texttt{argsort} operation is discrete; gradients bypass the discrete operation and flow through the gathered point coordinates. Boolean operations (\texttt{all}, \texttt{any}) are treated as non-differentiable gates in the forward pass, with gradients flowing through the point coordinates in the backward pass. Variable hull sizes across the batch are handled via zero-padding and masking in the shoelace summation.

\textbf{Relationship to frontal area:} For internal boxes, the convex hull area in the $yz$-plane (frontal view) provides a lower bound on usable payload volume. This differs from fuselage frontal area (Algorithm~\ref{alg:batched_frontal_area}), which computes the outer envelope for drag estimation.

\subsection{Non-Differentiable Feasibility Checks}

While the aforementioned methods are differentiable and used in training objectives, we also employ several discrete checks for design validation and filtering.

\paragraph{Convexity via Curvature Analysis}

To encourage smooth, manufacturable fuselage cross-sections, we check convexity of the primary segment B\'ezier curves. The signed curvature of a planar curve $\mathbf{B}(t) = [y(t), z(t)]$ is:
\begin{equation}
\kappa(t) = \frac{y'(t)z''(t) - y''(t)z'(t)}{(y'(t)^2 + z'(t)^2)^{3/2}}
\end{equation}
A perfectly convex curve has $\kappa(t)$ of constant sign. We sample $\kappa$ at 50 uniformly spaced $t \in [0,1]$ and count sign changes:
\begin{equation}
\mathrm{ConvexityViolation} = \sum_{i=1}^{49} \mathbb{1}[\mathrm{sign}(\kappa_{i}) \neq \mathrm{sign}(\kappa_{i-1})]
\end{equation}
This metric is an integer and non-differentiable, but provides interpretable feedback. Designs with $\mathrm{ConvexityViolation} > 0$ exhibit inflection points, which can lead to poor aerodynamic performance or manufacturing difficulties. This check is applied only to the primary fuselage segments, as bridge and nose segments inherently have controlled curvature variations.

\paragraph{Self-Intersection Detection}

Self-intersecting B\'ezier curves are geometrically invalid. To detect self-intersection, we solve the algebraic system $\mathbf{B}(t_1) = \mathbf{B}(t_2)$ for $t_1 \neq t_2$. This is formulated as a small linear system derived from the curve's control points, which is then solved efficiently using normal equations (\texttt{torch.linalg.solve}). The solution yields candidate parameter pairs $(t_1, t_2)$ which are then tested to ensure they fall within $[0, 1]$ and are distinct.
This procedure is implemented using batch matrix operations but returns a boolean mask indicating self-intersection presence. It is used for post-hoc filtering rather than training objectives.

%% file: references.bib
@inproceedings{arjovskyWassersteinGAN2017,
  title = {Wasserstein Generative Adversarial Networks},
  author = {Arjovsky, Martin and Chintala, Soumith and Bottou, L{\'e}on},
  booktitle = {International Conference on Machine Learning},
  pages = {214--223},
  year = {2017},
  organization = {PMLR},
  doi = {10.48550/arXiv.1701.07875}
}

@article{basirAdaptiveAugmentedLagrangian2023,
  title = {An Adaptive Augmented Lagrangian Method for Training Physics and Equality Constrained Artificial Neural Networks},
  author = {Basir, Shamsulhaq and Senocak, Inanc},
  year = {2023},
  doi = {10.48550/arXiv.2306.04904},
  langid = {english},
}

@article{benamaraMultifidelityPODSurrogateassisted2017,
  title = {Multi-Fidelity POD Surrogate-Assisted Optimization: Concept and Aero-Design Study},
  author = {Benamara, Tariq and Breitkopf, Piotr and Lepot, Ingrid and Sainvitu, Caroline and Villon, Pierre},
  year = {2017},
  journal = {Structural and Multidisciplinary Optimization},
  volume = {56},
  number = {6},
  pages = {1387--1412},
  doi = {10.1007/s00158-017-1730-4},
  langid = {english},
}

@misc{berzinsGeometryInformedNeuralNetworks2024,
  title = {Geometry-Informed Neural Networks},
  author = {Berzins, Arturs and Radler, Andreas and Volkmann, Eric and Sanokowski, Sebastian and Hochreiter, Sepp and Brandstetter, Johannes},
  year = {2025},
  doi = {10.48550/arXiv.2402.14009}
}

@article{chenBezierGANAutomaticGeneration2021,
  title = {Airfoil Design Parameterization and Optimization Using B{\'e}zier Generative Adversarial Networks},
  author = {Chen, Wei and Chiu, Kevin and Fuge, Mark D.},
  journal = {AIAA Journal},
  volume = {58},
  number = {11},
  pages = {4723--4735},
  year = {2020},
  publisher = {American Institute of Aeronautics and Astronautics},
  doi = {10.2514/1.J059317},
}

@inproceedings{chenInfoGANInterpretableRepresentation2016,
  title = {InfoGAN: Interpretable Representation Learning by Information Maximizing Generative Adversarial Nets},
  author = {Chen, Xi and Duan, Yan and Houthooft, Rein and Schulman, John and Sutskever, Ilya and Abbeel, Pieter},
  booktitle = {Advances in Neural Information Processing Systems},
  volume = {29},
  year = {2016},
  doi = {10.48550/arXiv.1606.03657}
}

@article{chenMOPaDGANReparameterizingEngineering2021,
  title = {MO-PaDGAN: Reparameterizing Engineering Designs for Augmented Multi-Objective Optimization},
  author = {Chen, Wei and Ahmed, Faez},
  year = {2021},
  journal = {Applied Soft Computing},
  volume = {113},
  doi = {10.1016/j.asoc.2021.107909},
}

@article{chenPaDGANLearningGenerate2020,
  title = {PaDGAN: Learning to Generate High-Quality Novel Designs},
  author = {Chen, Wei and Ahmed, Faez},
  year = {2020},
  journal = {Journal of Mechanical Design},
  volume = {143},
  number = {031703},
  doi = {10.1115/1.4048626},
}

@article{chenSynthesizingDesignsInterpart2019,
  title = {Synthesizing Designs with Interpart Dependencies Using Hierarchical Generative Adversarial Networks},
  author = {Chen, Wei and Fuge, Mark},
  year = {2019},
  journal = {Journal of Mechanical Design},
  volume = {141},
  number = {11},
  doi = {10.1115/1.4044076},
  langid = {english},
}

@article{cobbDesignUnmannedAir2022,
  title = {Design of Unmanned Air Vehicles Using Transformer Surrogate Models},
  author = {Cobb, Adam D. and Roy, Anirban and Elenius, Daniel and Jha, Susmit},
  year = {2022},
  doi = {10.48550/arXiv.2211.08138},
  langid = {english},
  pubstate = {prepublished},
}

@article{cobbDiverseSystemLevelDesign2022,
  title = {On Diverse System-Level Design Using Manifold Learning and Partial Simulated Annealing},
  author = {Cobb, A. and Roy, A. and Elenius, D. and Koneripalli, K. and Jha, S.},
  year = {2022},
  journal = {Proceedings of the Design Society},
  volume = {2},
  pages = {1541--1548},
  doi = {10.1017/pds.2022.156},
  langid = {english},
}

@inproceedings{CompressingLatentSpace2023,
title={Compressing Latent Space via Least Volume},
author={Qiuyi Chen and Mark Fuge},
booktitle={The Twelfth International Conference on Learning Representations},
year={2024},
doi={10.48550/arXiv.2404.17773}
}

@InProceedings{GDPPLearningDiverseGenerations2019,
  title = 	 {{GDPP}: Learning Diverse Generations using Determinantal Point Processes},
  author =       {Elfeki, Mohamed and Couprie, Camille and Riviere, Morgane and Elhoseiny, Mohamed},
  booktitle = 	 {Proceedings of the 36th International Conference on Machine Learning},
  pages = 	 {1774--1783},
  year = 	 {2019},
  volume = 	 {97},
  series = 	 {Proceedings of Machine Learning Research},
  month = 	 jun,
  publisher =    {PMLR},
  doi = 	 {10.48550/arXiv.1812.00068},
}

@article{debFastElitistMultiobjective2002,
  title = {A Fast and Elitist Multiobjective Genetic Algorithm: {NSGA-II}},
  author = {Deb, K. and Pratap, A. and Agarwal, S. and Meyarivan, T.},
  date = {2002-04},
  journaltitle = {IEEE Transactions on Evolutionary Computation},
  shortjournal = {IEEE Trans. Evol. Computat.},
  volume = {6},
  number = {2},
  pages = {182--197},
  doi = {10.1109/4235.996017},
  langid = {english},
}

@inproceedings{dinizOptimizingDiffusionDiffuse2024,
  title = {Optimizing Diffusion to Diffuse Optimal Designs},
  booktitle = {{{AIAA SCITECH}} 2024 {{Forum}}},
  author = {Diniz, Cashen and Fuge, Mark},
  year = {2024},
  publisher = {{American Institute of Aeronautics and Astronautics}},
  location = {Orlando, FL},
  doi = {10.2514/6.2024-2013},
  eventtitle = {{{AIAA SCITECH}} 2024 {{Forum}}},
  isbn = {978-1-62410-711-5},
  langid = {english},
}

@inproceedings{dongOneMoreContextual2025,
  title = {From One to More: Contextual Part Latents for 3D Generation},
  author = {Dong, Shaocong and Ding, Lihe and Chen, Xiao and Li, Yaokun and Wang, Yuxin and Wang, Yucheng and Wang, Qi and Kim, Jaehyeok and Gao, Chenjian and Huang, Zhanpeng and Wang, Zibin and Xue, Tianfan and Xu, Dan},
  booktitle = {Proceedings of the IEEE/CVF International Conference on Computer Vision (ICCV)},
  year = {2025},
  doi = {10.48550/arXiv.2507.08772}
}

@article{economonSU2OpenSourceSuite2016,
  title = {SU2: An Open-Source Suite for Multiphysics Simulation and Design},
  author = {Economon, Thomas D. and Palacios, Francisco and Copeland, Sean R. and Lukaczyk, Trent W. and Alonso, Juan J.},
  year = {2016},
  journal = {AIAA Journal},
  volume = {54},
  number = {3},
  pages = {828--846},
  doi = {10.2514/1.J053813},
  langid = {english},
}

@inproceedings{erikssonHighDimensionalBayesianOptimization2021,
  title = {High-Dimensional Bayesian Optimization with Sparse Axis-Aligned Subspaces},
  author = {Eriksson, David and Jankowiak, Martin},
  booktitle = {Uncertainty in Artificial Intelligence},
  pages = {493--503},
  year = {2021},
  organization = {PMLR},
  doi = {10.48550/arXiv.2103.00349}
}

@article{erikssonScalableGlobalOptimization2020,
  title = {Scalable Global Optimization via Local Bayesian Optimization},
  author = {Eriksson, David and Pearce, Michael and Gardner, Jacob and Turner, Ryan D. and Poloczek, Matthias},
  journal = {Advances in Neural Information Processing Systems},
  volume = {32},
  year = {2019},
  doi = {10.48550/arXiv.1910.01739}
}

@article{feltenEngiBenchFrameworkDataDriven2025,
  title = {EngiBench: A Framework for Data-Driven Engineering Design Research},
  author = {Felten, Florian and Apaza, Gabriel and Br{\"a}unlich, Gerhard and Diniz, Cashen and Dong, Xuliang and Drake, Arthur and Habibi, Milad and Hoffman, Nathaniel J. and Keeler, Matthew and Massoudi, Soheyl and VanGessel, Francis G. and Fuge, Mark},
  year = {2025},
  doi = {10.48550/arXiv.2508.00831},
  langid = {english},
  pubstate = {prepublished},
}

@software{ferretti_accelerated_optimization_2025,
  author = {Filippo Luca Ferretti and Diego Ferigo and Carlotta Sartore and Alessandro Croci and Omar G. Younis and Silvio Traversaro and Daniele Pucci},
  title = {Hardware-Accelerated Morphology Optimization via Physically Consistent Differentiable Simulation},
  year = {2025},
  url = {https://github.com/ami-iit/jaxsim}
}

@inproceedings{todorov2012mujoco,
  title = {MuJoCo: A Physics Engine for Model-Based Control},
  author = {Todorov, Emanuel and Erez, Tom and Tassa, Yuval},
  booktitle = {2012 IEEE/RSJ International Conference on Intelligent Robots and Systems},
  pages = {5026--5033},
  year = {2012},
  organization = {IEEE},
  doi = {10.1109/IROS.2012.6386109},
}

@article{regenwetter2023framed,
title = {FRAMED: An {AutoML} Approach for Structural Performance Prediction of Bicycle Frames},
journal = {Computer-Aided Design},
volume = {156},
pages = {103446},
year = {2023},
doi = {https://doi.org/10.1016/j.cad.2022.103446},
author = {Lyle Regenwetter and Colin Weaver and Faez Ahmed},
}

@inproceedings{feydyInterpolatingOptimalTransport,
  title = {Interpolating Between Optimal Transport and MMD Using Sinkhorn Divergences},
  author = {Feydy, Jean and S{\'e}journ{\'e}, Thibault and Vialard, Fran{\c{c}}ois-Xavier and Amari, Shun-ichi and Trouv{\'e}, Alain and Peyr{\'e}, Gabriel},
  booktitle = {The 22nd International Conference on Artificial Intelligence and Statistics},
  pages = {2681--2690},
  year = {2019},
  organization = {PMLR},
  langid = {english},
  doi = {10.48550/arXiv.1810.08278}
}

@inproceedings{genevayLearningGenerativeModels2017,
  title = {Learning Generative Models with Sinkhorn Divergences},
  author = {Genevay, Aude and Peyr{\'e}, Gabriel and Cuturi, Marco},
  booktitle = {International Conference on Artificial Intelligence and Statistics},
  pages = {1608--1617},
  year = {2018},
  organization = {PMLR},
  doi = {10.48550/arXiv.1706.00292}
}

@article{goodfellowGenerativeAdversarialNetworks2014,
  title = {Generative Adversarial Networks},
  author = {Goodfellow, Ian and Pouget-Abadie, Jean and Mirza, Mehdi and Xu, Bing and Warde-Farley, David and Ozair, Sherjil and Courville, Aaron and Bengio, Yoshua},
  journal = {Communications of the ACM},
  volume = {63},
  number = {11},
  pages = {139--144},
  year = {2020},
  publisher = {ACM New York, NY, USA},
  doi = {10.1145/3422622},
}

@article{heAerodynamicDesignOptimization2018,
  title = {An Aerodynamic Design Optimization Framework Using a Discrete Adjoint Approach with OpenFOAM},
  author = {He, Ping and Mader, Charles A. and Martins, Joaquim R.R.A. and Maki, Kevin J.},
  date = {2018-05},
  journaltitle = {Computers \& Fluids},
  shortjournal = {Computers \& Fluids},
  volume = {168},
  pages = {285--303},
  doi = {10.1016/j.compfluid.2018.04.012},
  langid = {english},
}

@article{heDAFoamOpenSourceAdjoint2020,
  title = {DAFoam: An Open-Source Adjoint Framework for Multidisciplinary Design Optimization with OpenFOAM},
  author = {He, Ping and Mader, Charles A. and Martins, Joaquim R. R. A. and Maki, Kevin J.},
  date = {2020-03},
  journaltitle = {AIAA Journal},
  shortjournal = {AIAA Journal},
  volume = {58},
  number = {3},
  pages = {1304--1319},
  doi = {10.2514/1.J058853},
  langid = {english},
}

@article{heyraninobariRANGEGANDesignSynthesis2021,
  title = {RANGE-GAN: Design Synthesis Under Constraints Using Conditional Generative Adversarial Networks},
  author = {Heyrani Nobari, Amin and Chen, Wei and Ahmed, Faez},
  year = {2021},
  journal = {Journal of Mechanical Design},
  pages = {1--16},
  doi = {10.1115/1.4052442},
}

@article{hoDenoisingDiffusionProbabilistic2020,
  title = {Denoising Diffusion Probabilistic Models},
  author = {Ho, Jonathan and Jain, Ajay and Abbeel, Pieter},
  journal = {Advances in Neural Information Processing Systems},
  volume = {33},
  pages = {6840--6851},
  year = {2020},
  doi = {10.48550/arXiv.2006.11239}
}

@inproceedings{karaliDesignDeepLearning2020,
  title = {Design of a Deep Learning Based Nonlinear Aerodynamic Surrogate Model for {UAVs}},
  booktitle = {{{AIAA Scitech}} 2020 {{Forum}}},
  author = {Karali, Hasan and Demirezen, Mustafa U. and Yukselen, Mahmut A. and Inalhan, Gokhan},
  year = {2020},
  publisher = {{American Institute of Aeronautics and Astronautics}},
  location = {Orlando, FL},
  doi = {10.2514/6.2020-1288},
  eventtitle = {{{AIAA Scitech}} 2020 {{Forum}}},
  isbn = {978-1-62410-595-1},
}

@inproceedings{kimOPENVSPBASEDAERODYNAMIC,
  title = {{OpenVSP} Based Aerodynamic Design Optimization Tool Building Method and Its Application to Tailless {UAV}},
  author = {Kim, Ju-Hoe and Tsuchiya, Takeshi},
  booktitle = {Proceedings of the 33rd Congress of the International Council of the Aeronautical Sciences (ICAS)},
  year = {2022},
  address = {Stockholm, Sweden},
}

@inproceedings{kingmaAutoEncodingVariationalBayes2022,
  title = {Auto-Encoding Variational Bayes},
  author = {Kingma, Diederik P. and Welling, Max},
  booktitle = {International Conference on Learning Representations},
  year = {2014},
  doi = {10.48550/arXiv.1312.6114}
}

@incollection{Kroo1997LargeScaleMDO,
  author = {Ilan M. Kroo},
  title = {MDO for Large-Scale Design},
  booktitle = {Multidisciplinary Design Optimization: State of the Art},
  pages = {22--44},
  year = {1997},
  publisher = {SIAM},
  address = {Philadelphia, PA},
  isbn = {978-0898713596},
}

@article{kontogiannisGeneralizedMethodologyMultidisciplinary2020,
  title = {A Generalized Methodology for Multidisciplinary Design Optimization Using Surrogate Modelling and Multifidelity Analysis},
  author = {Kontogiannis, Spyridon G. and Savill, Mark A.},
  year = {2020},
  journal = {Optimization and Engineering},
  volume = {21},
  number = {3},
  pages = {723--759},
  doi = {10.1007/s11081-020-09504-z},
  langid = {english},
}

@article{lethamReExaminingLinearEmbeddings,
  title = {Re-Examining Linear Embeddings for High-Dimensional Bayesian Optimization},
  author = {Letham, Benjamin and Calandra, Roberto and Rai, Akshara and Bakshy, Eytan},
  journal = {Advances in Neural Information Processing Systems},
  volume = {33},
  pages = {1546--1558},
  year = {2020},
  doi = {10.48550/arXiv.2001.11659}
}

@inproceedings{mahapatraMultiTaskLearningUser,
  title = {Multi-Task Learning with User Preferences: Gradient Descent with Controlled Ascent in Pareto Optimization},
  author = {Mahapatra, Debabrata and Rajan, Vaibhav},
  booktitle = {International Conference on Machine Learning},
  pages = {6597--6607},
  year = {2020},
  organization = {PMLR},
  langid = {english},
}

@book{Hoerner1965Drag,
  title = {Fluid-Dynamic Drag: Practical Information on Aerodynamic Drag and Hydrodynamic Resistance},
  author = {Hoerner, Sighard F.},
  year = {1965},
  publisher = {Hoerner Fluid Dynamics},
  address = {Bricktown, NJ},
}

@inproceedings{olson2025ax,
  title = {Ax: A Platform for Adaptive Experimentation},
  author = {
    Olson, Miles and Santorella, Elizabeth and Tiao, Louis C. and
    Cakmak, Sait and Garrard, Mia and Daulton, Samuel and
    Lin, Zhiyuan Jerry  and Ament, Sebastian and Beckerman, Bernard and
    Onofrey, Eric and Igusti, Paschal and Lara, Cristian and
    Letham, Benjamin and Cardoso, Cesar and Shen, Shiyun Sunny and
    Lin, Andy Chenyuan and Grange, Matthew and Kashtelyan, Elena and
    Eriksson, David and Balandat, Maximilian and Bakshy, Eytan.
  },
  booktitle = {AutoML 2025 ABCD Track},
  year = {2025},
  url = {https://ax.dev/},
}

@article{martinsMultidisciplinaryDesignOptimization2013,
  title = {Multidisciplinary Design Optimization: A Survey of Architectures},
  author = {Martins, Joaquim R. R. A. and Lambe, Andrew B.},
  year = {2013},
  journal = {AIAA Journal},
  volume = {51},
  number = {9},
  pages = {2049--2075},
  doi = {10.2514/1.J051895},
  langid = {english},
}

@article{massoudiIntegratedApproachDesigning2024,
  title = {An {{Integrated Approach}} to {{Designing Robust Gas-Bearing Supported Turbocompressors Through Surrogate Modeling}} and {{Constrained All-At-Once Multi-Objective Optimization}}},
  author = {Massoudi, Soheyl and Picard, Cyril and Schiffmann, J{\"u}rg},
  year = {2024},
  journal = {Journal of Mechanical Design},
  volume = {146},
  number = {121706},
  doi = {10.1115/1.4065823},
}

@article{mazeDiffusionModelsBeat2023,
  title = {Diffusion Models Beat GANs on Topology Optimization},
  author = {Maz{\'e}, Fran{\c{c}}ois and Ahmed, Faez},
  year = {2023},
  journal = {Proceedings of the AAAI Conference on Artificial Intelligence},
  volume = {37},
  number = {8},
  pages = {9108--9116},
  doi = {10.1609/aaai.v37i8.26093},
  langid = {english},
}

@inproceedings{paraGenerativeLayoutModeling2020,
  title = {Generative Layout Modeling Using Constraint Graphs},
  author = {Para, Wamiq and Guerrero, Paul and Kelly, Tom and Guibas, Leonidas and Wonka, Peter},
  booktitle = {2021 IEEE/CVF International Conference on Computer Vision (ICCV)},
  year = {2021},
  pages = {6670--6680},
  doi = {10.1109/ICCV48922.2021.00662},
  organization = {IEEE},
  address = {Montreal, QC, Canada}
}

@inproceedings{parrottMachineLearningSurrogates2023,
  title = {Machine Learning Surrogates for Optimal {2D} Spatial Packaging of Interconnected Systems with Physics Interactions ({SPI2})},
  booktitle = {{{AIAA AVIATION}} 2023 {{Forum}}},
  author = {Parrott, Corey and Peddada, Satya and Allison, James T. and James, Kai},
  year = {2023},
  publisher = {{American Institute of Aeronautics and Astronautics}},
  location = {San Diego, CA and Online},
  doi = {10.2514/6.2023-4375},
  eventtitle = {{{AIAA AVIATION}} 2023 {{Forum}}},
  isbn = {978-1-62410-704-7},
  langid = {english},
}

@inproceedings{paszkePyTorchImperativeStyle2019,
  title = {PyTorch: An Imperative Style, High-Performance Deep Learning Library},
  author = {Paszke, Adam and Gross, Sam and Massa, Francisco and Lerer, Adam and Bradbury, James and Chanan, Gregory and Killeen, Trevor and Lin, Zeming and Gimelshein, Natalia and Antiga, Luca and Desmaison, Alban and K{\"o}pf, Andreas and Yang, Edward and DeVito, Zach and Raison, Martin and Tejani, Alykhan and Chilamkurthy, Sasank and Steiner, Benoit and Fang, Lu and Bai, Junjie and Chintala, Soumith},
  booktitle = {Proceedings of the 33rd International Conference on Neural Information Processing Systems},
  articleno = {721},
}

@article{pymoo,
  author = {J. {Blank} and K. {Deb}},
  journal = {IEEE Access},
  title = {pymoo: Multi-Objective Optimization in Python},
  year = {2020},
  volume = {8},
  number = {},
  pages = {89497-89509},
  doi = {10.1109/ACCESS.2020.2990567},
}

@article{regenwetterBikeBenchBicycleDesign2025,
  title = {Bike-Bench: A Bicycle Design Benchmark for Generative Models with Objectives and Constraints},
  author = {Regenwetter, Lyle and Obaideh, Yazan Abu and Chiotti, Fabien and Lykourentzou, Ioanna and Ahmed, Faez},
  year = {2025},
  doi = {10.48550/arXiv.2508.00830},
  langid = {english},
  pubstate = {prepublished}
}

@article{regenwetterConstrainingGenerativeModels2024,
  title = {Constraining Generative Models for Engineering Design with Negative Data},
  author = {Regenwetter, Lyle and Giannone, Giorgio and Srivastava, Akash and Gutfreund, Dan and Ahmed, Faez},
  year = {2024},
  journal = {Transactions on Machine Learning Research},
  year = {2024},
  doi = {10.48550/arXiv.2306.15166},
}

@article{regenwetterDeepGenerativeModels2022,
  title = {Deep Generative Models in Engineering Design: A Review},
  author = {Regenwetter, Lyle and Nobari, Amin Heyrani and Ahmed, Faez},
  journal = {Journal of Mechanical Design},
  volume = {144},
  number = {7},
  year = {2022},
  publisher = {American Society of Mechanical Engineers},
  doi = {10.48550/arXiv.2110.10863}
}

@article{sanchez-lengelingInverseMolecularDesign2018,
  title = {Inverse Molecular Design Using Machine Learning: Generative Models for Matter Engineering},
  author = {Sanchez-Lengeling, Benjamin and Aspuru-Guzik, Al{\'a}n},
  year = {2018},
  journal = {Science},
  volume = {361},
  number = {6400},
  pages = {360--365},
  doi = {10.1126/science.aat2663},
  langid = {english},
}

@incollection{sobieszczanski-sobieskiMultidisciplinaryDesignOptimization1995,
  title = {Multidisciplinary Design Optimization: An Emerging New Engineering Discipline},
  booktitle = {Advances in {{Structural Optimization}}},
  author = {Sobieszczanski-Sobieski, Jaroslaw},
  editor = {Herskovits, Jos{\'e}},
  editora = {Gladwell, G. M. L.},
  editoratype = {redactor},
  date = {1995},
  volume = {25},
  pages = {483--496},
  publisher = {Springer Netherlands},
  location = {Dordrecht},
  doi = {10.1007/978-94-011-0453-1_14},
  isbn = {978-94-010-4203-1 978-94-011-0453-1},
  langid = {english},
}

@dataset{sungBlendedNetBlendedWing2025,
  title = {BlendedNet: A Blended Wing Body Aircraft Dataset and Surrogate Model for Aerodynamic Predictions},
  author = {Sung, Nicholas and Spreizer, Steven and Elrefaie, Mohamed and Samuel, Kaira and Jones, Matthew C. and Ahmed, Faez},
  publisher = {Harvard Dataverse},
  year = {2025},
  version = {V1},
  doi = {10.7910/DVN/VJT9EP},
}

@article{taoApplicationDeepLearning2019,
  title = {Application of Deep Learning Based Multi-Fidelity Surrogate Model to Robust Aerodynamic Design Optimization},
  author = {Tao, Jun and Sun, Gang},
  year = {2019},
  journal = {Aerospace Science and Technology},
  volume = {92},
  pages = {722--737},
  doi = {10.1016/j.ast.2019.07.002},
  langid = {english},
}

@article{waltherIntegrationAspectsCollaborative2020,
  title = {Integration Aspects of the Collaborative Aero-Structural Design of an Unmanned Aerial Vehicle},
  author = {Walther, J.-N. and Gastaldi, A.-A. and Maierl, R. and Jungo, A. and Zhang, M.},
  year = {2020},
  journal = {CEAS Aeronautical Journal},
  volume = {11},
  number = {1},
  pages = {217--227},
  doi = {10.1007/s13272-019-00412-2},
  langid = {english},
}

@article{xueJAXFEMDifferentiableGPUaccelerated2023,
  title = {JAX-FEM: A Differentiable GPU-Accelerated 3D Finite Element Solver for Automatic Inverse Design and Mechanistic Data Science},
  author = {Xue, Tianju and Liao, Shuheng and Gan, Zhengtao and Park, Chanwook and Xie, Xiaoyu and Liu, Wing Kam and Cao, Jian},
  year = {2023},
  journal = {Computer Physics Communications},
  volume = {291},
  doi = {10.1016/j.cpc.2023.108802},
  langid = {english},
}

@article{yaoSurrogateBasedMultistagemultilevel2012,
  title = {A Surrogate Based Multistage-Multilevel Optimization Procedure for Multidisciplinary Design Optimization},
  author = {Yao, Wen and Chen, Xiaoqian and Ouyang, Qi and Van Tooren, Michel},
  year = {2012},
  journal = {Structural and Multidisciplinary Optimization},
  volume = {45},
  number = {4},
  pages = {559--574},
  doi = {10.1007/s00158-011-0714-z},
  langid = {english},
}

@article{zhangLibMOONGradientbasedMultiObjective2024,
  title = {LibMOON: A Gradient-Based Multi-Objective Optimization Library in PyTorch},
  author = {Zhang, Xiaoyuan and Zhao, Liang and Yu, Yingying and Lin, Xi and Chen, Yifan and Zhao, Han and Zhang, Qingfu},
  journal = {Advances in Neural Information Processing Systems},
  volume = {37},
  pages = {2026--2044},
  year = {2024},
  doi = {10.48550/arXiv.2409.02969}
}

@article{zuoFastSparseFlow2022,
  title = {Fast Sparse Flow Field Prediction Around Airfoils via Multi-Head Perceptron Based Deep Learning Architecture},
  author = {Zuo, Kuijun and Bu, Shuhui and Zhang, Weiwei and Hu, Jiawei and Ye, Zhengyin and Yuan, Xianxu},
  journal = {Aerospace Science and Technology},
  volume = {130},
  year = {2022},
  doi = {10.1016/j.ast.2022.107942},
}

@article{banovicAlgorithmicDifferentiationOpen2018,
  title = {Algorithmic Differentiation of the Open CASCADE Technology CAD Kernel and its Coupling with an Adjoint CFD Solver},
  author = {Banovi{\'c}, Mladen and Mykhaskiv, Orest and Auriemma, Salvatore and Walther, Andrea and Legrand, Herv{\'e} and M{\"u}ller, Jens-Dominik},
  journal = {Optimization Methods and Software},
  volume = {33},
  number = {4-6},
  pages = {813--828},
  year = {2018},
  doi = {10.1080/10556788.2018.1431235}
}

@inproceedings{cascaval2022differentiable,
  title = {Differentiable {3D} {CAD} programs for bidirectional editing},
  author = {Cascaval, Dan and Shalah, Mira and Quinn, Phillip and Bodik, Rastislav and Agrawala, Maneesh and Schulz, Adriana},
  booktitle = {Computer Graphics Forum},
  volume = {41},
  number = {2},
  pages = {309--323},
  year = {2022},
  organization = {Wiley Online Library},
  doi = {10.1111/cgf.14476}
}

@article{prasad2022nurbsdiff,
  author = {Deva Prasad, Anjana and Balu, Aditya and Shah, Harshil and Sarkar, Soumik and Hegde, Chinmay and Krishnamurthy, Adarsh},
  title = {NURBS-Diff: A Differentiable Programming Module for NURBS},
  journal = {Computer-Aided Design},
  volume = {146},
  year = {2022},
  doi = {10.1016/j.cad.2022.103199},
}

@inproceedings{banovic2024pythonocc_ad,
  author = {Banovi{\'c}, Mladen and Hafemann, Thomas and St{\"u}ck, Arthur},
  title = {Algorithmic Differentiation of the {pythonOCC} Geometric Modeling Library},
  booktitle = {ECCOMAS Congress 2024 (9th European Congress on Computational Methods in Applied Sciences and Engineering)},
  address = {Lisbon, Portugal},
  month = jun,
  year = {2024},
  doi = {10.23967/eccomas.2024.197},
}

@inproceedings{park2019deepsdf,
  author = {Park, Jeong Joon and Florence, Peter R. and Straub, Julian and Newcombe, Richard A. and Lovegrove, Steven},
  title = {DeepSDF: Learning Continuous Signed Distance Functions for Shape Representation},
  booktitle = {Proceedings of the IEEE/CVF Conference on Computer Vision and Pattern Recognition (CVPR)},
  year = {2019},
  pages = {165--174},
  doi = {10.1109/CVPR.2019.00025}
}

@inproceedings{hao2020dualsdf,
  author = {Hao, Zekun and Averbuch-Elor, Hadar and Snavely, Noah and Belongie, Serge J.},
  title = {DualSDF: Semantic Shape Manipulation Using a Two-Level Representation},
  booktitle = {Proceedings of the IEEE/CVF Conference on Computer Vision and Pattern Recognition (CVPR)},
  year = {2020},
  pages = {7628--7638},
  doi = {10.1109/CVPR42600.2020.00765}
}

@inproceedings{vasu2022hybridsdf,
  author = {Vasu, Subeesh and Talabot, Nicolas and Lukoianov, Artem and Baqué, Pierre and Donier, Jonathan and Fua, Pascal},
  title = {HybridSDF: Combining Deep Implicit Shapes and Geometric Primitives for 3D Shape Representation and Manipulation},
  booktitle = {2022 International Conference on 3D Vision (3DV)},
  year = {2022},
  pages = {617--626},
  doi = {10.1109/3DV57658.2022.00072}
}

@article{cramerProblemFormulationMultidisciplinary1994,
  author = {Cramer, Evin J. and Dennis, Jr., J. E. and Frank, Paul D. and Lewis, Robert Michael and Shubin, Gregory R.},
  title = {Problem Formulation for Multidisciplinary Optimization},
  journal = {SIAM Journal on Optimization},
  volume = {4},
  number = {4},
  pages = {754-776},
  year = {1994},
  doi = {10.1137/0804044},
}

@article{talabot2025partsdf,
  title        = {{PartSDF}: Part-based Implicit Neural Representation for Composite {3D} Shape Parametrization and Optimization},
  author       = {Talabot, Nicolas and Clerc, Olivier and Demirtas, Arda Cinar and Goujon, Alexis and Le, Hieu and Oner, Doruk and Fua, Pascal},
  year         = {2025},
  journal      = {arXiv preprint arXiv:2502.12985},
  doi          = {10.48550/arXiv.2502.12985},
}

@article{balmerBridgeVAE,
title = {Design Space Exploration and Explanation via Conditional Variational Autoencoders in Meta-Model-Based Conceptual Design of Pedestrian Bridges},
journal = {Automation in Construction},
volume = {163},
pages = {105411},
year = {2024},
doi = {https://doi.org/10.1016/j.autcon.2024.105411},
author = {Vera Balmer and Sophia V. Kuhn and Rafael Bischof and Luis Salamanca and Walter Kaufmann and Fernando Perez-Cruz and Michael A. Kraus},
}

@article{chandrasekhar2021tounn,
  title={TOuNN: topology optimization using neural networks},
  author={Chandrasekhar, Aaditya and Suresh, Krishnan},
  journal={Structural and Multidisciplinary Optimization},
  volume={63},
  number={3},
  pages={1135--1149},
  year={2021},
  publisher={Springer}
}

@article{joglekar2024dmf,
  title={DMF-TONN: direct mesh-free topology optimization using neural networks},
  author={Joglekar, Aditya and Chen, Hongrui and Kara, Levent Burak},
  journal={Engineering with Computers},
  volume={40},
  number={4},
  pages={2227--2240},
  year={2024},
  publisher={Springer}
}

@article{bagazinski2023shipgen,
  title = {{ShipGen}: A Diffusion Model for Parametric Ship Hull Generation with Multiple Objectives and Constraints},
  author = {Bagazinski, Noah J. and Ahmed, Faez},
  year = {2023},
  journal = {Journal of Marine Science and Engineering},
  volume = {11},
  number = {12},
  pages = {2215},
  doi = {10.3390/jmse11122215}
}

@inproceedings{xu2022skexgen,
  title = {{SkexGen}: Autoregressive Generation of {CAD} Construction Sequences with Disentangled Codebooks},
  author = {Xu, Xiang and Willis, Karl D.D. and Lambourne, Joseph G. and Cheng, Chin-Yi and Jayaraman, Pradeep Kumar and Furukawa, Yasutaka},
  booktitle = {International Conference on Machine Learning},
  year = {2022},
  volume = {162},
  pages = {24698--24724},
  organization = {PMLR},
  doi = {10.48550/arXiv.2207.04632}
}

@inproceedings{etesam2025gearformer,
  title = {Deep Generative Model for Mechanical System Configuration Design},
  author = {Etesam, Yasaman and Cheong, Hyunmin and Ataei, Mohammadmehdi and Jayaraman, Pradeep Kumar},
  booktitle = {AAAI Conference on Artificial Intelligence},
  year = {2025},
  doi = {10.1609/aaai.v39i16.33812}
}

@inproceedings{wu2021deepcad,
  title = {{DeepCAD}: A Deep Generative Network for Computer-Aided Design Models},
  author = {Wu, Rundi and Xiao, Chang and Zheng, Changxi},
  booktitle = {Proceedings of the IEEE/CVF International Conference on Computer Vision},
  year = {2021},
  doi = {10.1109/ICCV48922.2021.00670}
}

@inproceedings{zhao2020robogrammar,
  title = {{RoboGrammar}: Graph Grammar for Terrain-Optimized Robot Design},
  author = {Zhao, Allan and Xu, Jie and Kober, Mina and Matusik, Wojciech},
  booktitle = {ACM SIGGRAPH Asia},
  year = {2020},
  doi = {10.1145/3414685.3417831}
}

@inproceedings{xu2024brepgen,
  title = {{BrepGen}: A {B-rep} Generative Diffusion Model with Structured Latent Geometry},
  author = {Xu, Xiang and Lambourne, Joseph G. and Jayaraman, Pradeep Kumar and Wang, Zhengqing and Willis, Karl D.D. and Furukawa, Yasutaka},
  booktitle = {ACM SIGGRAPH Conference Proceedings},
  year = {2024},
  doi = {10.1145/3658129}
}

@inproceedings{tracy2023dcol,
  title = {Differentiable Collision Detection for a Set of Convex Primitives},
  author = {Tracy, Kevin and Manchester, Zachary},
  booktitle = {IEEE International Conference on Robotics and Automation},
  year = {2023},
  doi = {10.1109/ICRA48891.2023.10160716}
}

@article{hoffmanJacobianRegularization2019,
  title = {Robust Learning with {Jacobian} Regularization},
  author = {Hoffman, Judy and Roberts, Daniel A. and Yaida, Sho},
  year = {2019},
  doi = {10.48550/arXiv.1908.02729}
}

@inproceedings{higginsBetaVAE2017,
  title = {$\beta$-{VAE}: Learning Basic Visual Concepts with a Constrained Variational Framework},
  author = {Higgins, Irina and Matthey, Lo{\"i}c and Pal, Arka and Burgess, Christopher P. and Glorot, Xavier and Botvinick, Matthew M. and Mohamed, Shakir and Lerchner, Alexander},
  booktitle = {International Conference on Learning Representations},
  year = {2017}
}

@article{ataei2024xlb,
  title = {{XLB}: A Differentiable Massively Parallel Lattice {Boltzmann} Library in {Python}},
  author = {Ataei, Mohammadmehdi and Salehipour, Hesam},
  year = {2024},
  journal = {Computer Physics Communications},
  doi = {10.1016/j.cpc.2024.109187}
}
